\documentclass[a4paper,twoside,openright,11pt]{report}

\usepackage{amsfonts,amssymb,amsmath}
\usepackage{graphicx}
\usepackage{psfrag}
\usepackage{fancyhdr}
\usepackage{eurosym}
\usepackage{enumerate}
\usepackage{appendix}

\newcommand{\id}{\mathbb{I}}
\newcommand{\im}{\mathrm{i}}
\newcommand{\e}{\mathrm{e}}
\newcommand{\tr}{\mathrm{tr}}
\newcommand{\ket}[1]{\vert #1 \rangle}
\newcommand{\bra}[1]{\langle #1 \vert}
\newcommand{\braket}[2]{\langle #1 \vert #2 \rangle}
\newcommand{\ketbra}[2]{\vert #1 \rangle \langle #2 \vert}
\newcommand{\exvalue}[2]{\langle \hat{#1} \rangle_{#2}}
\newcommand{\sandwich}[3]{\langle #1 \vert \hat{#2} \vert #3 \rangle}
\newcommand{\comm}[2]{[\hat{#1},\hat{#2}]}
\newcommand{\acomm}[2]{\{\hat{#1},\hat{#2}\}}
\newcommand{\ppoison}[2]{\{\mathcal{#1},\mathcal{#2} \}_{pp}}
\newcommand{\var}[1]{\left(\Delta#1\right)^2}

\newtheorem{theorem}{Theorem}[section]
\newtheorem{lemma}{Lemma}[section]

\begin{document}

\pagenumbering{roman}

\thispagestyle{empty}

\begin{center}
 \includegraphics[height=3.5cm,width=5cm]{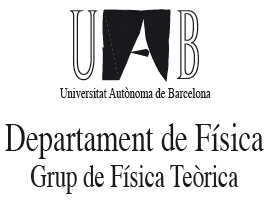}
\end{center}

\vspace{2cm}

\begin{center}
 {\LARGE {\bf Quantum Information with Continuous Variable systems}}\\
 \vspace{1cm}
 {\Large by}\\
 \vspace{1cm}
 {\Large Carles Rod\'o Sarr\'o}
\end{center}

\vspace{2cm}

\begin{center}
 Submitted in partial fulfillment\\
 to the requirements for the degree of\\
 \vspace{1cm}
 Doctor of Philosophy\\ 
 \vspace{0.5cm}
 at the\\
 \vspace{0.5cm}
 UNIVERSITAT AUT\`ONOMA DE BARCELONA\\
 08193 Bellaterra, Barcelona, Spain.\\
 April, 2010\\
 \vspace{1cm}
 Under the supervision of\\
 Prof. Anna Sanpera Trigueros
\end{center}

\newpage
\
\newpage


\vspace{8cm}

\begin{flushright}
 {\em ``Ethical axioms are found and tested not\\
 very differently from the axioms of science.\\
 Truth is what stands the test of experience''}.\\
 Albert Einstein.
\end{flushright}

\newpage
\
\newpage


\vspace{3cm}

\noindent {\Huge {\bf Acknowledgments}}\\

Han passat ja quasib\'e cinc anys. Durant aquest temps he tingut l'oportunitat de desenvolupar una important tasca docent al grup de F\'isica Te\`orica. Agraeixo en especial al Ram\'on Mu\~noz i a l'Eduard Mass\'o amb els qui he compartit la major part de la doc\`encia i amb els qui s'ha fet molt agradable treballar.

En segon lloc agraeixo a l'Albert Bramon l'entusiasme amb el que duia les seves classes de F\'isica Qu\`antica a la llicenciatura, aquella mat\`eria tan estramb\`otica... per\`o amb la que un s'encurioseix de seguida. Gr\`acies a l'Albert he tingut l'oportunitat de fer la recerca del doctorat amb l'Anna Sanpera. Qu\`e dir de l'Anna? Doncs una brillant investigadora entregada i molt involucrada no nom\'es en la recerca en aquest camp tan d'actualitat de la Informaci\'o Qu\`antica sin\'o en la seva difusi\'o a estudiants m\'es joves. Gr\`acies Anna, per la paci\`encia i per trobar sempre temps i esfor\c c en el meu treball.

Fer el doctorat amb l'Anna m'ha obert la possibilitat d'assistir i ``actuar'' a congressos d'arreu i aix\'i poder treballar amb investigadors de relev\`ancia. En particular amb el Gerardo Adesso de qui guardo un grat record, gener\'os i molt emprenedor amb ell gran part de la tesis he desenvolupat.

Tamb\'e vull recordar molt especialment als companys i ex-companys de despatx, amb ells he passat la major part del temps Javi, \`Alex, Marc, Simone i Joan Antoni; gr\`acies per les variopintes converses dins i fora la feina.

Finalment agraeixo a la fam\'ilia que m'ha recolzat en aquesta etapa, a la Mirka que ha canviat el meu rumb i al meu pare sobretot. Crec que es deu al teu afan de coneixement i als teves inquietuds que vaig comen\c car a estudiar F\'isica i aix\'i, \'es a tu a qui et dedico la Tesis.

\begin{flushright}
 Carles
\end{flushright}

\newpage


\noindent {\Large{\bf List of Publications}}\\

\noindent The articles published during the completion of the PhD which are enclosed in this thesis are:

\begin{enumerate}
\item {\sf Efficiency in Quantum Key Distribution Protocols with Entangled Gaussian States.}\\
C. Rod{\'o}, O. Romero-Isart, K. Eckert, and A. Sanpera.\\
{\em Pre-print version}: arXiv:quant-ph/0611277\\
{\em Journal-ref}: {\scshape Open Systems \& Information Dynamics \textbf{14}, 69 (2007)}

\item {\sf Operational Quantification of Continuous-Variable Correlations.}\\
C. Rod{\'o}, G. Adesso, and A. Sanpera.\\
{\em Pre-print version}: arXiv:0707:2811\\
{\em Journal-ref}: {\scshape Physical Review Letters \textbf{100}, 110505, (2008)}

\item {\sf Multipartite continuous-variable solution for the Byzantine agreement problem.}\\
R. Neigovzen, C. Rod{\'o}, G. Adesso, and A. Sanpera.\\
{\em Pre-print version}: arXiv:0712.2404\\
{\em Journal-ref}: {\scshape Physical Review A \textbf{77}, 062307, (2008)}

\item {\sf Manipulating mesoscopic multipartite entanglement with atom-light interfaces.}\\
J. Stasi{\'n}ska, C. Rod{\'o}, S. Paganelli, G. Birkl, and A. Sanpera.\\
{\em Pre-print version}: arXiv:0907.4261\\
{\em Journal-ref}: {\scshape Physical Review A \textbf{80}, 062304, (2009)}
\end{enumerate}

\noindent There is also one article in preparation which is within the framework of this PhD thesis:

\begin{enumerate}
\item {\sf A covariance matrix formalism for atom-light interfaces.}\\
J. Stasi{\'n}ska, S. Paganelli, C. Rod{\'o} and A. Sanpera.\\
{\em Journal-ref}: Submitted to {\scshape New Journal of Physics}
\end{enumerate}

\noindent Finally, one other article has been published during the PhD which is not related to the topic of this thesis:

\begin{enumerate}
\item {\sf Transport and entanglement generation in the Bose-Hubbard model.}\\
O. Romero-Isart, K. Eckert, C. Rod{\'o}, and A. Sanpera.\\
{\em Pre-print version}: quant-ph/0703177\\
{\em Journal-ref}: {\scshape Journal of Physics A: Mathematical and Theoretical \textbf{40}, 8019 (2007)}
\end{enumerate}

\newpage


\tableofcontents

\pagestyle{fancy}
\fancyhf{}
\renewcommand{\chaptermark}[1]{\markboth{\textbf{#1}}{}}
\renewcommand{\sectionmark}[1]{\markright{\textbf{\thesection. #1}}{}}
\fancyhead[LE,RO]{\bfseries\thepage}
\fancyhead[RE]{\bfseries\leftmark}
\fancyhead[LO]{\bfseries\rightmark}


\chapter{Introduction}

\pagenumbering{arabic}

The theory of Quantum Information emerges as an effort to generalize classical information theory into the quantum realm, by using the laws of quantum physics to encode, process and extract information. It was Rolf Landauer, known among other things for his remarkable contributions to the theory of electrical conductivity, who at the end of 1960 coined the idea that information rather than an abstract mathematical concept, is at the very end a physical process governed by physical laws. At the same time, Richard Feynman in ``Simulating Physics with Computers'' realized that in order to simulate intrinsically quantum physical processes, a classical computer will necessarily fail by its mere deterministic continuous nature while he showed how a quantum one would succeed efficiently in simulating and computing.  At the end of the 80s, David Deutsch in his seminal paper ``Quantum theory, the Church-Turing principle and the universal quantum computer'' set the ground for the theory of quantum information. He considered a universal classical Turing machine, which is the prototype of any classical computer, and propose the first universal quantum Turing machine paving the way towards quantum circuit modeling.

An algorithm is nothing else than a set of rules to solve efficiently a given problem in a finite number of steps. Problems are usually classified in different complexity classes. An important class of problems are the so-called, non-deterministic polynomial (NP) problems, whose efficient solution using classical computers grows exponentially with the input size of the problem. The cornerstone example of a NP problem is factorization. A big leap in the theory of Quantum Information was done by Peter Shor when he proposed in 1994 a quantum algorithm to factorize a prime number in an efficient way. Precisely the lack of an efficient classical solution to this problem sustains present cryptographic methods. The importance of his discovery, had led to a revolution on the emergent field of quantum information, involving for the first time non-academic interests as Quantum Cryptography with already successful implementations. Simultaneously, in 1995 Ignacio Cirac and Peter Zoller proposed the first realizable model of a quantum computer, using an array of trapped ions, an available technology since the beginning of the 1990. 

Since the end of the 1990, until now, very different models, involving solid states devices, quantum dots, quantum optics, ... have emerged as possible candidates to process information in a reliable and powerful way. At the same time, novel problems and solutions have arise by exploiting the properties of quantum laws in many different scenarios. Without the impressive experimental advance in manipulation of quantum systems, either using ultra-cold atoms, photons or ions, Quantum Information would have not reach the enormous current interest.\\

This thesis deals with the study of quantum communication protocols with Continuous Variable (CV) systems. Continuous Variable systems are those described by canonical conjugated coordinates $x$ and $p$ endowed with infinite dimensional Hilbert spaces, thus involving a complex mathematical structure. A special class of CV states, are the so-called Gaussian states. With them, it has been possible to implement certain quantum tasks as quantum teleportation, quantum cryptography and quantum computation with fantastic experimental success. The importance of Gaussian states is two-fold; firstly, its structural mathematical description makes them much more amenable than any other CV system. Secondly, its production, manipulation and detection with current optical technology can be done with a very high degree of accuracy and control. Nevertheless, it is known that in spite of their exceptional role within the space of all Continuous Variable states, in fact, Gaussian states are not always the best candidates to perform quantum information tasks. Thus non-Gaussian states emerge as potentially good candidates for communication and computation purposes. This dissertation is organized as follows.

In chapter~\ref{cvf}, we review the formalism of Continuous Variable systems focussing on Gaussian states. We show that Gaussian states admit an easy mathematical description based on phase-space Wigner functions as well as with covariance matrices. We introduce also the basic ingredients to describe multipartite systems with entanglement together with the most relevant well known results and finally we detail the description of light as a CV system.

In chapter~\ref{cryptography}, we present a protocol that permits to extract quantum keys from an entangled Continuous Variable system. Differently from discrete systems, Gaussian entangled states cannot be distilled with Gaussian operations only {\em i.e.} entangled Gaussian states are always bound entangled. However it was already shown, that it is still possible to extract perfectly correlated classical bits to establish secret random keys using an ``entanglement based'' approach. Differently from previous attempts where the realistic implementation was not considered, we properly modify the protocol using bipartite Gaussian entanglement to perform quantum key distribution in an efficient and realistic way. We describe and demonstrate security in front of different possible attacks on the communication, detailing the resources demanded while quantifying and relating the efficiency of the protocol with the entanglement shared between the parties involved. Our results are reported in \cite{Rodo2007OSYD}.

In chapter~\ref{byzantine}, we move to multipartite Gaussian states. There, we consider a simple 3-partite protocol known as Byzantine Agreement (detectable broadcast). The Byzantine Agreement is an old classical communication problem in which parties (with possible traitors among them) can only communicate pairwise, while trying to reach a common decision. Classically, there is a bound in the number of possible traitors that can be involved in the game if only classical secure channels are used. In the simplest case where three parties are involved, one of them being a traitor, no classical solution exists. Nevertheless, a quantum solution exist, {\em i.e.} letting a traitor being involved and using as a fundamental resource multipartite entanglement it is permitted to reach a common agreement. We demonstrate that detectable broadcast is also solvable within Continuous Variable using multipartite entangled Gaussian states and Gaussian operations (homodyne detection). Furthermore, we show under which premises concerning entanglement content of the state, noise, inefficient homodyne detectors, our protocol is efficient and applicable with present technology. Our results are reported in \cite{Rodo2008PRA}.

In chapter~\ref{nongaussian}, we move to the problem of quantification of correlations (quantum and/or classical) between two Continuous Variable modes. We propose to define correlations between the two modes as the maximal number of correlated bits extracted via local quadrature measurements on each mode. On Gaussian states, where entanglement is accessible via their covariance matrix our quantification majorizes entanglement, reducing to an entanglement monotone for pure states. For mixed Gaussian states we provide an operational receipt to quantify explicitly the classical correlations presents in the states. We then address non-Gaussian states with our operational quantification that is based on and up to second moments only in contrast to the exact detection of entanglement that generally involves measurements of high-order moments. For non-Gaussian states, such as photonic Bell states, photon subtracted states and mixtures of Gaussian states, the bit quadrature correlations are shown to be also a monotonic function of the negativity. This quantification yields a feasible, operational way to measure non-Gaussian entanglement in current experiments by means of direct homodyne detection, without needing a complete state tomography. Our analysis demonstrates the rather surprising feature that entanglement in the considered non-Guassian examples can thus be detected and experimentally quantified with the same complexity as if dealing with Gaussian states. Our results are reported in \cite{Rodo2008PRL}.

In chapter~\ref{mesoscopic}, we focus to atomic ensembles described as CV systems. Entanglement between distant mesoscopic atomic ensembles can be induced by measuring an ancillary light system. We show how to generate, manipulate and detect mesoscopic entanglement between an arbitrary number of atomic samples through a quantum non-demolition matter-light interface. Measurement induced entanglement between two macroscopical atomic samples was reported experimentally in 2001. There, the interaction between a single laser pulse propagating through two spatially separated atomic samples combined with a final projective measurement on the light led to the creation of pure EPR entanglement between the two samples. Due to the quantum non-demolition character of the measurement, verification of the EPR state was done by passing a second pulse and measuring variances on light. Our proposal extends this idea in a non-trivial way for multipartite entanglement (GHZ and cluster-like) without needing local magnetic fields. We propose a novel experimental realization of measurement induced entanglement. Moreover, we show quite surprisingly that given the irreversible character of a measurement, the interaction of the atomic sample with a second pulse light can modify and even reverse the entangling action of the first one leaving the samples in a separable state. Our results are reported in \cite{Stasinska2009PRA} and \cite{Stasinska2010arXiv}.

Finally, in chapter~\ref{conclusions}, we conclude summarizing our results, listing some open questions and giving future directions of research.


\chapter{Continuous Variable formalism}\label{cvf}

Continuous Variable (CV) systems are those systems described by two canonical conjugated degrees of freedom {\em i.e.} there exist two observables that fulfill Canonical Commutation Relations (CCR). This chapter comprises a detailed description of Continuous Variable systems. It provides also the mathematical framework needed to analyze the problems treated within this thesis. After introducing a phase-space formalism and the corresponding quasi-probability distributions I shall restrict first to Gaussian states which describe, among others, coherent, squeezed and thermal states. As a cornerstone example, I will shortly develop the canonical quantization of light, ending by showing how to deal with Gaussian states of light. Presently, these states are the preferred resources in experiments of QI using Continuous Variable systems. For further background information the interested reader is referred to \cite{Braunstein2005RMP,vanLoock2002FP,Zhang1990RMP,Adesso2007JPA,Ferraro2005arXiv,Hillery1984PRep,Lee1995PRep}.

\section{Continuous Variable systems}

The Canonical Commutation Relations for two canonical observables $\hat q$ and $\hat p$ read~\footnote{Quadrature operators are chosen adimensional in such a way that $\hbar$ is not going to appear in any formula.}

\begin{equation}\label{ccr}
 \comm{q}{p} = \im \id.
\end{equation}
So a direct consequence to the fact that two hermitian operators $\hat q$ and $\hat p$ fulfill the CCR is that

(i) the underlying Hilbert space cannot be finite dimensional. This can be seen by applying the trace into Eq.~\eqref{ccr}. Using finite dimensional operator algebra one would obtain, on one hand, $\im \, {\rm dim} \mathcal{H}$, while on the other $\tr(\hat q \hat p)-\tr(\hat p \hat q) = 0$.

(ii) $\hat q$ and $\hat p$ cannot be bounded, since the relation $[\hat q^m, \hat p] = m \hat q^{m-1} \im$ (obtained from $[f(\hat A), B] = \frac{d f(\hat A)}{d \hat A}\comm{A}{B})$ implies that~\footnote{Using $||\hat A||\cdot||\hat B|| \geq ||\hat A\hat B|| = \frac{1}{2}||\comm{A}{B}+\acomm{A}{B}|| \geq \frac{1}{2}||\comm{A}{B}||$.} $||\hat q||^m \cdot ||\hat p|| \geq \frac{1}{2} ||[\hat q^m, \hat p]|| = \frac{1}{2}m|| \hat q||^{m-1}$, which means that $||\hat q|| \cdot ||\hat p|| \geq \frac{1}{2} m$ has to be true for all $m$.
\\
This is a direct consequence of the fact that $\hat q$ and $\hat p$ possess a continuous spectra and act in an infinite dimensional Hilbert space.

Examples of CV systems we can think of are the position-momentum of a massive particle, the quadratures of an electromagnetic field, the collective spin of a polarized ensemble of atoms \cite{Hammerer2008arXiv} or the radial modes of trapped ions \cite{Serafini2009NJP}. In all of the examples above, there exist two observables fulfilling \eqref{ccr}. As we will show, these observables obey the standard bosonic commutation relations and so we call these systems bosonic modes. We can deal with several modes, and in this case, by ordering the operators in canonical pairs through $\hat R^T = (\hat q_1,\hat p_1,\hat q_2,\hat p_2,\ldots,\hat q_N,\hat p_N)$ we can compactly state CCR as

\begin{equation}\label{nccr}
 \comm{R_i}{R_j} = \im \id (\mathcal{J}_N)_{ij}
\end{equation}
where $i,j=1,2,\ldots,2N$ and $\mathcal{J}_N = \oplus_{\mu=1}^N \mathcal{J}$ accounts for all modes. $\mathcal{J}$ is the so-called symplectic matrix which corresponds to an antisymmetric and non-degenerate form fulfilling (i) $\forall \eta, \zeta\in \mathbb{R}^{2N}: \langle \eta \vert \mathcal{J} \vert \zeta \rangle = - \langle \zeta \vert \mathcal{J} \vert \eta \rangle$ and (ii) $\forall \eta: \langle \eta \vert \mathcal{J} \vert \zeta \rangle = 0 \Rightarrow \zeta=0$. In the appropriate choice of basis (canonical coordinates) the symplectic matrix is brought to the standard form
\begin{equation}
\mathcal{J}=\begin{pmatrix}
 0 & 1 \\
 -1 & 0
\end{pmatrix}.
\end{equation}

\section{Canonical Commutation Relations}

Canonical Commutation Relations\footnote{The CCR are related with the classical Poisson brackets via the $1^{\rm st }$ quantization transcription:
$\ppoison{A}{B} \equiv \sum_\mu (\frac{\partial \mathcal A}{\partial Q_\mu} \frac{\partial \mathcal B}{\partial P_\mu} - \frac{\partial \mathcal B}{\partial Q_\mu} \frac{\partial \mathcal A}{\partial P_\mu}) \longrightarrow - \im \comm{A}{B} \equiv -\im(\hat A \hat B - \hat B \hat A)$ and $\mathcal A \longrightarrow \hat A$.} can also be expressed using annihilation and creation operators $\hat a_\mu$ and $\hat a_\mu^\dag$ which obey standard bosonic commutation relations
\begin{equation}\label{accr}
 [\hat{a}_\mu,\hat{a}_\nu^\dag] = \delta_{\mu \nu}, \quad
 [\hat{a}_\mu,\hat{a}_\nu] = [\hat{a}_\mu^\dag,\hat{a}_\nu^\dag]=0
\end{equation}
$\mu, \nu = 1,2,\ldots,N$. The CCR expressed in Eq.~\eqref{nccr} and \eqref{accr} are related by a unitary matrix $U=1/\sqrt{2}\begin{pmatrix}
 \id_N & \im \id_N \\
 \id_N & - \im \id_N
\end{pmatrix}$ such that if we define $\hat O^T = (\hat a_1,\hat a_2,\ldots,\hat a_N,\hat a_1^\dag,\hat
a_2^\dag,\ldots$ $,\hat a_N^\dag)$ then $\hat O_i = U_{ij} \hat R_j$.

Notice that the representation of the CCR up to unitaries is not unique. For instance, for a single mode in the Schr\"odinger representation each degree of freedom is embedded in $\mathcal{H}=\mathcal
L^2(\mathbb{R}^2)$, while the operators $\hat q$ and $\hat p$ act multiplicative and derivative respectively
\begin{equation}
 \left. \begin{array}{ccc}
 \hat q &=& q\\
 \hat p &=& -\im \frac{\partial}{\partial q}
 \end{array} \right\}
\end{equation}
but also $\hat q = +\im \frac{\partial}{\partial p}, \quad \hat p = p$ is equally possible. In both representations the operators are unbounded.
A way to remove ambiguities (up to unitaries) and to treat with bounded operators is by using Weyl operators.
The Weyl operator is defined as
\begin{equation}
 \hat{W}_\zeta \equiv e^{\im \zeta^T \cdot \mathcal{J} \cdot \hat{R}}
\end{equation}
where $\zeta^T = (\zeta_1, \zeta_2,\ldots, \zeta_{2N}) \in \mathbb{R}^{2N}$.

The Weyl operator acts in the states as a translation in the phase-space (displacements $e^{\im \eta \hat{p}} \ket{q} = \ket{q - \eta}$ and kicks $e^{\im \zeta \hat{q}} \ket{q} = e^{\im \zeta q} \ket{q}$) as it can be checked by looking to its action onto an arbitrary position-momentum operator
\begin{equation}
 \hat{W}_\zeta^\dag \hat{R_i} \hat{W}_\zeta = \hat{R_i} - \zeta_i \id.
\end{equation}
It satisfies the Weyl relation
\begin{equation}
 \hat{W}_\zeta \hat{W}_\eta = e^{-\frac{\im}{2}
 \zeta^T \cdot \mathcal{J} \cdot \eta} \hat{W}_{\zeta+ \eta},
\end{equation}
analogously, it fulfills
\begin{equation}
 \hat{W}_\zeta \hat{W}_\eta = \hat{W}_\eta \hat{W}_\zeta e^{-\im
 \zeta^T \cdot \mathcal{J} \cdot \eta},
\end{equation}
showing the non-commutative character of the canonical observables.
There exists only one equivalent representation of the Weyl relation according to the following theorem.
\begin{theorem}
 {\em (Stone-von Neumann theorem)}
Let $\hat{W}_1$ and $\hat{W}_2$ be two Weyl systems over a finite dimensional phase-space $(N < \infty)$. If the two Weyl systems are strongly continuous~\footnote{$\forall \ket{\psi} \in \mathcal{H}:
\lim_{\zeta \to 0} ||\, \ket{\psi} - \hat{W}_\zeta \ket{\psi}|| = 0$.} and irreducible~\footnote{$\forall \zeta \in \mathbb{R}^{2N}:
[\hat{W}_\zeta, \hat A] = 0 \Rightarrow \hat A \propto \id$.} then they are equivalent (up to an unitary).
\end{theorem}

\section{Phase-space}

Phase-space formulations of Quantum Mechanics, offers a framework in which quantum phenomena can be described using as much classical language as allowed. There are various formulations of non-relativistic Quantum Mechanics see {\em e.g.} \cite{Styer2002AJP}. These formulations differ in mathematical description, yet each one makes identical predictions for all experimental results.

Phase-space formulations can often provide useful physical insights. Furthermore, it requires dealing only with constant number equations and not with operators, which can be of significant practical advantage. This mathematical advantage arises here from the fact that the infinite-dimensional complex Hilbert space structure which is, in principle, a difficult object to work with, can be mapped into the linear algebra structure of the finite-dimensional real phase-space. We will extend this map and show how to characterize states and operations in sections~\ref{qstates} and \ref{soperations} respectively.

\subsection{Phase-space geometry}\label{psgeometry}

A system of $N$ canonical degrees of freedom is described classically by a $2N$-dimensional real vector space~\footnote{They are isomorphic (there exist a bijective morphism between the two groups).} $V \simeq \mathbb{R}^{2N}$. Together with the symplectic form it defines a symplectic real vector space (the phase-space) $\Omega \simeq \mathbb{R}^{2N}$. The phase-space is naturally
equipped with a complex structure and can be identified with a complex Hilbert space $\mathcal{H}_\Omega \simeq \mathbb{C}^{N}$.
If $\braket{\,\,}{\,\,}$ stands for the scalar product in $\mathcal{H}_\Omega$ and $\braket{\,\,}{\,\,}_\mathcal{J}$ for the symplectic scalar product in $\Omega$ their connection reads
\begin{equation}
 \braket{\eta}{\zeta} = \braket{\mathcal{J} \eta}{\zeta}_\mathcal{J} + \im
 \braket{\eta}{\zeta}_\mathcal{J}.
\end{equation}
Notice that $\eta = (q,p) \in \Omega$ while $\eta = q + \im p \in \mathcal{H}_\Omega$ such that any orthonormal basis in $\mathcal{H}_\Omega$ leads to a canonical basis in $\Omega$. Moreover, any Gaussian unitary operator (which preserves the scalar product) acting on $\mathcal{H}_\Omega$, leads to a symplectic operation $S$ in the phase-space in such a way that the symplectic scalar product is also preserved. The inverse is also true provided that the symplectic operation commutes with the symplectic matrix $\mathcal{J}$.

\subsection{Symplectic operations}\label{soperations}

Gaussian operations \cite{Giedke2002PRA} are completely positive maps, thus preserving the Gaussian character on states, that can be implemented by means of Gaussian unitary operators (symplectic operations) plus Bell measurements (homodyne measurements). Homodyne/heterodyne detection is a fundamental Gaussian operation, that is, the physical measurement of one/two of the canonical conjugated coordinates. Nevertheless we will concentrate for the moment with symplectic operations, hence canonical transformations $S$ that preserve the CCR and therefore leave the basic kinematic rules unchanged. That is, if we transform our canonical operators $\hat R_S = S \cdot \hat R$, still equation \eqref{nccr} is fulfilled. In a totally equivalent way, we can define symplectic transformation as the ones which preserve the symplectic scalar product and therefore~\footnote{From now on we neglect the subscript $N$ in symplectic matrix.}
\begin{equation}
 S^T \cdot \mathcal{J} \cdot S = \mathcal{J}.
\end{equation}
The set of real $2N\times2N$ matrices $S$ satisfying the above condition form the symplectic group $Sp(2N,\mathbb{R})$. To construct the affine symplectic group we just need to add also the phase-space translations $s$ that transform $\hat R_S = S \cdot \hat R + s$ and whose group generators are $\hat G^{(0)}_i = \mathcal{J}_{ij}\hat R_j$. Apart from that, the group generators of the representation of $Sp(2N,\mathbb{R})$ which physically corresponds to the Hamiltonians which perform the symplectic transformations on the states, are of the form $\hat G_{ij} = \frac{1}{2}\{ \hat R_i,\hat R_j \}$. This corresponds to hermitian Hamiltonians of quadratic order in the canonical
operators. When rewriting them in terms of creation / annihilation operators we can divide it into two groups.
Passive generators (compact):\\\\ $\hat G^{(1)}_{\mu \nu} = \im \frac{(\hat a^\dag_\mu\hat a_\nu-\hat a^\dag_\nu\hat a_\mu)}{2}$,
\quad {$\hat G^{(2)}_{\mu \nu} = \frac{(\hat a^\dag_\mu\hat a_\nu+\hat a^\dag_\nu\hat a_\mu)}{2}$,\\\\
and active generators (non-compact):\\\\ $\hat G^{(3)}_{\mu \nu} = \im \frac{(\hat a^\dag_\mu\hat a^\dag_\nu-\hat a_\nu\hat a_\mu)}{2}$,
\quad $\hat G^{(4)}_{\mu \nu} = \frac{(\hat a^\dag_\mu\hat a^\dag_\nu+\hat a_\nu\hat a_\mu)}{2}$.\\
The passive ones, are generators which commute with all number operators $\hat n_\mu \equiv \hat a_\mu^\dag \hat a_\mu$, and so, they preserve the total number, in this sense they are passive. If the system under study is the electromagnetic field, what is being preserved under passive transformations is the total number of photons. In such a case, passive transformations can be implemented optically by only using beam splitters, phase shifts and mirrors. Conversely, only by using them, we can implement any Hamiltonian constructed by a linear combination of the compact generators. Finally, with all the five classes of generators, we can generate all the Gaussian unitaries, $\hat U_\lambda = e^{\im \lambda \cdot \hat G}$.

For one mode ($N=1$) the simplest passive generator is the phase shift operator and its corresponding symplectic operation in phase-space
\begin{equation}
 \hat U_\theta = e^{\im \theta \hat a^{\dag} \hat a} \Leftrightarrow S_\theta = \begin{pmatrix}
 \cos{\theta} & \sin{\theta} \\
 -\sin{\theta} & \cos{\theta}
 \end{pmatrix}.
\end{equation}
On the other hand, we have the active generators that change the energy of the state. The most important one is the single mode squeezing operator, whose unitary expression (for a squeezing parameter $r>0$) and symplectic operation in phase-space reads
\begin{equation}
 \hat U_r = e^{\frac{r}{2}(\hat a^2 - \hat a^{\dag 2})} \Leftrightarrow S_r = \begin{pmatrix}
 e^{-r} & 0 \\
 0 & e^r
\end{pmatrix}.
\end{equation}
For experimental reasons, instead of the parameter $r$ one uses a decibel expression $10 \log e^{2r}$ in dBs~\footnote{All logarithms are in basis 10, unless differently specified.}. The squeezing operator squeezes the uncertainties on $q$ and $p$ in a complementary way {\em i.e.} it squeezes position while stretches the momentum with the same factor in such a way that the state remains as close to the uncertainty limit as it was before. When the squeezing parameter $r$ is positive we call it a q-squeezer (amplitude squeezer). Analogously p-squeezer or (phase squeezer) occurs for negatives squeezing parameters.

Finally, phase-space translations for one mode are described by the unitary and symplectic operation in phase-space
\begin{equation}
\hat U_\alpha = e^{\alpha \hat a^\dag - \alpha^* \hat a}  \Leftrightarrow s_\alpha = \begin{pmatrix}
 q_0 \\
 p_0
\end{pmatrix}
\end{equation}
where $\alpha = \frac{q_0 + \im p_0}{\sqrt2}$.

For two modes ($N=2$) the most important non-trivial unitaries are beam splitters (reflectivity $R=\sin^2{\theta/2}$ and transmitivity $T=\cos^2{\theta/2}$) and two mode squeezers that amounts respectively to
\begin{equation}
 \hat U_{BS} = e^{\frac{\theta}{2}(\hat a_1 \hat a_2^\dag - \hat a_1^\dag \hat a_2)} \Leftrightarrow S_{BS} = \begin{pmatrix}
 \cos{\theta/2} & 0 & \sin{\theta/2} & 0\\
 0 & \cos{\theta/2} & 0 & \sin{\theta/2}\\
 -\sin{\theta/2} & 0 & \cos{\theta/2} & 0\\
 0 & -\sin{\theta/2} & 0 & \cos{\theta/2}
\end{pmatrix}
\end{equation}
and
\begin{equation}
 \hat U_{TMS} = e^{r(\hat a_1 \hat a_2 - \hat a_1^\dag \hat a_2^\dag)} \Leftrightarrow S_{TMS} = \begin{pmatrix}
 \cosh{r} & 0 & \sinh{r} & 0\\
 0 & \cosh{r} & 0 & -\sinh{r}\\
 \sinh{r} & 0 & \cosh{r} & 0\\
 0 & -\sinh{r} & 0 & \cosh{r}
\end{pmatrix}.
\end{equation}

\section{Probability distribution functions}

One of the most important tools of the phase-space formulation of Quantum Mechanics are the phase-space probability distribution functions. The best known and widely used is the Wigner distribution function, but there is not a unique way of defining a quantum phase-space distribution function. In fact, several distribution functions with different properties, rules of association and operator ordering can also be well defined \cite{Klauder2007JPA}. For instance sometimes normal ordered ($\mathcal{P}$-function), antinormal ordered ($\mathcal{Q}$-function), generalized antinormal ordered (Husimi-function), standard ordered, antistandard ordered,\ldots distributions can be more convenient depending on the problem being considered. In this thesis we are only going to work with the totally symmetrical ordered one (Weyl ordered); the Wigner distribution function.

Due to the fact that joint probability distributions at a fixed position $\hat q$ and momentum $\hat p$ are not allowed by Quantum Mechanics (Heisenberg uncertainty theorem), the quantum phase-space
distribution functions should be considered simply as a useful mathematical tool. Joint probabilities can be negative, so that one deals rather with quasi-probability distribution. As long as it yields a correct description of physically observable quantities, their use is accepted.

\subsection{Quantum states and probability distributions}\label{qstates}

We would like to see here now how to describe quantum systems of CV using probability distributions functions. Let's recall that a density operator $\hat \rho$, defines a quantum state iff it satisfies the following properties
\begin{equation}
 \tr{\hat \rho} = 1, \quad \hat \rho \geq 0 \quad [\Rightarrow \quad \hat \rho^\dag = \hat \rho].
\end{equation}
Such operator belongs to the bounded linear operators Hilbert space $\mathcal B(\mathcal H)$. If the state is pure $\hat\rho=\ketbra{\psi}{\psi}$ where $\ket{\psi}$ belongs to a Hilbert space $\mathbb{C}^d$ for qu{\em d}its, (discrete variables system) or $\mathcal L^2(\Omega)$ for $N$ modes, (Continuous Variable system). Notice the inconvenience of density operator formalism for CV since the states belong to an infinite dimensional Hilbert space.

For systems of Continuous Variable, the Wigner distribution function gives a complete description of the state. Given a state $\hat \rho$ corresponding to a single mode, the Wigner distribution function is defined as~\footnote{For pure states the definition gets simplified to $\mathcal{W}_\rho(q,p) = \frac{1}{\pi} \int dx \, e^{-2\im px} \psi^*(q-x) \psi(q+x) =$ $= \int dx \, \tilde{\mathcal{W}}_\rho(q,x) e^{-2\im px}$. One can see a direct link between the wave function and the Symplectic-Fourier transform of the Wigner distribution via $\psi(q) = \sqrt{\frac{\pi}{\tilde{\mathcal{W}}(q_0,0)}} \tilde{\mathcal{W}}(\frac{1}{2}(q+q_0),\frac{1}{2}(q-q_0))$.}
\begin{equation}\label{wigner}
 \mathcal{W}_{\rho}(q,p) = \frac{1}{\pi} \int dx
 \sandwich{q+x}{\rho}{q-x} e^{-2\im px}.
\end{equation}
This transformation is called Weyl-Fourier transformation and it gives the bridge between density operators and distribution functions. Sometimes, for computational reasons it is better to compute first the characteristic distribution function which is obtained through
\begin{equation}\label{characteristic}
 \chi_\rho(\zeta, \eta) = \tr \{\hat{\rho}\hat{W}_{(\zeta, \eta)} \}.
\end{equation}
The above two distribution functions are fully equivalent in the sense of describing completely our quantum state and are related by a Symplectic-Fourier transform
\begin{equation}
 \mathcal{W}_\rho(q,p) = \frac{1}{(2 \pi)^2} \int d\zeta \int d\eta \,
 \chi_\rho(\zeta, \eta) e^{-\im \zeta p + \im \eta q} = \frac{1}{2\pi}
 \mathcal {SFT} \{ \chi_\rho(\zeta, \eta) \},
\end{equation}
\begin{equation}
 \chi_\rho(\zeta, \eta) = \int dq \int dp \, \mathcal{W}_\rho(q,p)
 e^{\im \zeta p - \im \eta q} = 2\pi \mathcal {SFT}^{-1} \{
 \mathcal{W}_\rho(q,p) \}.
\end{equation}
The Weyl-Fourier transformation is invertible and it provides a way to recover our density operator from both distribution functions
\begin{equation}
 \begin{split}
 \hat{\rho} &= \frac{1}{2 \pi} \int dq \int dp
 \int d\zeta \int d\eta \, \mathcal{W}_\rho(q,p) e^{-\im \zeta p + \im
 \eta q} \hat{W}_{(-\zeta, -\eta)} =\\
 &= \frac{1}{2 \pi} \int d\zeta \int d\eta \, \chi_\rho(\zeta,\eta)
 \hat{W}_{(-\zeta, -\eta)}.
 \end{split}
\end{equation}
At this level, $\mathcal{W}$ (and $\chi$) defines a quantum state iff they satisfy the following properties:

(i) the state is normalized
\begin{equation}
 \int dq \int dp \, \mathcal{W}(q,p) = 1, \quad \chi(0,0) = 1
\end{equation}
and, 

(ii) the state is non-negative defined
\begin{equation}
 \int dq \int dp \,
 \mathcal{W}(q,p) \mathcal{W}_p(q,p) \geq 0 \quad [\Rightarrow \quad \mathcal{W}^*(q,p) = \mathcal{W}(q,p)]
\end{equation}
\begin{equation}
 \sum_{i,j=1}^{2N}a_i^*a_j \chi(\zeta_j-
 \zeta_i) e^{\frac{\im}{2}(\zeta_i^T\cdot \mathcal{J} \cdot \zeta_j)} \geq 0
 \quad [\Rightarrow \quad \chi^*(\zeta,\eta) = \chi(-\zeta,-\eta)]
\end{equation}
for all pure states $\mathcal{W}_p$ and for all $a_{i,j} \in \mathbb{R}$.
This can be shown using the following theorem:

\begin{theorem}
 {\em (Quantum Bochner-Khinchin theorem)}
 For $\chi(\eta)$ to be a characteristic function of a quantum state the following conditions are necessary and sufficient\\
 1.) $\chi(0) = 1$ and $\chi(\eta)$ is continuous at $\eta = 0$,\\
 2.) $\chi(\eta)$ is $\mathcal{J}-positive$ (symplectic-positive defined).
\end{theorem}

\subsection{Properties of the Wigner distribution}

{\em Properties}

\begin{enumerate}[i)]
\item {\em Quasidistribution:} It is a real valued quasi-distribution because it admits negatives values.\\
\item {\em T-symmetry:} It has time symmetry
\begin{equation}
 t \rightarrow -t \Longleftrightarrow \mathcal{W}(q,p,t) \rightarrow \mathcal{W}(q,-p,t).
\end{equation}
\item {\em X-symmetry:} It has space symmetry
\begin{equation}
 q \rightarrow -q \Longleftrightarrow
 \mathcal{W}(q,p,t) \rightarrow \mathcal{W}(-q,-p,t).
\end{equation}
\item {\em Galilei invariant:} It is Galilei invariant
\begin{equation}
 q \rightarrow q-a \Longleftrightarrow \mathcal{W}(q,p,t) \rightarrow \mathcal{W}(q + a,p,t).
\end{equation}
\item {\em T-evolution:} The equation of motion for each point in the phase-space is classical in the absence of forces~\footnote{Remember that when we are speaking about states of light $m$ has to be interpret as permittivity of vacuum $\epsilon_0$. Notice there is a minus sign difference with
Heisenberg's equation of motion.
\begin{equation}
 \frac{d\hat{A}}{dt} = \frac{1}{\im}\comm{A}{H}.
\end{equation}}
\begin{equation}
 \frac{d\hat{\rho}}{dt} = -\frac{1}{\im}\comm{\rho}{H} \Longleftrightarrow \frac{\partial \mathcal{W}(q,p,t)}{\partial t} = -\frac{p}{m} \frac{\partial \mathcal{W}(q,p,t)}{\partial q}.
\end{equation}
\item {\em Bounded:} It is bounded
\begin{equation}\label{bounded}
 |\mathcal{W}(q,p)| \leq \frac{1}{\pi}.
\end{equation}
For pure states~\footnote{For mixed states, use Schwarz's inequality $|\braket{\psi_1}{\psi_2}|^2 \leq \braket{\psi_1}{\psi_1} \braket{\psi_2}{\psi_2}$ at the density operator level {\em i.e.} $0 \leq \tr (\rho_1 \rho_2)^n \leq \tr (\rho_1)^n \tr (\rho_2)^n$ for $n=1$.} the demonstration reads
\begin{equation}
\begin{split}
|\mathcal{W}(q,p)|^2 &= \frac{1}{\pi^2} \left| \int dx
\, e^{-2\im px}
\psi^*(q-x) \psi(q+x) \right|^2 \leq\\
& \leq \frac{1}{\pi^2} \int dx \left|e^{\im px}
\psi(q-x)\right|^2 \int dx \left| e^{-\im px}
\psi(q+x)\right|^2 = \frac{1}{\pi^2}.
\end{split}
\end{equation}
\item {\em Normalized:} It is well normalized
\begin{equation}
 \int dq \int dp \, \mathcal{W}(q,p) = 1.
\end{equation}
\item {\em Quantum marginal distributions:} It possesses good marginal distributions~\footnote{For pure states they correspond to the square modulus of the wave function in position $|\psi(q)|^2$
and in momentum $|\tilde{\psi}(p)|^2$ representation.}
\begin{equation}
 \bar{\mathcal{W}}(q) = \int dp \, \mathcal{W}(q,p) = \sandwich{q}{\rho}{q} \geq 0,
\end{equation}
\begin{equation}
 \bar{\mathcal{W}}(p) = \int dq \, \mathcal{W}(q,p) = \sandwich{p}{\rho}{p} \geq 0.
\end{equation}
\item {\em Orthonormal:} The orthonormality is preserved
\begin{equation}
\left|\int dq \, \psi^*(q)\phi(q)\right|^2 = 2\pi \int dq \int dp \, \mathcal{W}_\psi(q,p) \mathcal{W}_\phi(q,p).
\end{equation}
If the distributions are equal $\psi=\phi$ we conclude $\int dq \int dp \, \mathcal{W}^2(q,p) = \frac{1}{2\pi}$ for all pure states, which is a lower bound in general see appendix~\ref{appendix2} and Eq.~\eqref{purityGauss}, also it excludes classical distributions such $\mathcal{W}(q,p) = \delta(q-q_0) \delta(p-p_0)$. If they are orthogonal, $\psi \perp \phi$, we conclude $\int dq \int dp \, \mathcal{W}_\psi(q,p) \mathcal{W}_\phi(q,p) = 0$ which tells us that the Wigner distribution function, in general, cannot be everywhere positive.
\item {\em Complete orthonormal set:} The set of functions $\mathcal{W}_{nm}(q,p)$ form a complete orthonormal set (if $\psi_{n}(q)$ are already a set)
\begin{equation}2
 \int dq \int dp \, \mathcal{W}^*_{nm}(q,p)
 \mathcal{W}_{n'm'}(q,p) = \frac{1}{2 \pi} \, \delta_{nn'}
 \delta_{mm'},
\end{equation}
\begin{equation}
 \sum_{n,m} \mathcal{W}^*_{nm}(q,p) \mathcal{W}_{nm}(q',p') = \frac{1}{2 \pi} \delta(q-q') \delta(p-p'),
\end{equation}
where
\begin{equation}
 \mathcal{W}_{nm}(q,p) = \frac{1}{\pi} \int dx \, e^{-2\im px} \psi^*_n(q-x) \psi_m(q+x).
\end{equation}
\end{enumerate}

\subsection{The generating function of a Classical probability distribution}

Denoting by $y$ ($x$) a random variable which can be discrete $y \in \{ y_i \}$ (or continuous $x \in [a,b]$) and its corresponding (density) probability $p(y_i)$ $(p(x))$, we can establish the normalization constrain as

\begin{equation}
 \left. \begin{array}{ccc}
 \sum_i p(y_i) &=& 1\\
 \int_a^b p(x) dx &=& 1
 \end{array} \right\}.
\end{equation}

\noindent{Of relevant importance given a probability distribution are the following quantities:}

\begin{enumerate}[i)]
\item {\em Mean value of $u(x)$}: \quad $E[u(x)] = \int u(x) p(x) dx$.
\item {\em Moment of order $m$ respect point $c$ of $x$}: \quad $\alpha^m_c = E[(x-c)^m]$.
\item {\em Mean value of $x$}: \quad $\mu = \alpha^1_0 = E[x]$.
\item {\em Standard deviation of $x$:} \quad $\sigma = \sqrt{{\rm var}(x)} = \sqrt{\alpha^2_\mu} = \sqrt{E[(x-\mu)^2]}$.
\item {\em Covariance of $x_i$ and $x_j$:}~\footnote{Here subindex $i,j$ labels all the possible variables of the distribution, when they are equal, $C_{ii}$ corresponds to the variance ${\rm var}(x_i)$ of the variable $x_i$.} \quad $C_{ij} = {\rm cov}(x_i,x_j) = E[(x_i-\mu_i)(x_j-\mu_j)]$,
\end{enumerate}
where $i,j=1,2,\ldots,2N$.

\begin{theorem}
 {\em (Taylor's theorem)}
 Any well behaved distribution function can be reconstructed by its (in general) infinite moments.
\end{theorem}

This theorem, of considerable importance, tell us that any distribution $p(x)$ can be retrieved by its moments $\alpha^m_c$ only. We define the vector $d$ and the matrix $C$ called mean vector and covariance matrix by

\begin{equation}
 \left. \begin{array}{ccl}
 C &=& [[{\rm cov}(x_i,x_j)]]\\
 d &=& [[\mu_i]]
 \end{array} \right\}.
\end{equation}
What is more important is that $d$ and $C$ encode all the information of $1^{\rm st }$ and $2^{\rm nd}$ moments.

If we define the generating function of a distribution function by a Laplace transformation (provided it exists)
\begin{equation}
 M(\eta) = \mathcal {LT}\{ p(x) \} = E[e^{x \eta}],
\end{equation}
all moments can be obtained by subsequently differentiating the generating function

\begin{equation}
 \left.\alpha^m_0 = \frac{\partial^{(m)} M(\eta)}{\partial \eta^m} \right|_{\eta=0}.
\end{equation}

\subsection{The generating function of a quasi-probability distribution}

In the same way as in Classical Probability where all the moments of a distribution characterize the distribution, the Wigner quasi-distribution function is fully characterized by its moments.
To adapt the classical formalism to the quantum Wigner quasi-distribution function we have to introduce the following transcription $\eta \longrightarrow \im \eta, M \longrightarrow \chi, \mathcal{LT} \longrightarrow \mathcal{FT}$~\footnote{$\mathcal{LT}\{ f(x) \} = \int_0^\infty f(x)\e^{-y x} dx$ while $\mathcal{FT}\{ f(x) \} = \frac{1}{\sqrt{2\pi}}\int_{-\infty}^\infty f(x)\e^{- \im y x} dx = \tilde f(y)$.}.
We then define the generating function of the Wigner distribution (characteristic function) by a Fourier transformation, which always exists, since the Wigner distribution is an integrable function.
In general it is complex and reads

\begin{equation}
 \chi(\eta) = \mathcal {FT}\{ \mathcal{W}(x) \} = E[e^{\im x \eta}],
\end{equation}
then all moments can be obtained by subsequently differentiating the generating function

\begin{equation}
 \left.\beta^m_0 = \frac{1}{\im^m} \frac{\partial^{(m)} \chi(\eta)}
 {\partial \eta^m} \right|_{\eta=0}.
\end{equation}

Analogously, we define, given a quantum Wigner distribution function the displacement vector (DV) $d$ (a $2N$ real vector) and the covariance matrix (CM) $\gamma$ (a $2N \times 2N$ real symmetric matrix). The DV contains the information of the first moments. By the space symmetry only relative DVs are of physical meaning. The CM is much more richer, it contains information (up to second moments) about the purity, entanglement,\ldots The CM $\gamma$, corresponding to a physical state must be symplectic-positive defined, {\em i.e.}

\begin{equation}\label{positivity}
 \gamma + \im \mathcal{J} \geq 0.
\end{equation}

Such a constrain also implies that, because the symplectic matrix is antisymmetric, that $\gamma \geq 0$. Positivity implies hermiticity which translates here to $\gamma^T=\gamma$. Its useful also to introduce here the symplectic spectrum~\footnote{Every positive matrix $M$ can be diagonalized under symplectics with a symplectic spectrum ${\rm symspec}(M)$ of the form $\{\mu_i\}$ and degenerated with multiplicity 2. The values $\{\mu_i\}$ are obtained throw ${\rm spec}(-\im \mathcal{J} M)=\{\pm \mu_i\}$, $i=1, 2, \ldots, \frac{\dim(M)}{2}$.} of the CM because the positivity condition reads, in terms of the symplectic eigenvalues, as $\mu_i\geq1$, $i=1, 2, \ldots, N$.

\section{Gaussian states}

Gaussian states are those states with a Gaussian Wigner distribution function. Among all the CV systems, Gaussian states, are of the greatest importance. A Gaussian distribution appears as the limit of many others and occurs in a great variety of different conditions. This fact is reflected in the Central Limit Theorem which is one of the cornerstone of the Classical Probability and Statistics Theory together with the Law of Large Numbers.

\begin{theorem}
 {\em (Central limit theorem)}
 Suppose we have $n$ independent random variables $x_1, x_2,\ldots, x_n$ which are all distributed with a mean value $\mu$ and a standard deviation $\sigma$ (each of them can have different arbitrary distribution functions $p_i(x_i)$). In the limit $n \to \infty$ the arithmetic mean $\bar x = \frac{1}{n}\sum_{i=1}^n x_i$ is Gaussian (or normal) distributed with mean value $\mu$ and standard deviation
$\frac{\sigma}{\sqrt{n}}$ {\em i.e.} $\bar p(\bar x) = \frac{1}{\sqrt{2 \pi}\sigma}e^{-\frac{(\bar x - \mu)^2}{2 \sigma^2}}$.
\end{theorem}

The importance of Gaussian probability distributions is also encoded in the following theorem

\begin{theorem}
 {\em (Marcinkiewicz's theorem)}
 \label{mt}
 If we define the cumulant generating function as $K(\eta) = \ln{M(\eta)}$, then, either the cumulant is a polynomial of order 2 or it has infinite terms.
\end{theorem}

\begin{lemma}
 {\em (Gaussianity lemma)}\label{gauss}
 $p(x)$ is a Gaussian(non-Gaussian) distribution iff the cumulant is a polynomial of order 2($\infty$).
\end{lemma}

In general, to describe a probability distribution, all moments are necessary but as long as we are concern with Gaussian distributions, $1^{\rm st }$ and $2^{\rm nd}$ moments are sufficient. In fact, all other higher moments can be rewritten in terms of them. This is a consequence of the theorem~\ref{mt}.

\subsection{Displacement Vector and Covariance Matrix}\label{DViCM}

The Displacement Vector (DV) and the Covariance Matrix (CM) are enough to describe Gaussian states. Analogously to the classical case, it is straightforward to obtain the moments of order $\beta^m_0$ of a distribution by differentiating the generating function. Computing $1^{\rm st}$ and $2^{\rm nd}$ moments, through \eqref{characteristic} we get~\footnote{Notice: $ \left.(\frac{\partial}{\partial \eta_i} e^{\im \eta^T \cdot \hat{R}}) \right|_{\eta_i = 0} = \im \hat R_i$ and $ \left.(\frac{\partial^2}{\partial \eta_i \partial \eta_j} e^{\im \eta^T \cdot \hat{R}}) \right|_{\eta_{i,j} = 0} = \frac{1}{2} \comm{R_i}{R_j} - \hat R_i \hat R_j = -\frac{1}{2} \acomm{R_i}{R_j}$ where we have used Cambell-Hausdorff formula $e^{\hat A + \hat B} = e^{\hat A} e^{\hat B} e^{-\frac{1}{2} \comm{A}{B}}$ (when $\comm{A}{B} \propto \id$).}

\begin{equation}
 \left.\beta^1_{0,i} = -\im \frac{\partial}{\partial \eta_i} \chi(\eta) \right|_{\eta = 0} = \tr (\hat{\rho} \hat{R'}_i),
\end{equation}

\begin{equation}
 \left.\beta^2_{0,ij} = (-\im)^2 \frac{\partial^2}{\partial \eta_i \partial \eta_j} \chi(\eta) \right|_{\eta = 0} = \frac{1}{2} \tr (\hat{\rho} \{ \hat{R'}_i, \hat{R'}_j \}) = \tr (\hat{\rho} \hat{R'}_i \hat{R'}_j) - \frac{\im}{2} \mathcal{J}_{ij},
\end{equation}
where $\hat{R'}_i = \mathcal{J} \cdot \hat R_i$.

Finally, we define the DV and the CM as~\footnote{For pure states $d_i = \exvalue{R_i}{\rho}$ and $\gamma_{ij} = \langle \{ \hat{R}_i-d_i \hat{\id}, \hat{R}_j-d_j \hat{\id} \} \rangle_{\rho} = \langle \acomm{R_i}{R_j} \rangle_{\rho} - 2 \exvalue{R_i}{\rho} \exvalue{R_j}{\rho}$, where, by the anticommutator definition, we see a factor 2 of difference with the classical analog and so $\gamma_{ij} \sim 2 C_{ij}$. The diagonal terms can be rewritten in terms of the uncertainties as $\gamma_{ii} = 2 (\Delta R_i)_\rho^2$ where as usual $(\Delta A)_\psi = \sqrt{\exvalue{A^2}{\psi} - (\exvalue{A}{\psi})^2}$.}

\begin{eqnarray}
 d_i &=& \tr (\hat{\rho} \hat{R}_i)\\
 \gamma_{ij} &=& \tr (\hat{\rho} \{ \hat{R}_i-d_i \hat{\id}, \hat{R}_j-d_j \hat{\id} \}) =\nonumber \\
 &=&2 \tr [\hat{\rho}
 (\hat{R}_i-d_i \hat{\id}) (\hat{R}_j-d_j \hat{\id})] - \im \mathcal{J}_{ij} =\\
 &=& 2 {\rm Re} \{ \tr [\hat{\rho} (\hat{R}_i-d_i \hat{\id}) (\hat{R}_j-d_j \hat{\id})] \}. \nonumber
\end{eqnarray}

\subsection{Phase-space representation of states}

It is important to remark here that symplectic operations at the level of the DV and CM act in such a way that any Gaussian unitary $\hat{U}_S$ maps to the following transformation $\gamma_S = S \cdot \gamma \cdot S^T $ and $d_S = S \cdot d + s$ where $S$ stands for an element of the symplectic group, while $s$ stands for a phase-space translation.

With these definitions it can be shown that the Wigner distribution of any Gaussian state can be written in terms of the DV and CM through~\footnote{We see here that ${\rm max} \left[\mathcal{W}(\zeta)\right] = \mathcal{W}(d) = \frac{1}{\pi^N \sqrt{\det \gamma}} \leq \frac{1}{\pi^N}$ where the equality holds for pure states only \eqref{bounded}.}

\begin{equation}\label{GWigner}
 \mathcal{W}(\zeta) = \frac{1}{\pi^N \sqrt{\det \gamma}} e^{-(\zeta-d)^T \cdot \frac{1}{\gamma} \cdot (\zeta -d)},
\end{equation}
while its symplectic-Fourier transform reads

\begin{equation}\label{GCharacteristic}
 \chi(\eta) = e^{\im \eta^T \cdot \mathcal{J} \cdot d - \eta^T \cdot \mathcal{J}^T\frac{\gamma}{4}\mathcal{J} \cdot \eta} = e^{\im \eta^T
 \cdot d' - \eta^T \cdot \frac{\gamma'}{4} \cdot \eta},
\end{equation}
where $d'_i = \mathcal{J}_{ij} d_j$ and $\gamma'_{ij} = \mathcal{J}^T_{ik} \gamma_{kl} \mathcal{J}_{lj}$.

It is very useful when dealing with Gaussian states to represent them pictorically in the phase-space. As a example we plot here the Wigner distribution function of a rotated squeezed coherent Gaussian state in the phase-space as seen in Fig.~\ref{psgeneric}(a). In Fig.~\ref{psgeneric}(b), we plot also its pictorical representation, obtained by an horizontal cut of the Wigner function at a factor $\e^{-1/2}$ of the maximum. This closed curve fulfills the following expression $\frac{\mathcal{W}(\zeta)}{\mathcal{W}(d)} = \e^{-1/2}$ (for Gaussian states is nothing else than an ellipse).
The area $A=\frac{\pi}{2}\frac{1}{\mathcal{P}}=\pi\Delta\tilde q \Delta\tilde p$ is closely related with the purity of the state see~\eqref{purity} and naturally constrained by the uncertainty principle in an appropriate frame $(\tilde q,\tilde p)$, the one which uncertainties coincide with the major/minor semiaxes of the ellipse. This can be casted in the following theorem.

\begin{figure}[h]
 \centering
 \psfrag{A}{$q$}\psfrag{B}{$p$}\psfrag{C}{$ \mathcal{W}(\zeta) $}\psfrag{D}{$q_0$}\psfrag{E}{$p_0$}\psfrag{F}{$ \mathcal{W}(d)$}
a) \includegraphics[width=7cm]{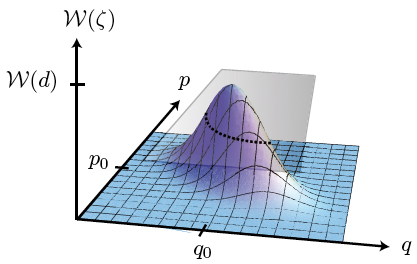}
 \psfrag{A}{$q$}\psfrag{B}{$p$}\psfrag{C}{$q_0$}\psfrag{D}{$p_0$}\psfrag{E}{$\Delta q$}\psfrag{F}{$\Delta p$} \hfill
b) \includegraphics[width=5cm]{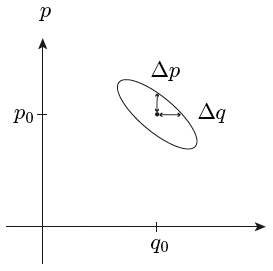}
 \caption{\small{A rotated squeezed coherent Gaussian state with rotating angle $\phi=50^\circ$, squeezing parameter $r=0.4$ and displacement $\alpha=\frac{q_0+\im p_0}{\sqrt2}$. a) Wigner distribution function of the state where $d^T=(q_0,p_0)$. b) Gaussian state pictorical representation in the phase-space containing the DV and CM information where $2(\Delta q)^2 = \cosh{2r}-\sinh{2r}\cos{2\phi}$ and $2(\Delta p)^2 = \cosh{2r}+\sinh{2r}\cos{2\phi}$.}}\label{psgeneric}
\end{figure}

\begin{theorem}
 {\em (Minimum uncertainty states theorem)}\label{must}
 Equality in Heisenberg's uncertainty theorem is attained iff the state is a pure Gaussian state {\em i.e.} a rotated squeezed coherent state, $\ket{\psi} = \hat U_\theta \hat U_r \hat U_\alpha \ket{0}$.
\end{theorem}

All pure Gaussian states of one mode, characterized by its $\gamma$ (and if necessary by $d$), can be obtained from the vacuum state by an appropriate squeezing+rotation plus displacement in the phase-space. These states, by virtue of theorem~\ref{must}, are the minimum uncertainty states. Instead, mixed Gaussian states of one mode can be all obtained from a thermal state through a squeezing+rotation plus displacement.

As the cornerstone examples of Gaussian states, we have the vacuum, coherent, squeezed and thermal states. One can compute their first and second moments and construct the corresponding DV and CM shown below.

\begin{itemize}
\item {\em Vacuum}: $\ket{0}$ s.t. $\hat a \ket{0} = 0$
\begin{equation}
 \gamma_0 = \begin{pmatrix}
 1 & 0\\
 0 & 1
 \end{pmatrix}, \quad d_0 = \begin{pmatrix}
 0\\
 0
 \end{pmatrix}.
\end{equation}
\item {\em (Pure) Coherent}:~\footnote{A Coherent state can alternatively be defined as the eigenstate of the annihilation operator, $\hat a \ket{\alpha} = \alpha \ket{\alpha}$. Coherent states form an {\em overcomplete} non-orthogonal ($|\braket{\alpha}{\alpha'}|^2=\e^{-|\alpha-\alpha'|^2}$ because $\braket{\alpha}{\alpha'} = \exp{[-(|\alpha|^2+|\alpha'|^2)/2+\alpha^* \alpha']}$) set basis ($\frac{1}{\pi}\int d^2\alpha \ketbra{\alpha}{\alpha}= \id$) of vectors of the Hilbert space.} $\ket{\alpha} = \mathcal{\hat D}(\alpha) \ket{0} = \hat U_\alpha \ket{0}$
\begin{equation}
 \gamma_\alpha = S_\alpha \gamma_0 S_\alpha^T = \begin{pmatrix}
 1 & 0\\
 0 & 1
 \end{pmatrix}, \quad d_\alpha = S_\alpha d_0 + s_\alpha =
 \begin{pmatrix}
 q_0\\
 p_0
 \end{pmatrix},
\end{equation}
where $\alpha = \alpha_R + \im \alpha_I = \frac{q_0 + \im p_0}{\sqrt2}$.
\item {\em (Pure) Squeezed}: $\ket{r} = \mathcal{\hat S}(r) \ket{0} = \hat U_r \ket{0}$
\begin{equation}
 \gamma_r = S_r \gamma_0 S_r^T = \begin{pmatrix}
 e^{-2r} & 0\\
 0 & e^{2r}
 \end{pmatrix}, \quad d_r = S_r d_0 + s_r = \begin{pmatrix}
 0\\
 0
 \end{pmatrix}.
\end{equation}
\item {\em (Mixed) Thermal}: $\hat \rho_\beta = \frac{1}{\pi M} \int d^2\alpha \ketbra{\alpha}{\alpha} \e^{-|\alpha|^2/M}$
\begin{equation}
 \gamma_\beta = \begin{pmatrix}
 2M+1 & 0\\
 0 & 2M+1
 \end{pmatrix}, \quad d_\beta = \begin{pmatrix}
 0\\
 0
 \end{pmatrix},
\end{equation}
where $M = \frac{1}{e^{\beta\hbar\omega} - 1} \geq 0$ being $\beta$ the inverse temperature (we fix units such that $k_{B}=1$).
\end{itemize}
In Fig.~\ref{psexamples} we plot pictorically the above examples.

\begin{figure}[h!]
 \centering
a) \psfrag{A}{$q$}\psfrag{B}{$p$}\psfrag{C}{$\frac{1}{\sqrt2}$}\psfrag{D}{$\frac{1}{\sqrt2}$}
 \includegraphics[width=6cm]{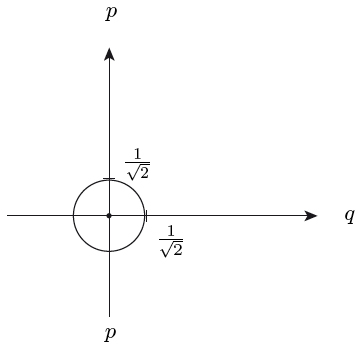} \hfill
b) \psfrag{A}{$q$}\psfrag{B}{$p$}\psfrag{C}{$q_0$}\psfrag{D}{$p_0$}\psfrag{E}{$\frac{1}{\sqrt2}$}\psfrag{F}{$\frac{1}{\sqrt2}$}
 \includegraphics[width=6cm]{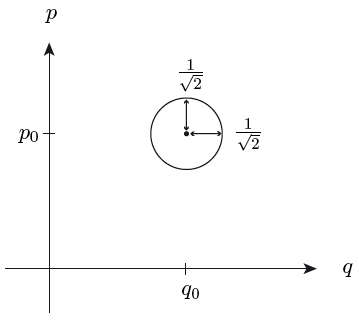}\\
c) \psfrag{A}{$q$}\psfrag{B}{$p$}\psfrag{C}{$\frac{\e^{-r}}{\sqrt2}$}\psfrag{D}{$\frac{\e^r}{\sqrt2}$}
 \includegraphics[width=6cm]{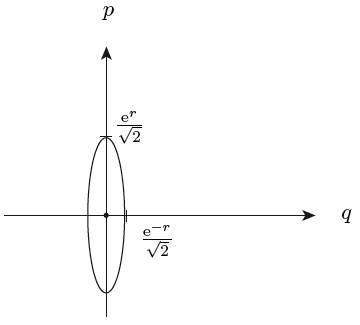} \hfill
d) \psfrag{A}{$q$}\psfrag{B}{$p$}\psfrag{C}{$\frac{\e^r}{\sqrt2}$}\psfrag{D}{$\frac{\e^{-r}}{\sqrt2}$}
 \includegraphics[width=6cm]{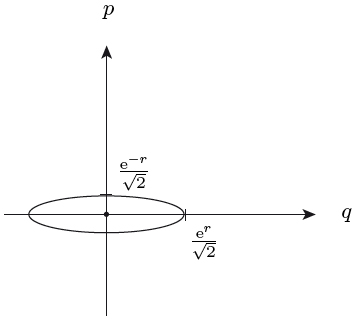}\\
e) \psfrag{A}{$q$}\psfrag{B}{$p$}\psfrag{C}{$\sqrt{M+1/2}$}\psfrag{D}{$\sqrt{M+1/2}$}
 \includegraphics[width=6cm]{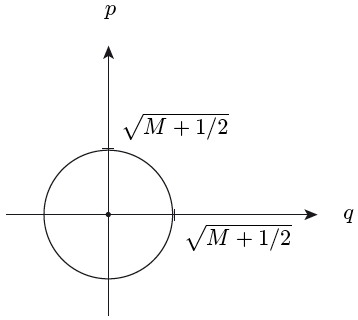}\\
 \caption{\small{a) Vacuum state. b) Coherent state being $\alpha=\frac{q_0+\im p_0}{\sqrt2}$. c) Squeezed state in position. d) Squeezed state in momentum. e) Thermal state of inverse temperature $\beta=\frac{1}{\hbar\omega}\ln(\frac{M+1}{M})$. All states except the thermal state are pure and are also minimal uncertainty states ($A=\frac{\pi}{2}$).}}\label{psexamples}
\end{figure}

\subsection{Hilbert space, phase-space and DV-CM connection}

We have already shown how to describe quantum states and operations at different levels {\em i.e.} Hilbert space, phase-space and DV-CM. Two main connections are needed still to perform calculations in the phase-space: the ordering of the operators and the metric between them.

The Weyl association rule tells us about the ordering operators. Provided we are using the Wigner distribution, which is symmetrical ordered, when working with observables we have to take into account that as we are in the phase-space (we have avoided its operator character) we have to symmetrize them. The way we have to symmetrize operators is

\begin{equation}
 e^{\im \zeta \hat{q} + \im \eta \hat{p}} \longrightarrow :e^{\im \zeta \hat{q} + \im \eta \hat{p}}: = e^{\im \zeta \hat{q} + \im \eta \hat{p}} \longleftrightarrow e ^{\im \zeta q + \im \eta p},
\end{equation}
where $:\,\,\,:$ stands for the symmetrical order. In general for a polynomial on $q$ and $p$

\begin{equation}
 \hat{q}^n\hat{p}^m \longrightarrow :\hat{q}^n\hat{p}^m: = \frac{1}{2^n}\sum_{r=0}^{n}\binom{n}{r}\hat{q}^r\hat{p}^m\hat{q}^{n-r} = \frac{1}{2^m}\sum_{r=0}^{m}\binom{m}{r}\hat{p}^r\hat{q}^n\hat{p}^{m-r} \longleftrightarrow q^np^m.
\end{equation}

Lets us present an example, consider the observable $\mathcal{QP}$, its quantum associated operator is of course $\hat q \hat p$. We know that $\hat q$ and $\hat p$ do not commute but in the phase-space $qp$ and $pq$ are functionally treated in the same way. Imagine we need to find its average value, we have then to remove the ambiguity by symmetrizing. The recipe is $\hat q \hat p \longrightarrow
:\hat{q}\hat{p}: = \frac{\hat{q}\hat{p}+\hat{p} \, \hat{q}}{2} \longleftrightarrow qp$. And so the average to be performed is
\begin{equation}
 <\mathcal{QP}> = <\hat q \hat p >_\rho = <\frac{\hat{q} \hat{p} + \hat{p} \hat{q}}{2} + \im/2 >_\rho 
 = <qp + \im/2>_{\mathcal{W}}.
\end{equation}
Operationally the averages on phase-space (with respect to the Wigner) correspond to the averages of symmetrical ordered operators on the Hilbert space.

More important and relevant averages, concern the moments, which can be obtained via the Wigner distribution as

\begin{equation}
 d_i = \tr (\hat{\rho} \hat{R}_i)= \int d^{2N}\zeta \,\, \left[ \zeta_i \right] \mathcal{W}(\zeta)
\end{equation}
and
\begin{equation}
 \gamma_{ij} = \tr (\hat{\rho} \{ \hat{R}_i-d_i \hat{\id}, \hat{R}_j-d_j \hat{\id} \}) = \int d^{2N}\zeta \,\, \left[ 2(\zeta_i - d_i)(\zeta_j - d_j) \right] \mathcal{W}(\zeta)
\end{equation}
where one see explicitly the symmetrization.

\begin{theorem}
 {\em (Quantum Parseval theorem)}\label{qpt}
 Let $\hat{W}_\zeta$ be a strongly continuous and irreducible Weyl system acting on the Hilbert space $\mathcal{H}_\Omega$ with phase-space $\Omega$. Then $\hat A \mapsto A^\chi(\eta) = \tr \{\hat A \hat{W}_\eta \}$, with $\eta \in \Omega$, is an isometric map from the Hilbert space $\mathcal{H}_\Omega$ (Hilbert-Schmidt operators) onto the Hilbert space $\mathcal{L}^2(\Omega)$ (square-integrable measurable functions on $\Omega$) such that\\
 \begin{equation}
 \tr (\hat A^\dag \hat B) = \frac{1}{(2\pi)^N} \int d^{2N}\eta \,\, \tr \{\hat A \hat{W}_\eta \}^*\tr \{\hat B \hat{W}_\eta \}.
 \end{equation}
\end{theorem}

From this theorem of capital importance, follows how to compute the scalar product between operators

\begin{equation}
 \begin{split}
 \tr \{\hat{A}^{\dag} \hat{B}\} &= \frac{1}{(2\pi)^N} \int d^{2N}\eta \, \, {A^\chi}^*(\eta) B^\chi(\eta) =\\
 &= \frac{1}{(2\pi)^N} \int d^{2N}\zeta \, \, A^\mathcal{W}(\zeta) B^\mathcal{W}(\zeta),
 \end{split}
\end{equation}
where the trace of an operator is defined via~\footnote{Use that $\id^\mathcal{W}=1$ and $\id^\chi=(2 \pi)^N \delta^{(2N)}(\eta)$ computed from Eq.~\eqref{gwigner} and Eq.~\eqref{gcharacteristic}.}
\begin{equation}
 \tr \{\hat{A}\} = A^\chi(0,0) = \frac{1}{(2\pi)^N} \int d^{2N}\zeta \, \, A^\mathcal{W}(\zeta),
\end{equation}
and the expectation value of an observable
\begin{equation}
 \begin{split}
 \langle \hat{A} \rangle_{\rho} &= \tr \{\hat{\rho} \hat{A}\} = \frac{1}{(2 \pi)^N} \int d^{2N}\eta \, \, \chi^*(\eta) A^\chi(\eta) =\\
 &= \int d^{2N}\zeta \,\,\mathcal{W}(\zeta)A^\mathcal{W}(\zeta).
 \end{split}
\end{equation}
To justify the above expression we just need to define properly the Fourier-Weyl transform for an arbitrary operator as
\begin{equation}
 \begin{split}
 \hat{A} &= \mathcal{FWT} \{A^\chi(\eta)\} =\frac{1}{(2 \pi)^N} \int d^{2N}\eta \,\, A^\chi(\eta) \hat{W}_{-\eta} =\\
 &= \frac{1}{(2 \pi)^{2N}} \int d^{2N}\eta \int d^{2N}\zeta \,\, A^\mathcal{W}(\zeta) e^{\im \zeta^T \cdot \mathcal{J} \cdot \eta} \hat{W}_{-\eta},
 \end{split}
\end{equation}
and its inverse

\begin{equation}\label{gcharacteristic}
 A^\chi(\eta) = \mathcal{FWT}^{-1} \{\hat{A}\} = \tr \{\hat{A} \hat{W}_{\eta} \},
\end{equation}

\begin{equation}\label{gwigner}
 A^\mathcal{W}(\bar\zeta,\bar\eta) = 2^N \int d^N\lambda \, \sandwich{\bar\zeta+\lambda}{A}{\bar\zeta-\lambda} e^{-2\im \bar\eta \lambda}
\end{equation}
where $\bar\zeta^T = (\zeta_1,\zeta_2,\ldots,\zeta_N$) idem for $\bar\eta$.

The two equivalent representations characteristic ($\chi$) and Wigner ($\mathcal{W}$) are $\mathcal {SFT}$ related~\footnote{Notice that $A^\mathcal{W} = (2\pi)^N\mathcal{W}$ if $\hat A = \hat\rho$ see Eq.~\eqref{wigner} (for normalization convenience) while $A^\chi = \chi$ if $\hat A = \hat\rho$ see Eq.~\eqref{characteristic}.}

\begin{equation}
 A^\mathcal{W}(\zeta) = \mathcal {SFT} \{A^\mathcal{\chi}(\eta) \} = \frac{1}{(2 \pi)^N} \int d^{2N}\eta \,\, A^\chi(\eta) e^{- \im \zeta^T \cdot \mathcal{J} \cdot \eta},
\end{equation}

\begin{equation}
 A^\chi(\eta) = \mathcal {SFT}^{-1} \{A^\mathcal{W}(\zeta)\} = \frac{1}{(2 \pi)^N} \int d^{2N}\zeta \,\, A^\mathcal{W}(\zeta) e^{\im \zeta^T \cdot \mathcal{J} \cdot \eta}.
\end{equation}

With the above transformations everything is now prepared to be translated in the phase-space.

\subsection{Fidelity and purity of Continuous Variable systems}

An important concept in statistical physics is to know how close/similar two distributions are. The natural solution is to construct a distance between distribution by choosing a proper metric in the distribution space. A metric $M$ is well defined if it satisfies the following properties:

\begin{enumerate}[i)]
\item {\em Symmetric:} $M(X,Y)=M(Y,X)$.
\item {\em Triangle inequality:} $M(X,Z) \leq M(X,Y)+M(Y,Z)$.
\item {\em Identity:} $M(X,Y)=0$ iff $X=Y$.
\item $\Leftarrow$ i)+ii)+iii) {\em Non-Negativity:} $M(X,Y) \geq 0$. The demonstration reads $2M(X,Y) = M(X,Y)+M(Y,X) \geq M(X,X) = 0$.
\end{enumerate}

Classically given two probability distributions $\{q_x\}$ and $\{p_x\}$ one can define the trace distance $D$ and the fidelity $F$ between them as follows
\begin{eqnarray}
D(\{q_x\},\{p_x\}) = \frac{1}{2}\sum_x |q_x-p_x|,\\
F(\{q_x\},\{p_x\}) = \sum_x\sqrt{q_x p_x},
\end{eqnarray}
while the trace distance is a proper metric this is not the case for the fidelity since it fails to agree with iii).

The quantum analogues quantities are the quantum trace distance~\footnote{Trace distance is constructed throw the trace norm $||\,\,||_1$ defined for an arbitrary matrix $M$ as $||M||_1 = \tr |M| = \tr \sqrt{M^T M}=\sum {\rm singular values}(M)=\sum {\rm eigenvalues}(\sqrt{M^T M})=\sum {\rm spec}(\sqrt{M^T M})$.} and the quantum fidelity 
\begin{eqnarray}
\mathcal{D}(\hat \rho, \hat \sigma)=\frac{1}{2}\tr |\hat \rho-\hat \sigma|,\\
\mathcal{F}(\hat \rho, \hat \sigma)=\left[ \tr \sqrt{\hat{\rho}^{1/2} \hat{\sigma} \hat{\rho}^{1/2}} \right]^2.
\end{eqnarray}
They can be related to the classical ones by considering the probability distributions obtained by a measurement
\begin{eqnarray}
\mathcal{D}(\hat \rho, \hat \sigma) = \max_{\{\hat E_n\}} D(\{q_n\},\{p_n\}),\\
\sqrt{\mathcal{F}(\hat \rho, \hat \sigma)} = \min_{\{\hat E_n\}} F(\{q_n\},\{p_n\}),
\end{eqnarray}
where $\{q_n\}=\tr(\hat \rho \hat E_n)$, $\{p_n\}=\tr(\hat \sigma \hat E_n)$ are the probability distributions of an arbitrary positive-operator-valued measurement (POVM) {\em i.e.} fulfilling $\sum_n\hat E_n=\id$.

Not only $\mathcal{D}(\hat \rho, \hat \sigma)$ is a proper metric, but also are the so-called Bures distance and Bures angle defined as $\mathcal{B}(\hat \rho, \hat \sigma) = \sqrt{2-2\sqrt{\mathcal{F}(\hat \rho, \hat \sigma)}}$ and $\mathcal{A}(\hat \rho, \hat \sigma) = \arccos(\sqrt{\mathcal{F}(\hat \rho, \hat \sigma)})$ respectively. From now on, we consider exclusively the properties of the quantum fidelity (also-called Bures-Uhlmann fidelity).

It's not obvious but it is symmetric and normalized between $1$ (equal states) and $0$ (orthogonal states). Its definition is simplified when one of the two states is pure (say $\hat\rho_1$), in this case it converges to the Hilbert-Schmidt fidelity

\begin{equation}
 \mathcal{F}(\hat{\rho}_1, \hat{\rho}_2) = \tr(\hat{\rho}_1 \hat{\rho}_2) = \sandwich{\psi_1}{\rho_2}{\psi_1}.
\end{equation}

In case both states are pure, then, the fidelity becomes simply the overlap (transition probability) between the two states

\begin{equation}
 \mathcal{F}(\hat{\rho}_1, \hat{\rho}_2) = |\braket{\psi_1}{\psi_2}|^2.
\end{equation}

It is useful here to use theorem~\ref{qpt} to evaluate the Hilbert-Schmidt fidelity between two Gaussian state (when at least one is pure)~\footnote{The second and third equality is true for all CV states.}

\begin{equation}
 \begin{split}
 \mathcal{F} (\hat{\rho}_1, \hat{\rho}_2) &= \tr (\hat{\rho}_1 \hat{\rho}_2) = \left(\frac{1}{2 \pi}\right)^N \int d^{2N}\eta \, \chi_1^*(\eta) \chi_2(\eta) = (2 \pi)^{N} \int d^{2N}\zeta \, \mathcal{W}_1(\zeta) \mathcal{W}_2(\zeta) =\\
 &= \frac{1}{\sqrt{\det (\frac{\gamma_1 + \gamma_2}{2})}} e^{-d^T (\frac{1}{\gamma_1 + \gamma_2}) d}
 \end{split}
\end{equation}
where $\gamma_{1(2)}$ and $d_{1(2)}$ belongs to $\hat{\rho}_{1(2)}$, while $d=d_2-d_1$.\\

Another important concept in Quantum Information is the purity $\mathcal{P}$ of a quantum state. 
In general, a pure state is a state which can be written in a suitable basis as a ket ($\hat \rho=\ketbra{\psi}{\psi}$) in the Hilbert space, and so $\hat \rho^2 = \hat \rho \, [\Rightarrow \tr \hat\rho^2=1]$.
A mixed state, on the contrary, cannot be written as a ket but only as a density operator and then $\hat \rho^2 \neq \hat \rho$.
In Continuous Variable a generic mixed state can always be written (for example using the $\mathcal{Q}$-function representation~\footnote{$\mathcal{Q}$-function is defined as $\mathcal{Q}(\alpha)=\frac{1}{\pi}\sandwich{\alpha}{\rho}{\alpha}$ and normalized as $\int d^2\alpha \mathcal{Q}(\alpha)=1$.}) as $\hat \rho = \int d^2\alpha \mathcal{Q}(\alpha)\ketbra{\alpha}{\alpha}$.

The purity, which measures how close is the state from a pure one, is defined as follows
\begin{equation}\label{purity}
 \mathcal{P}(\hat{\rho}) = \tr(\hat{\rho}^2).
\end{equation}
While for qu{\em d}its it is normalized between $1$ (pure states) and $\frac{1}{d}$ (maximally mixed states), for Continuous Variable systems (``$d\to\infty$'') maximally mixed states have purity 0. Using theorem~\ref{qpt} we can evaluate the purity of a Gaussian state~\footnote{The first equality is true for all CV states.}

\begin{equation}\label{purityGauss}
 \mathcal{P}(\hat{\rho}) = (2 \pi)^{N} \int d^{2N}\zeta \, [\mathcal{W} (\zeta)]^2 = \frac{1}{\sqrt{\det \gamma}}.
\end{equation}

\subsection{Bipartite Gaussian states, Schmidt decomposition, purification}

At the level of density operators, multipartite systems are described on a tensorial Hilbert space structure. This means that we have to ``tensor product'' $\otimes$, the Hilbert space of each party {\em i.e.} $\mathcal{H} = \bigotimes_{k=1}^N \mathcal{H}_k$. Notice however that multipartite Gaussian CV systems have a covariance matrix corresponding to a ``direct sum'' $\oplus$, of each party's associated phase-space {\em i.e.} $\Omega =
\bigoplus_{k=1}^N \Omega_k$. This is reminiscent of the Quantum Parseval theorem, which transforms tensor product between density matrices to products of Wigner functions (and Characteristic functions) and at the same time direct sums of covariance matrices and displacements vectors. Therefore, an advantage on Gaussian states is that we fully describe a state by a finite dimensional $N \times N$ matrix plus a $N\times1$ vector instead of its corresponding infinite dimensional density matrix. Additionally, dimensionality of the phase-space increases slower, as dimensions are added instead of multiplied.~\footnote{Remember that $\dim(\hat\rho_1\otimes\hat\rho_2)=\dim(\hat\rho_1)\dim(\hat\rho_2)$ while $\dim(\gamma_1\oplus\gamma_2)=\dim(\gamma_1)+\dim(\gamma_2)$.}

An important property of Gaussian states is that their reductions are again Gaussians. Imagine we have a bipartite Gaussian state $\hat\rho$ with covariance matrix $\gamma$ composed by $N_A+N_B=N$ modes~\footnote{From now on we suppose $N_A\geq N_B$.}, then tracing the $N_B$ modes corresponds to the reduced state $\hat\rho_A = \tr_B \hat\rho$ with covariance matrix $\gamma_A$ obtained by the upper left $N_A\times N_A$ block matrix of $\gamma$ (vice versa with $B$).
Therefore any bipartite Gaussian state can be written in a block structure as
$\gamma = \begin{pmatrix}
 A & C\\
 C^T & B
\end{pmatrix}$, where $A=A^T(=\gamma_A)$ and $B=B^T(=\gamma_B)$ are the reductions while $C$ amounts for the correlations between the modes.

For discrete variables, the Schmidt decomposition asserts that every pure bipartite state $\ket{\psi}$ (supposing $N_A \geq N_B$) can be transformed by local unitary operations to the normal form $\ket{\psi} = \sum_{i=1}^{N_B}\sqrt{\lambda_i}\ket{e_i}\otimes\ket{f_i}$ where $\{e_i\}$($\{f_i\}$) are orthonormal basis of $A$($B$) and $\{\lambda_i\}$ is the spectrum of the reduced state for $B$, $\hat\rho_B = \tr_A \ketbra{\psi}{\psi}$ satisfying $\lambda_i\geq0$ and $ \sum_{i=1}^{N_B}\lambda_i^2=1$.

\begin{theorem}
 {\em (Schmidt decomposition)}
 Every pure bipartite Gaussian states of $N=N_A+N_B$ modes $N_A \geq N_B$, by local symplectic transformations can be brought to the normal form
\begin{equation}
\gamma=\gamma_0 \oplus \bigoplus_{i=1}^{N_B}\gamma_i,
\end{equation}
where $\gamma_0=\id_{2(N_A-N_B)}$, $\gamma_i=\begin{pmatrix}
 \lambda_i & 0 & c_i & 0\\
 0 & \lambda_i & 0 & -c_i\\
 c_i & 0 & \lambda_i & 0\\
 0 & -c_i & 0 & \lambda_i
 \end{pmatrix}$, being $c_i=\sqrt{\lambda_i^2-1}$ and $\{\lambda_i\}$ the symplectic spectrum of the reduced covariance matrix for $B$.
\end{theorem}

We see a very peculiar behavior here, each mode of $B$ is entangled with at most one mode of $A$. The remaining $N_A-N_B$ modes of $A$ are uncorrelated.

\begin{lemma}
 {\em (Standard form I)}\label{sform}
 Every $1\times1$ mode (mixed) Gaussian state can be transformed, by local symplectic transformations to the standard form
 
 \begin{equation}\gamma=\begin{pmatrix}
 \lambda_a & 0 & c_x & 0\\
 0 & \lambda_a & 0 & c_p\\
 c_x & 0 & \lambda_b & 0\\
 0 & c_p & 0 & \lambda_b
 \end{pmatrix}.
 \end{equation}
\end{lemma}
A Gaussian state in the standard form is called symmetric if $\lambda_a = \lambda_b$, and fully symmetric if it is symmetric and in addition $c_x = -c_p$.

There is a simple way to construct the standard form if one uses the following four local symplectic invariants,~\footnote{These four invariants can be written in terms of the symplectic spectrum of the covariance matrix of the state and its reductions as $\mathcal{P}=1/\sqrt{\prod_i\mu_i^2}$, $\Delta=\sum_i\mu_i^2$, $\mathcal{P}_A=1/\sqrt{\prod_i\mu_{A,i}^2}$, $\mathcal{P}_B=1/\sqrt{\prod_i\mu_{B,i}^2}$.} the purities $\mathcal{P}_A = 1/\sqrt{\det A}=1/\lambda_a$, $\mathcal{P}_B = 1/\sqrt{\det B}=1/\lambda_b$, $\mathcal{P} = 1/\sqrt{\det \gamma} = \left[(\lambda_a \lambda_b-c_x^2)(\lambda_a \lambda_b-c_p^2)\right]^{-1/2}$, and the serelian $\Delta = \det A+\det B+2\det C = \lambda_a^2+\lambda_b^2+2c_xc_p$ because they can be inverted as follows

\begin{equation}
 \begin{split}
 \lambda_a &=1/\mathcal{P}_A,\\
 \lambda_b &=1/\mathcal{P}_B,\\
 c_x &=\frac{\sqrt{\mathcal{P}_A\mathcal{P}_B}}{4}(a_- +a_+),\\
 c_p &=\frac{\sqrt{\mathcal{P}_A\mathcal{P}_B}}{4}(a_- -a_+),\\
 a_{\pm} &=\sqrt{\left[ \Delta-(\mathcal{P}_A\pm\mathcal{P}_B)^2/(\mathcal{P}_A\mathcal{P}_B)^2 \right]^2 - 4/\mathcal{P}^2}.
 \end{split}
\end{equation}

The local and global purities, $\mathcal{P}_A$, $\mathcal{P}_B$ and $\mathcal{P}$ of the state are constrained to be less or equal to one, {\em i.e.} $\lambda_a,\lambda_b \geq 1$ and $(\lambda_a \lambda_b-c_x^2)(\lambda_a \lambda_b-c_p^2)\geq1$. The symplectic positivity \eqref{positivity} implies, in terms of the invariants, that $1+\frac{1}{{\mathcal{P}}^2} \geq \Delta$ or equivalently $1+(\lambda_a \lambda_b-c_x^2)(\lambda_a \lambda_b-c_p^2)\geq\lambda_a^2+\lambda_b^2+2c_xc_p$.

We stress here that all pure bipartite Gaussian states are symmetric ($\lambda_a=\lambda_b=\lambda$) and fulfills $c_x=-c_p=\sqrt{\lambda^2-1}$ being $\lambda \geq 1$. Introducing the change of parameters, $\cosh{2r}=\lambda$ we can write any pure bipartite $1\times1$ Gaussian state as a two mode squeezed state $\gamma_{TMS}=S_{TMS}\id S^T_{TMS}=\begin{pmatrix}
 \cosh{2r} & 0 & \sinh{2r} & 0\\
 0 & \cosh{2r} & 0 & -\sinh{2r}\\
 \sinh{2r} & 0 & \cosh{2r} & 0\\
 0 & -\sinh{2r} & 0 & \cosh{2r}
 \end{pmatrix}$ with positive $r$.
 
\begin{lemma}
 {\em (Purification)}\label{puri}
 Every (mixed) Gaussian state of $N_A$ modes represented by $\gamma_A$ admits a purification, {\em i.e.} there exist a pure Gaussian state of $2 N_A$ modes whose reduction is $\gamma_A$ and reads
\begin{equation}
 \gamma = \begin{pmatrix}
 \gamma_A & C\\
 C^T & \theta_A\gamma_A\theta_A^T
\end{pmatrix},
\end{equation}
with $C=\mathcal{J}\sqrt{-(\mathcal{J}\gamma_A)^2 - \id} \, \theta_A$ where $\theta_A=\begin{pmatrix}
 1 & 0\\
 0 & -1
\end{pmatrix}^{\oplus N_A}$.
\end{lemma}

\subsection{States and operations}\label{staandope}

By virtue of the Choi-Jamio\l kowski isomorphism between completely positive (CP) maps acting on $\mathcal B(\mathcal H)$ (physical actions) and positive operators belonging to $\mathcal B(\mathcal H)\otimes\mathcal B(\mathcal H)$ (unnormalized states), Gaussian operations were fully characterized in \cite{Giedke2002PRA}. In there the authors showed that each Gaussian operation $\mathcal G$ can be associated to a corresponding Gaussian state $G$ {\em i.e.} there exist an isomorphism between Gaussian CP maps and Gaussian states. Since all Gaussian states can be generated from vacuum state by Gaussian unitary operations and discarding subsystems, then symplectic transformations plus homodyne measurements complete all Gaussian operations.

\begin{lemma}\label{iso}
 {\em (State-operation's isomorphism lemma)}
If a $N_A \times N_B$-mode Gaussian state $G$ is defined by its first and second moments through
\begin{equation}
G: \quad
\Gamma=\left(\begin{array}{cc}
\Gamma_A&\Gamma_{AB}\\
\Gamma_{AB}^T&\Gamma_B
\end{array}\right), \quad
\Delta=\left(\begin{array}{c}
\Delta_A\\
\Delta_B
\end{array}\right),
\end{equation}
then the application of $\mathcal G$ on a $N_B$-mode Gaussian state $(\gamma,d)$ produces a $N_A$-mode Gaussian state $(\gamma',d')$ such that
\begin{eqnarray}
\mathcal{G}: \quad
\gamma \mapsto \gamma' &=& \tilde{\Gamma}_A-\tilde{\Gamma}_{AB}\frac{1}{\tilde{\Gamma}_B+\gamma}\tilde{\Gamma}_{AB}^T,\\
d \mapsto d' &=& \Delta_A+\tilde{\Gamma}_{AB}\frac{1}{\tilde{\Gamma}_B+\gamma}(\Delta_B+d),
\end{eqnarray}
where $\tilde{\Gamma}=(\id_N \oplus \theta_N)\Gamma(\id_N \oplus \theta_N)$ with $N=N_A+N_B$.
\end{lemma}

Homodyne detection is a typical example of a Gaussian operation, which realizes a projective (or von Newmann) measurement of one quadrature operator, say $\hat x$, thus with associated POVM $\ketbra{x}{x}$.
Take a Gaussian state $\gamma$ of $N$ modes, it can always be divided into $N_A \times N_B$ modes as
\begin{equation}
\gamma=\left(\begin{array}{cc}
\gamma_A&C\\
C^T&\gamma_B
\end{array}\right),
\end{equation} and with zero displacement vector. Then, a homodyne measurement of $\hat x$ on the last $N_B$ modes, by the lemma~\ref{iso}, can be described by a Gaussian operator $\hat \rho_x$ with corresponding DV and CM given by

\begin{equation}
\Delta_x^T=\left(0,0,\ldots,x,x,\ldots\right)\nonumber
\end{equation}
and

\begin{equation}
\Gamma_x=\lim_{r\to\infty}\left(\begin{array}{ccc}
\cosh{r}\,\id_{N_A} & \sinh{r}\,\theta_{N_A} & 0\\
\sinh{r}\,\theta_{N_A} & \cosh{r}\,\id_{N_A} & 0\\
0 & 0 & \left(\begin{array}{cc}
1/r&0\\
0&r
\end{array}\right)\id_{N_B}
\end{array}\right)\nonumber.
\end{equation}

Summarizing, if we measure the $x$ component of the last $N_B$ modes corresponding to $B$, obtaining the result $(x_1,x_2,\ldots,x_{N_B})$, system $A$ will turn into a Gaussian state with covariance matrix
\begin{equation}
\gamma_A'=\gamma_A-C^T(X \gamma_B X)^{\mathcal{MP}}C,
\end{equation}
and displacement vector

\begin{equation}
d'_A=C^T(X \gamma_B X)^{\mathcal{MP}}d'_B,
\end{equation}
where $d'_B=(x_1,0,x_2,0,\ldots,x_{N_B},0)$, ${\mathcal{MP}}$ denotes Moore Penrose or pseudo-inverse (inverse on the support whenever the matrix is not of full rank) and $X$ is the projector with diagonal entries ${\rm diag}(1,0,1,0,\ldots)$.

Heterodyne measurement, whose POVM corresponds to $\frac{1}{\pi}\ketbra{\alpha}{\alpha}$, and in general~\footnote{All pure Gaussian states can be obtained from $\ket{\alpha}$ by Gaussian unitaries.} all POVM of the form $\ketbra{\gamma,d}{\gamma,d}$, can be achieved with homodyne measurement by the use of ancillary systems and beam splitters.

\section{Entanglement in Continuous Variable: criteria and measures}

Concerning bipartite entanglement, for discrete variable systems an important separability criterium based on the partial transpose (time reversal) exists. If $\hat\rho$ is a generic state, then $\hat\rho^{T_A}$ represents the state after one perform time reversal on subsystem $A$.

\begin{lemma}
 {\em (NPPT Peres criterium)}\label{NPPTP}
 Given a bipartite state $\hat\rho$, if it has non-positive partial transpose ($\hat\rho^{T_A} \ngeq 0 \Rightarrow \hat\rho^{T_B} \ngeq 0$), then $\hat\rho$ is entangled {\em \cite{Peres1996PRL}}.
\end{lemma}

\begin{lemma}
 {\em (NPPT Horodecki criterium)}\label{NPPTH}
 In $\mathbb{C}^2\otimes\mathbb{C}^2$ and $\mathbb{C}^2\otimes \mathbb{C}^3$ given a bipartite state $\hat\rho$, it is entangled iff it has non-positive partial transpose ($\hat\rho^{T_A} \ngeq 0
 \Rightarrow \hat\rho^{T_B} \ngeq 0$) {\em \cite{Horodecki1996PLA}}.
\end{lemma}
For Continuous Variable states, Peres criterium also holds while Horodecki criterium is true provided our state is composed of $1 \times N$ modes. In particular for Gaussian states, time reversal is very easy to implement at the covariance matrix level. If $\hat T$ is the reversal operator then $S_T =\theta= \begin{pmatrix}
 1 & 0\\
 0 & -1
\end{pmatrix}$ is the corresponding symplectic operations in phase-space. So we can rewrite the lemma~\ref{NPPTH} for Gaussian states.

\begin{lemma}
 {\em (NPPT Simon criterium)}\label{GNPPTH}
 For $1 \times N$ modes given a bipartite Gaussian state $\gamma$, it is entangled iff it has non-positive partial transpose ($\theta_A \gamma \theta_A^T + \im \mathcal{J} \ngeq 0 \Rightarrow \theta_B \gamma \theta_B^T + \im \mathcal{J} \ngeq 0$) {\em \cite{Simon2000PRL,Werner2001PRL}}.
\end{lemma}
Under partial transposition the serelian changes as $\Delta \to \tilde\Delta=\Delta-4\det C=\lambda_a^2+\lambda_b^2-2c_xc_p$ and so positivity of the partial transpose can be written, in terms of invariants, as $1+\frac{1}{{\mathcal{P}}^2} \geq \tilde\Delta$ while in terms of the symplectic eigenvalues of the partial transpose, as $\tilde\mu_i \geq1$, $i=1,2,\ldots,N$.

\begin{lemma}
 {\em (CV Inseparability Duan criterium)\footnote{For all CV states a less restrictive upper bound is imposed by the uncertainty principle, {\em i.e.} $\var{\hat u}_\rho+\var{\hat v}_\rho \geq \left|a^2-\frac{1}{a^2}\right|$ due to the sum uncertainty relation $(\Delta \hat A)^2 + (\Delta \hat B)^2 \geq 2\Delta \hat A \Delta \hat B \geq |\langle\comm{A}{B}\rangle|$.}}\label{INSEPD}
 We define the EPR-like operators, $\hat u = |a|\hat x_1 + \frac{1}{a} \hat x_2$ and $\hat v = |a|\hat p_1 - \frac{1}{a} \hat p_2$ where $a \in \mathbb{R}$. If a bipartite state $\hat \rho$ (Gaussian or non-Gaussian) is separable, then $\var{\hat u}_\rho+\var{\hat v}_\rho \geq a^2+\frac{1}{a^2}$, for all $a \in \mathbb{R}$ {\em \cite{Duan2000PRL}}.
\end{lemma}

\begin{lemma}
 {\em (GS Inseparability Duan criterium)}\label{GINSEPD}
 Any bipartite Gaussian state $\hat \rho$ can be written in standard form II (by two local squeezings on standard form I) as $\gamma=\left(
 \begin{array}{cccc}
 n_1 & 0 & c_1 & 0\\
 0 & n_2 & 0 & c_2\\
 c_1 & 0 & m_1 & 0\\
 0 & c_2 & 0 & m_2
 \end{array}
 \right)$ where $\frac{n_1-1}{m_1-1} = \frac{n_2-1}{m_2-1}$, $|c_1|-|c_2|=\sqrt{(n_1-1)(m_1-1)}- \sqrt{(n_2-1)(m_2-1)}$, then the state is separable iff $\var{\hat u}_\rho+\var{\hat v}_\rho \geq a_0^2+\frac{1}{a_0^2}$, for all $a_0 \in \mathbb{R}$, where $a_0^2=\sqrt{\frac{m_1-1}{n_1-1}}=\sqrt{\frac{m_2-1}{n_2-1}}$ and EPR-like operators $\hat u = a_0\hat x_1 - \frac{c_1}{|c_1|}\frac{1}{a_0} \hat x_2$ and $\hat v = a_0\hat p_1 - \frac{c_2}{|c_2|}\frac{1}{a_0} \hat p_2$ {\em \cite{Duan2000PRL}}.
\end{lemma}

Concerning entanglement measures it is usual to deal, as an entanglement measure for pure state, with the entropy of entanglement and for mixed ones with the logarithmic negativity.

\begin{itemize}
\item Entropy of entanglement:
\begin{equation}
 E_S(\hat\rho) = S(\hat\rho_A) = -\tr(\hat\rho_A \log_2 \hat\rho_A),
\end{equation}
where $S$ is the von Neumann Entropy $S(\hat\rho) = -\tr(\hat\rho \log_2 \hat\rho)$, and $\hat\rho_A$ is the reduction of $A$. For any CV state it reduces (in terms of the Schmidt coefficients) to
\begin{equation}
 E_S(\hat\rho) = -\sum_{i=1}^\infty \lambda_i^2 \log_2 \lambda_i^2,
\end{equation}
while for Gaussian states (in terms of the symplectic eigenvalues),
\begin{equation}
 E_S(\gamma) = -\sum_{i=1}^{N_A}[(\frac{\mu_i+1}{2}) \log_2 (\frac {\mu_i+1}{2}) - (\frac{\mu_i-1}{2}) \log_2 (\frac{\mu_i-1}{2})],
\end{equation}
where $\{\pm\mu_i\} = {\rm spec}(-\im \mathcal{J} \gamma_A)$.

The entropy of entanglement is the ``unique'' measure of entanglement for pure states. It depends only on the Schmidt coefficients and not on the choice of basis, therefore it's invariant under local unitary operations. Furthermore it's unique because all other entanglement measures are in direct correspondence with the entropy of entanglement.

\item Logarithmic negativity \cite{Vidal2002PRA}:
\begin{equation}
 E_N(\hat\rho) = LN(\hat\rho) = \log_2 ||\hat\rho^{T_A}||_1,
\end{equation}
where $||\,\,||_1$ is the trace norm~\footnote{The trace norm for hermitian operators is simple because $||\hat M||_1 = \tr \sqrt{\hat M^2}=\sum |{\rm spec}(\hat M)|=1+2\sum|{\rm negativevalues}(\hat M)|$.}. For any CV state it can be written (in terms of the negative eigenvalues of the partial transpose) as
\begin{equation}
 E_N(\hat\rho) = \log_2 [1 + 2\sum_{i=1}^\infty |\min(\tilde\lambda_i,0)|],
\end{equation}
while for Gaussian states (in terms of the symplectic eigenvalues of the partial transpose),
\begin{equation}\label{lognega}
 E_N(\gamma) = -\sum_{i=1}^N\log_2 [\min(\tilde\mu_i,1)],
\end{equation}
where $\{\pm\tilde \mu_i\} = {\rm spec}(-\im \mathcal{J} \gamma^{T_A})$.

The negativity $N(\hat\rho) = \frac{2^{LN(\hat\rho)}-1}{2}$ is also a computable measure of entanglement for mixed states. It quantifies the violation of the NPPT criterium {\em i.e.} how much the partial transposition of a density matrix fails to be positive. It's invariant under local unitary operations and an entanglement monotone. For Gaussian states reads

\begin{equation}\label{nega}
 N(\gamma) = \frac{1}{2}[\prod_{i=1}^N\frac{1}{\min(\tilde\mu_i,1)}-1].
\end{equation}

\end{itemize}

\newpage

\begin{subappendices}

\section{Appendix: Quantum description of light}\label{appendix1}

\subsection{Classical harmonic oscillator}

A 1$D$ classical harmonic oscillator obeys a second order ordinary differential equation $\ddot x = -\omega^2 x$ in the position $x$ where $\omega$ is the angular frequency. This equation can be derived in the Lagrangian formalism from the Lagrangian $L=\frac{m}{2}\dot x^2-\frac{m \omega^2}{2}x^2$ using Euler-Lagrange equations. Here the constant $m$ turns to be the mass.

To treat this system quantum mechanically one need first to derive the Hamiltonian of the system, say~\footnote{As usual $p$ is the canonical momentum conjugated to $x$, {\em i.e.}, $p=\frac{\partial L}{\partial \dot x}=m\dot x$.} $H=-\sum_i\frac{\partial L}{\partial \dot x_i}\dot x_i - L = \frac{p^2}{2m} + \frac{m \omega^2}{2}x^2$, and perform the $1^{\rm st}$ quantization transcription. One ends up with the Hamiltonian operator

\begin{equation}
\hat H = \frac{m \omega^2}{2}\hat x^2 + \frac{\hat p^2}{2m}
\end{equation}
and a commutation relation that says $\comm{x}{p} = \im \hbar$.

\subsection{Light modes as independent harmonic oscillators}

Since Maxwell, we know that classically, light can be relativistically described~\footnote{Compactly in covariant form they read $\partial_\mu F^{\mu\nu} = -J^\nu$ and $\partial_\mu \tilde F^{\mu\nu} = 0$
where $J^\mu = (c \rho,\vec J)$ is the 4-vector current and $F_{\mu\nu} = \partial_\mu A_\nu - \partial_\nu A_\mu$ the 4-tensor Faraday with $A^\mu = (\phi,c \vec A)$.} by the so-called Maxwell field equations. As we will show, each mode of the electromagnetic field is an independent harmonic oscillator with the same classical dynamics. This is shown by constructing the classical Hamiltonian from the Maxwell equations as follows.

We start with Maxwell equations for the electric and magnetic field in presence of sources ($\rho$) and currents ($\vec J$)
\begin{equation}
 \begin{array}{ccl}
 \vec \nabla \cdot \vec E &=& \frac{\rho}{\varepsilon_0},\\
 \vec \nabla \cdot \vec B &=& 0,\\
 \vec \nabla \times \vec E &=& -\frac{\partial \vec B}{\partial t},\\
 \vec \nabla \times \vec B &=& \mu_0 \varepsilon_0 \frac{\partial \vec E}{\partial t} + \mu_0 \vec J,
 \end{array}
\end{equation}
where $\varepsilon_0$ is the permittivity and $\mu_0$ the permeability of the vacuum.
It's here useful to introduce the scalar, $\phi$, and vectorial potential, $\vec A$, fields throw $\vec E = - \vec \nabla \phi - \frac{\partial \vec A}{\partial t}$ and $\vec B = \vec \nabla \times \vec A$. They are more fundamental although not so very ``physical'' (it is not gauge invariant, for example). Nevertheless it helps to simplify the calculations, because in this way we end up with only one scalar and one vectorial second order partial differential equations, being the other two remaining equations trivial identities

\begin{equation}
 \begin{array}{lcc}
 \vec \nabla^2 \phi + \frac{\partial}{\partial t}(\vec \nabla \cdot \vec
 A)= -\frac{\rho}{\varepsilon_0},\\
 \vec \nabla^2 \vec A - \vec \nabla \cdot (\vec \nabla \cdot \vec A) = \mu_0 \varepsilon_0 \frac{\partial}{\partial t} (\vec \nabla \phi + \frac{\partial \vec A}{\partial t})- \mu_0 \vec J,
 \end{array}
\end{equation}
From now on we work in free space~\footnote{Recall that $\mu_0 \varepsilon_0 = 1/c^2$.} $\vec J = \rho = 0 [\Rightarrow \phi = 0]$. We still have the freedom to fix the gauge. The appropriate gauge is the radiation (Coulomb) gauge $\vec \nabla \cdot \vec A = 0$. As a result we end up with the wave equation~\footnote{In covariant form it reads $\Box A^\nu = 0$.}

\begin{equation}
 \left( \vec \nabla ^2 - \frac{1}{c^2}\frac{\partial^2}{\partial t^2} \right) \vec A = \vec 0.
\end{equation}
An ansatz for the solution of the vector potential can be written as a superposition with polarizations $\alpha=x,y$ and frequencies $\vec k$ motivated by the linearity plus a separation on time-space variables

\begin{equation}
 \vec A(\vec r, t) = \sum_{k,\alpha} \vec A_{k, \alpha}(\vec r, t) = \sum_{k,\alpha} q_{k,\alpha}(t) \vec u_{k,\alpha}(\vec r).
\end{equation}
With such an ansatz, the wave equation decouples in a Helmholtz equation $\vec \nabla^2 \vec u_{k,\alpha}(\vec r) = -k^2 \vec u_{k,\alpha}(\vec r)$ and a harmonic oscillatory equation $\ddot{q}_{k,\alpha}(t) = -c^2 k^2 q_{k,\alpha}(t)$.
For the spatial solution, it is conventional to define a ``box'' of volume $L^3$ in such a way that the normalized solution ($\int_{L^3} d^3 \vec r \, \vec u^*_{k,\alpha}(\vec r) \vec u_{k',\alpha'}(\vec r) = \delta_{k,k'} \delta_{\alpha,\alpha'}$) correspond to traveling plane waves

\begin{equation}
 \vec u_{k,\alpha}(\vec r) = \frac{\vec e_\alpha}{\sqrt{L^3}} e^{\im \vec k \cdot \vec r},
\end{equation}
where $\vec e_\alpha$ is the polarization vector. This is so because we have imposed periodic boundary conditions on the walls of the quantization box in order to mimic the behavior of the electromagnetic field in free space (in order to find expressions in the continuum one has to perform the limit $L\to\infty$ appropriately).
For the temporal solution, we recall that the corresponding classical Hamiltonian is $H_{k,\alpha} = \frac{1}{2} m \omega_k^2 q_{k,\alpha}^2+\frac{1}{2m}p_{k,\alpha}^2$. The solution is nothing else than a oscillatory movement in the $q_{k,\alpha}(t)$ coordinate. We introduce the conjugate momentum~\footnote{As usual $p$ is the canonical momentum conjugated to $x$, {\em i.e.}, $p=\frac{\partial L}{\partial \dot x}=m\dot x$.}, $p_{k,\alpha}(t) = m \dot{q}_{k,\alpha}(t)$. The angular frequency is $\omega_k = c \, k$ while the ``mass'', which is a free parameter in the Lagrangian formalism, has to be identified with the permittivity $\varepsilon_0$ of the vacuum. Thus one has to solve the coupled system

\begin{equation}
 \begin{array}{lcc}
 \dot{p}_{k,\alpha} = -\varepsilon_0 \omega_k^2 q_{k,\alpha},\\
 \dot{q}_{k,\alpha} = \frac{1}{\varepsilon_0} p_{k,\alpha},
 \end{array}
\end{equation}
which decouples with the change of variables~\footnote{The inverse transformation results in $q = \sqrt{\frac{\hbar}{2 \varepsilon_0\omega}} ( a^* + a )$ and $p = \im \sqrt{\frac{\varepsilon_0 \hbar \omega}{2}} ( a^* - a )$.} $a = \sqrt{\frac{\varepsilon_0\omega}{2\hbar}} ( q + \frac{\im p}{\varepsilon_0 \omega} )$, $a^* = \sqrt{\frac{\varepsilon_0\omega}{2\hbar}} ( q - \frac{\im p}{\varepsilon_0 \omega} )$
to

\begin{equation}
 \begin{array}{lcc}
 \dot a_{k,\alpha} = -\im \omega_k a_{k,\alpha},\\
 \dot a_{k,\alpha}^* = \im \omega_k a_{k,\alpha}^*,
 \end{array}
\end{equation}
with trivial oscillatory solutions $a_{k,\alpha}(t) = a_{k,\alpha} e^{-\im \omega_k t}$ and $a_{k,\alpha}^*(t) = a_{k,\alpha}^* e^{\im \omega_k t}$. Finally

\begin{equation}
 q_{k,\alpha}(t) = \sqrt{\frac{\hbar}{2 \varepsilon_0 \omega_k}}(a_{k,\alpha} e^{-\im \omega_k t} + a^*_{k,\alpha} e^{\im \omega_k t}).
\end{equation}
Grouping the space and time solution and summing over ($k,\alpha$) one obtain the electromagnetic waves solution $\vec A_{k,\alpha} (\vec r, t)=f(\vec k \cdot \vec r \pm \omega_k t)$.

\subsection{Quantization of the electromagnetic field}

If we quantize the solution for the potential vector we can now write it in terms of creation and annihilation operators with bosonic commutation rules ${[ \hat{a}_{k,\alpha},\hat{a}_{k,\alpha}^\dag ]} = \delta_{k,k'} \delta_{\alpha,\alpha'}$ as

\begin{equation}
 \hat{\vec A}(\vec r, t) = \sum_{k,\alpha} \sqrt{\frac{\hbar}{2 \omega_k \varepsilon_0 L^3}} \left( \vec e_\alpha \hat a_{k,\alpha} e^{\im (\vec k \cdot \vec r - \omega_k t)} + \vec e_\alpha^* \hat a_{k,\alpha}^\dag e^{-\im( \vec k \cdot \vec r - \omega_k t)} \right).
\end{equation}
Quantized electric and magnetic fields can be straightforward recovered via definition of ${\vec A}$. In the Heisenberg picture, the following expressions are obtained

\begin{equation}
 \hat{\vec E}(\vec r, t) = - \frac{\partial}{\partial t}\hat{\vec A}(\vec r, t) = \im \sum_{k,\alpha} \sqrt{\frac{\hbar \omega_k}{2 \varepsilon_0 L^3}} \left( \vec e_\alpha \hat a_{k,\alpha} e^{\im (\vec k \cdot \vec r - \omega_k t)} - \vec e_\alpha^* \hat a_{k,\alpha}^\dag e^{-\im( \vec k \cdot \vec r - \omega_k t)} \right),
\end{equation}
where the unitary electric field polarization vector $\vec e_\alpha \perp \vec k$ because $\vec \nabla \cdot \hat{\vec E} = 0$ being $\vec e_\alpha^* \cdot \vec e_{\alpha'}=\delta_{\alpha,\alpha'}$ and

\begin{equation}
 \hat{\vec B}(\vec r, t) = \vec \nabla \times \hat{\vec A}(\vec r, t) = \im \sum_{k,\alpha} \sqrt{\frac{\mu_0 \hbar \omega_k}{2 L^3}} \left( \vec f_\alpha \hat a_{k,\alpha} e^{\im (\vec k \cdot \vec r - \omega_k t)} - \vec f_\alpha^* \hat a_{k,\alpha}^\dag e^{-\im( \vec k \cdot \vec r - \omega_k t)} \right),
\end{equation}
where the unitary magnetic field polarization vector $\vec f_\alpha = \vec e_\alpha \times \frac{\vec k}{|\vec k|} \perp \vec e_\alpha, \vec k$ fulfills $\vec f_\alpha^* \cdot \vec f_{\alpha'}=\delta_{\alpha,\alpha'}$.

\begin{figure}[h]
 \centering
 \psfrag{A}{$x$}\psfrag{B}{$y$}\psfrag{C}{$z$}\psfrag{D}{$\vec k$}\psfrag{E}{$<\hat{\vec E}(\vec r, t_0)>$}\psfrag{F}{$<\hat{\vec B}(\vec r, t_0)>$}
 \includegraphics[width=8cm]{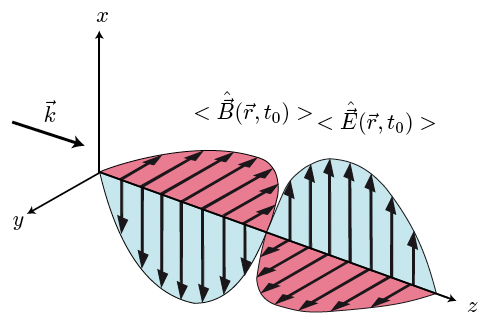}
 \caption{\small{Electric, magnetic and wave vectors of an electromagnetic traveling plane wave at a fixed time $t_0$.}}\label{emwave}
\end{figure}
One can compute the Hamiltonian for a single mode
\begin{equation}
 \hat H_{k,\alpha} = \frac{1}{2} \int_{L^3} d^3 \vec r \left( \varepsilon_0 |\hat{\vec E}_{k,\alpha}|^2 + \frac{1}{\mu_0} |\hat{\vec B}_{k,\alpha}|^2 \right ) = \frac{\hbar \omega_k}{2} \left( \hat a_{k, \alpha}^\dag \hat a_{k, \alpha} + \hat a_{k, \alpha} \hat a_{k, \alpha}^\dag \right),
\end{equation}
then the total Hamiltonian reads
\begin{equation}
 \hat H_{EM} = \frac{1}{2} \int_{L^3} d^3 \vec r \left( \varepsilon_0|\hat{\vec E}|^2 + \frac{1}{\mu_0}|\hat{\vec B}|^2 \right ) = \sum_{k, \alpha} \hbar \omega_k \left( \hat a_{k, \alpha}^\dag \hat a_{k, \alpha} + \frac{1}{2} \right).
\end{equation}
It is in this moment when one can, by comparing the above Hamiltonian with the one of the quantized harmonic oscillator
\begin{equation}
 \hat H_{HO} = \sum_{k, \alpha} \hbar \omega_k \left( \hat a_{k, \alpha}^\dag \hat a_{k, \alpha} + \frac{1}{2} \right),
\end{equation}
associate each mode of the electromagnetic field to an independent harmonic oscillator with the same frequency.

\subsection{Quadratures of the electromagnetic field}

To end with the comparison between the light and the harmonic oscillator we present here -in analogy to the position-momentum coordinates of the massive particle in the harmonic oscillator- the amplitude-phase quadratures of light modes.
Q-quadrature (amplitude quadrature):

\begin{equation}
 \hat Q_{k,\alpha} = \frac{\hat a_{k,\alpha}^\dag + \hat a_{k,\alpha}}{\sqrt2},
\end{equation}
P-quadrature (phase quadrature):

\begin{equation}
 \hat P_{k,\alpha} = \im \frac{(\hat a_{k,\alpha}^\dag - \hat a_{k,\alpha})}{\sqrt2},
\end{equation}
with bossonic commutation rules
\begin{equation}
 {[ \hat{Q}_{k,\alpha},\hat{P}_{k',\alpha'} ]} = \im \delta_{k,k'} \delta_{\alpha,\alpha'},
\end{equation}
and Heisenberg uncertainty principle
\begin{equation}
 \Delta Q \Delta P \geq \frac{1}{2}|<\comm{Q}{P}>|= \frac{1}{2}.
\end{equation}
In terms of the quadratures the Hamiltonian reads

\begin{equation}
 \hat H_{EM} = \hat H_{HO} = \sum_{k, \alpha} \frac{\hbar \omega_k}{2} \left( \hat Q_{k, \alpha}^2 + \hat P_{k, \alpha}^2 \right),
\end{equation}
while the electric field~\footnote{The classical values for the fields correspond to the expectation values for the quantum fields.}

\begin{equation}
 |\hat{\vec E}(\vec r, t)| = \sum_{k,\alpha} \sqrt{\frac{\hbar \omega_k}{\varepsilon_0 L^3}} \left( \hat Q_{k,\alpha} \sin{(\omega_k t-\vec k \cdot \vec r)} - \hat P_{k,\alpha} \cos{(\omega_k t-\vec k \cdot \vec r)} \right).
\end{equation}
The uncertainty principle is encoded between the non-commuting observables potential field $\hat{\vec A}$ and electric $\hat{\vec E}$ (or magnetic field $\hat{\vec B}$). Quadrature operators are nothing else than observables proportionals to $\hat{\vec A}$ and $\hat{\vec E}$ at the same time, and so they contain the same information about the light. But in an harmonic oscillator the position and momentum are always one quarter cycle out of phase, therefore the amplitude of the electric field ``now'' and a quarter cycle later effectively describe both the electric and potential field.

\subsection{Quantum states of the electromagnetic field}

We have seen that the Hamiltonian can be written as an harmonic oscillator for each mode, thus the eigenstates of the Hamiltonian are Fock states, which turn out to be a very suitable basis to express generic quantum states of light. In section~\ref{DViCM} we have described many examples of Gaussian states at the phase-space level. Here the aim is to describe them by looking at the expectation value of the time dependent electric field as well as its fluctuations. To this aim, we compute and plot (see Fig.~\ref{electricexamples}), explicitly the expectation value of the electric field. In general

\begin{equation}
<|\hat{\vec E}(\vec r, t)|> \sim \, <\hat Q> \sin(\omega t - \vec k \cdot \vec r) - <\hat P> \cos(\omega t - \vec k \cdot \vec r),
\end{equation}
and its corresponding uncertainty (fluctuations)
\begin{equation}
\begin{split}
(\Delta E(\vec r, t))^2 \sim (\Delta Q)^2 \sin^2(\omega t - \vec k \cdot \vec r) + (\Delta P)^2 \cos^2(\omega t - \vec k \cdot \vec r) -\\
-V(\hat Q,\hat P) \sin(\omega t - \vec k \cdot \vec r) \cos(\omega t - \vec k \cdot \vec r),
\end{split}
\end{equation}
where $V(\hat Q,\hat P) = <\acomm{Q}{P}> - 2 <\hat Q><\hat P>$. We list here the most relevant states of the electromagnetic field. Among them, Gaussian states produced in the lab with coherent laser light and linear optical devices. We write them in the Fock basis as it is a very convenient basis for all calculations needed (see Tabs.~\ref{dades} and \ref{dades2} for detailed calculations).

\begin{itemize}
\item {\em Vacuum state}, $\ket{0}$, is the only Gaussian state of all Fock states. It can be defined as the ground state of the harmonic oscillator.

$<|\hat{\vec E}(\vec r, t)|> = 0$,

$\Delta E(\vec r, t) \sim \sqrt{\frac{1}{2}}$.

\item {\em Fock states~\footnote{Complete orthonormal set basis $\braket{n}{m} = \delta_{n,m}$.}}, $\ket{n}$, are eigenstates of the number operator, and thus, so of the energy because $\hat H \ket{n}=E_n \ket{n}$. They are non-Gaussian and can be written as the $n$-th excitation (photon) from the vacuum, $\ket{n} = \frac{(\hat a^\dag)^n}{\sqrt{n!}} \ket{0}$ via the ladder operators that act on them raising and lowering the number of photons presents in the state $\hat a^\dag \ket{n} = \sqrt{n+1}\ket{n+1}$, $\hat a \ket{n} = \sqrt{n}\ket{n-1}$.

$<|\hat{\vec E}(\vec r, t)|> = 0$,

$\Delta E(\vec r, t) \sim \sqrt{n+\frac{1}{2}}$.

\item {\em Fock superpositions}, $\ket{\psi} = \sum_i \psi_i\ket{n_i}$ are non-Gaussians. The simplest one, is $\ket{\psi} = \frac{1}{\sqrt{2}} (\ket{0}+\ket{1})$ with balanced amplitudes.

$<|\hat{\vec E}(\vec r, t)|> \sim \sqrt{\frac{1}{2}} \sin(\omega t - \vec k \cdot \vec r)$, 

$\Delta E(\vec r, t) \sim \sqrt{\frac{1}{2}\sin^2(\omega t - \vec k \cdot \vec r) + \cos^2(\omega t - \vec k \cdot \vec r)}$.

\item {\em Coherent states}, $\ket{\alpha}$, are defined as the eigenstates of the annihilation operator $\hat a\ket{\alpha} = \alpha \ket{\alpha}$ or as $\ket{\alpha} = e^{-\frac{|\alpha|^2}{2}} \sum_{n=0}^\infty \frac{\alpha^n}{\sqrt{n!}}\ket{n}$. Coherent state have not a well defined the number of photons ($\hat n = \hat a^\dag \hat a$) because $\exvalue{n}{\alpha} = |\alpha|^2$ and $(\Delta n)_\alpha = |\alpha|$, nevertheless, as $\alpha$ increases the relative indetermination $\frac{(\Delta n)_\alpha}{\exvalue{n}{\alpha}} = |\alpha|^{-1}$ tends to zero. Coherent state have equal variance and mean, therefore, its photon distribution is Poissonian. It is immediate to see that if one computes the probability to find $n$ photons in a coherent state {\em i.e.} $P_n = |\braket{n}{\alpha}|^2 = \frac{|\alpha|^{2n}}{n!}\e^{-|\alpha|^2}$ which is clearly a Poissonian distribution with mean and variance $|\alpha|^2$.

$<|\hat{\vec E}(\vec r, t)|> \sim \sqrt{2}{\rm Re}(\alpha) \sin(\omega t - \vec k \cdot \vec r) - \sqrt{2}{\rm Im}(\alpha) \cos(\omega t - \vec k \cdot \vec r)$,

$\Delta E(\vec r, t) \sim \sqrt{\frac{1}{2}}$.

\item {\em Squeezed states}, $\ket{r}$, are minimal uncertainty states with arbitrary small uncertainty in one quadrature while increased uncertainty in the orthogonal one. Squeezed states of the light fields are used to enhance precision measurements. They read in Fock basis as $\ket{r} = \frac{1}{\sqrt{\cosh{r}}}\sum_{n=0}^\infty (-\frac{1}{2}\tanh{r})^n \frac{\sqrt{(2n)!}}{n!} \ket{2n}$.

$<|\hat{\vec E}(\vec r, t)|> = 0$,

$\Delta E(\vec r, t) \sim \sqrt{\frac{1}{2} \e^{-2 r} \sin^2(\omega t - \vec k \cdot \vec r)+\frac{1}{2} \e^{2 r} \cos ^2(\omega t - \vec k \cdot \vec r)}$.

\item {\em Thermal states}, $\hat \rho_\beta$, are a superposition of all Fock states occupied with a probability $P_n=(1-\e^{-\beta\hbar\omega})\e^{-\beta\hbar\omega n}$. It's the quantum distribution function which describes the thermal state statistics of the blackbody radiation. At the Fock level they read as $\hat \rho_\beta = (1-\e^{-\beta\hbar\omega})\sum_{n=0}^{\infty} \e^{-\beta\hbar\omega n}\ketbra{n}{n}$.

$<|\hat{\vec E}(\vec r, t)|> = 0$,

$\Delta E(\vec r, t) \sim \sqrt{M+\frac{1}{2}}$.
\end{itemize}

\newpage

We can summarize the above results in the following tables:

\begin{equation}
 \begin{tabular}{c||c|c|c|c|c|c}
 & $<\hat Q>$ & $<\hat P>$ & $\Delta Q$ & $\Delta P$ & $<\hat H>$ & $\Delta Q \Delta P \geq \frac{1}{2}$\\
 \hline
 \hline
 $\ket{0}$ & {\tiny 0} & {\tiny 0} & {\tiny $\sqrt{1/2}$} & {\tiny $\sqrt{1/2}$} & {\tiny $\frac{\hbar \omega}{2}$} & {\tiny 1/2}\\
 \hline
 $\ket{n}$ & {\tiny 0} & {\tiny 0} & {\tiny $\sqrt{n + 1/2}$} & {\tiny $\sqrt{n + 1/2}$} & {\tiny $\frac{\hbar \omega}{2}(1 + 2n)$} & {\tiny n + 1/2}\\
 \hline
 $\ket{\psi}$ & {\tiny $\sqrt{1/2}$} & {\tiny 0} & {\tiny $\sqrt{1/2}$} & {\tiny 1} & {\tiny $\hbar \omega$} & {\tiny $\sqrt{1/2}$}\\
 \hline
 $\ket{\alpha}$ & {\tiny $\sqrt{2} \, {\rm Re}(\alpha)$} & {\tiny $\sqrt{2} \, {\rm Im}(\alpha)$} & {\tiny $\sqrt{1/2}$} & {\tiny $\sqrt{1/2}$} & {\tiny $\frac{\hbar \omega}{2}(1 + 2|\alpha|^2)$} & {\tiny $1/2$}\\
 \hline
 $\ket{r}$ & {\tiny 0} & {\tiny 0} & {\tiny $\sqrt{\frac{\e^{-2r}}{2}}$} & {\tiny $\sqrt{\frac{\e^{2r}}{2}}$} & {\tiny $\frac{\hbar \omega}{2}\cosh{2r}$} & {\tiny $1/2$}\\
 \hline
 $\hat \rho_\beta$ & {\tiny 0} & {\tiny 0} & {\tiny $\sqrt{M+1/2}$} & {\tiny $\sqrt{M+1/2}$} & {\tiny $\frac{\hbar \omega}{2}(1 + 2M)$} & {\tiny $M+1/2$}
 \end{tabular}\label{dades}
\end{equation}

\begin{equation}
 \begin{tabular}{c||c|c|c|c|c}
 & $<\hat Q^2>$ & $<\hat P^2>$ & $<\hat Q \hat P>$ & $<\hat P \hat Q>$ & $V(\hat Q,\hat P)$\\
 \hline
 \hline
 $\ket{0}$ & {\tiny $\frac{1}{2}$} & {\tiny $\frac{1}{2}$} & {\tiny $\frac{\im}{2}$} & {\tiny $-\frac{\im}{2}$} & {\tiny $0$}\\
 \hline
 $\ket{n}$ & {\tiny $n + \frac{1}{2}$} & {\tiny $n + \frac{1}{2}$} & {\tiny $\frac{\im}{2}$} & {\tiny $-\frac{\im}{2}$} & {\tiny $0$}\\
 \hline
 $\ket{\psi}$ & {\tiny $1$} & {\tiny $1$} & {\tiny $\frac{\im}{2}$} & {\tiny $-\frac{\im}{2}$} & {\tiny $0$}\\
 \hline
 $\ket{\alpha}$ & {\tiny $\frac{1}{2}(1 + 2|\alpha|^2 + \alpha^2 + \alpha^{*2})$} & {\tiny $\frac{1}{2}(1 + 2|\alpha|^2 - \alpha^2 - \alpha^{*2})$} & {\tiny $\frac{\im}{2}(1-\alpha^2+\alpha^{*2})$} & {\tiny $\frac{\im}{2}(-1-\alpha^2+\alpha^{*2})$} & {\tiny $0$}\\
 \hline
 $\ket{r}$ & {\tiny $\frac{\e^{-2r}}{2}$} & {\tiny $\frac{\e^{2r}}{2}$} & {\tiny $\frac{\im}{2}$} & {\tiny $-\frac{\im}{2}$} & {\tiny $0$}\\
 \hline
 $\hat \rho_\beta$ & {\tiny $M+\frac{1}{2}$} & {\tiny $M+\frac{1}{2}$} & {\tiny $\frac{\im}{2}$} & {\tiny $-\frac{\im}{2}$} & {\tiny $0$}\\
 \end{tabular}\label{dades2}
\end{equation}
Recalling section~\ref{DViCM}, on one hand the expectation values of the two quadratures are encoded in the displacement vector which gives the center of the diagrams in the phase-space. On the other, the two uncertainties are encoded in the diagonal terms of the covariance matrix and express the width of the phase-space points. The off diagonal terms (symmetrics) contain the information of the $Q$-$P$ correlations.
We observe then, that the components of the displacement vector are closely related with the amplitude of the electric field while the covariance matrix contains the fluctuations of the electric field. Notice the correspondence $V(\hat Q,\hat P) = \gamma_{12}$. This important term amounts for the correlations in fluctuations and is zero for all minimum uncertainty states.

Analyzing the plots we see two important behaviors. First, as a general feature a superposition of eigenstates of the energy (Fock states) produces a $t$-dependence in any observable, as one sees in the electric field which oscillates in time. Second, coherent states are the states of an harmonic oscillator system which mimic in the best possible way the classical motion of a particle in a quadratic potential, in this sense, they are the most classical states of light that one can produce. As seen in the plot the expectation value of the electric field oscillates classically while the uncertainty is minimum and coherent with the field.

\begin{figure}[h]
 \centering
a) \psfrag{X}{$t$}\psfrag{Y}{$<|\hat{\vec E}(\vec r, t)|>$}
 \includegraphics[width=6cm,height=4.5cm]{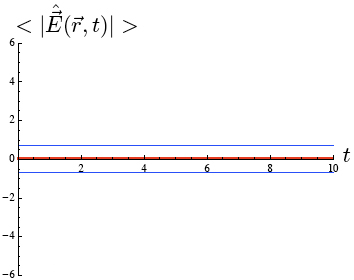} \hfill
b) \psfrag{X}{$t$}\psfrag{Y}{$<|\hat{\vec E}(\vec r, t)|>$}
 \includegraphics[width=6cm,height=4.5cm]{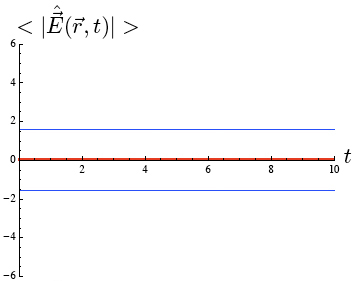}\\
c) \psfrag{X}{$t$}\psfrag{Y}{$<|\hat{\vec E}(\vec r, t)|>$}
 \includegraphics[width=6cm,height=4.5cm]{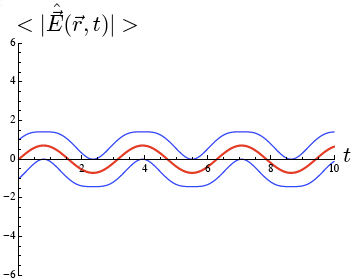} \hfill
d) \psfrag{X}{$t$}\psfrag{Y}{$<|\hat{\vec E}(\vec r, t)|>$}
 \includegraphics[width=6cm,height=4.5cm]{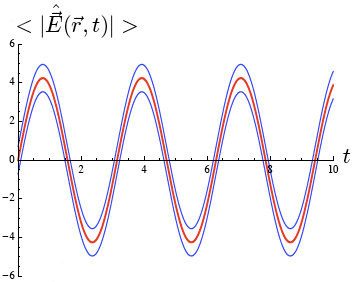}\\
e) \psfrag{X}{$t$}\psfrag{Y}{$<|\hat{\vec E}(\vec r, t)|>$}
 \includegraphics[width=6cm,height=4.5cm]{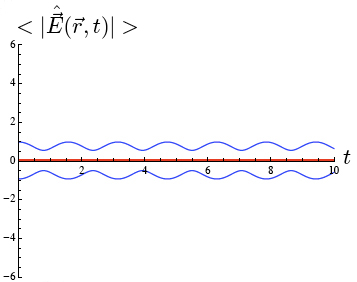} \hfill
f) \psfrag{X}{$t$}\psfrag{Y}{$<|\hat{\vec E}(\vec r, t)|>$}
 \includegraphics[width=6cm,height=4.5cm]{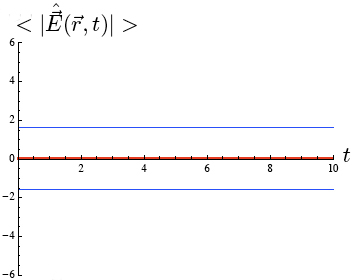}\\
g) \psfrag{X}{$t$}\psfrag{Y}{$<|\hat{\vec E}(\vec r, t)|>$}
 \includegraphics[width=6cm,height=4.5cm]{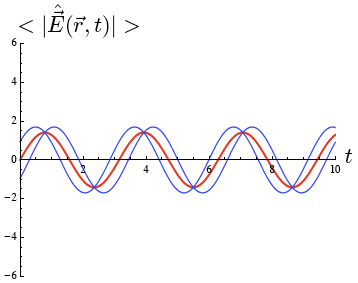} \hfill
h) \psfrag{X}{$t$}\psfrag{Y}{$<|\hat{\vec E}(\vec r, t)|>$}
 \includegraphics[width=6cm,height=4.5cm]{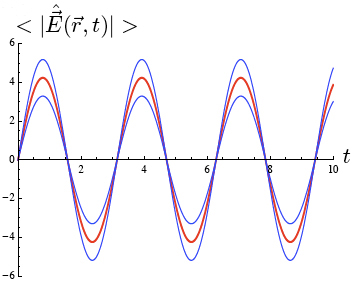}\\
 \caption{\small{In red the average electric field, in blue the associated uncertainty. a) Vacuum, b) Fock, c) Superposition, d) Coherent, e) Squeezed, f) Thermal, g) Coherent squeezed in position (amplitude), h) Coherent squeezed in momentum (phase).}}\label{electricexamples}
\end{figure}

\newpage

\section{Appendix: Wigner integrals}\label{appendix2}

As it has been shown the Wigner distribution function is Gaussian whenever the state is Gaussian, with straightforward calculations on can deduce two extremely useful Gaussian integrals 

\begin{equation}
 \int e^{-\zeta^T \cdot A \cdot \zeta} d^{2N}\zeta = \frac{\pi^N}{\sqrt{\det A}},
\end{equation}

\begin{equation}
 \int e^{-\zeta^T \cdot A \cdot \zeta - b^T \cdot \zeta} d^{2N}\zeta = \frac{\pi^N}{\sqrt{\det A}} e^{b^T \cdot \frac{1}{4A} \cdot b}.
\end{equation}

In the case we are dealing with non-Gaussian states the Wigner distribution function is not any more Gaussian and one is forced to perform non-Gaussian integrals. We give here some very useful formulas that convert the problem of integrating non-Gaussian functions into a derivative problem

\begin{equation}
 \begin{split}
 \int& \zeta_i^{m_i} \zeta_j^{m_j} \zeta_k^{m_k}{\cdot}{\cdot}{\cdot} e^{-\zeta^T \cdot A \cdot \zeta} d^{2N}\zeta =\\
 &= (-1)^{m_i + m_j + m_k + {\cdot}{\cdot}{\cdot}} \left. \frac{\pi^N}{\sqrt{\det A}} \frac
{\partial^{m_i}}{\partial b_i^{m_i}} \frac{\partial^{m_j}}{\partial b_j^{m_j}} \frac{\partial^{m_k}}{\partial b_k^{m_k}} {\cdot}{\cdot} {\cdot} e^{b^T \cdot \frac{1}{4A} \cdot b} \right|_{b_i = 0, b_j = 0,
b_k = 0, \ldots},\\
 \end{split}
\end{equation}

\begin{equation}
 \begin{split}
 \int& \zeta_i^{m_i} \zeta_j^{m_j} \zeta_k^{m_k}{\cdot}{\cdot}{\cdot} e^{-\zeta^T \cdot A \cdot \zeta - c^T \cdot \zeta} d^{2N}\zeta =\\
 &= (-1)^{m_i + m_j + m_k + {\cdot}{\cdot}{\cdot}} \left. \frac{\pi^N} {\sqrt{\det A}} \frac{\partial^{m_i}}{\partial b_i^{m_i}} \frac{\partial^{m_j}}{\partial b_j^{m_j}} \frac{\partial^{m_k}}{\partial b_k^{m_k}} {\cdot}{\cdot}{\cdot} e^{(b + c)^T \cdot \frac{1}{4A} \cdot (b+ c)} \right|_{b_i = 0, b_j = 0, b_k = 0, \ldots}.\\
 \end{split}
\end{equation}

Extra useful integrals are

\begin{equation}
 \int \mathcal{W}(\zeta) d^{2N} \zeta = 1,
\end{equation}

\begin{equation}
 \int (\mathcal{W}(\zeta))^2 d^{2N} \zeta = \frac{1}{(2 \pi)^{N} \sqrt{\det \gamma}},
\end{equation}

\begin{equation}
 \int \chi(\eta) d^{2N} \eta = \frac{(4 \pi)^{N}}{\sqrt{\det \gamma}} e^{-d^T \frac{1}{\gamma} d},
\end{equation}

\begin{equation}
 \int (\chi(\eta))^2 d^{2N} \eta = \frac{(2 \pi)^{N}}{\sqrt{\det \gamma}} e^{-d^T \frac{4}{\gamma} d},
\end{equation}

\begin{equation}
 \int |\chi(\eta)|^2 d^{2N} \eta = \frac{(2 \pi)^{N}}{\sqrt{\det \gamma}}.
\end{equation}

\end{subappendices}


\chapter{Quantum Cryptography protocols with Continuous Variable}\label{cryptography}

Cryptography refers to strategies which permit the secure communication between two distant parties (traditionally denoted by Alice and Bob) that wish to communicate secretly. So its purpose is to design new communication algorithms being sure that secrecy is preserved. In Classical Cryptography there is only one cryptographic protocol, known as the ideal Vernam cypher method, which is absolutely secure. The Vernam cypher consists of a {\em random} secret key (private key), shared between the sender (Alice) and the receiver (Bob), used to encode and decode messages with absolute security. However, this method suffers from two drawbacks. First, if Alice wants to communicate $N$ bits to Bob, they will need to share in advance a key with at least $N$ bits. Second, this key can only be used once to make the method unbreakable. This implies that, the key must originally be exchanged by hand before the communication to keep secrecy and it is essential that the key is totally random and secret. This problem is known as key distribution.

To solve the problem of a private key distribution, classical strategies exist to distribute keys between them albeit being this distribution only partially secure. For example, by designing algorithms which permit distribution of private keys in a ``practical secure'' way. In such algorithms, security relies on the difficulty to invert some mathematical operations. Since the time needed to break the security of these algorithms to obtain the key is long, then practical security is achievable.

The second problem, the necessity of a new key for each message, is also solved using Public Key Cryptosystems. They work as follows: assume that Alice and Bob possess, in advance, a common secret key. This secret key must be exchanged by hand once, and it has to be random and secure. Using this secret key, Alice and Bob can distribute among them, several private keys. Then, they could use as many distributed private key as necessary to encrypt/decrypt messages through {\em e.g.} the Vernam cypher method. In other words, any time Alice wants to communicate with Bob, she only needs to use her secret key to distribute as many private keys as needed to encrypt later on messages. This distribution works as follows, Alice makes one key publicly available, referred as the public key that encodes a private key. From this public key any receiver could in principle extract the private key. But only Bob (the receiver in possession of the secret key) can extract it in an efficient way. Any receiver without the secret key needs to decrypt a problem with NP complexity. This way of distributing private keys can be done even when the secret key's length is much smaller than the private keys needed for the Vernam cypher encryption and even if the secret key is used many times, solving both issues.

In this chapter we have studied the possibility of using Continuous Variable to perform quantum cryptography protocols by means of bipartite entanglement. Like in any practical implementation of a protocol, efficiency is an important issue since resources are not unlimited, thus a special attention will be devoted to the efficient implementation of the protocol. Before explaining how secure cryptography with Continuous Variable can be realized, let us introduce the basic concepts in both classical and quantum scenario.

\section{Classical Cryptography}

\subsection{Vernam cypher}

The best and most well known classical private key (or one-time-pad) cryptosystem is the so-called Vernam cypher. To achieve absolute security, the Vernam cypher requires the prior distribution of a random classical private key, denoted as $k$. The basic steps of the protocol are the followings:

{\em Encoding}:
if Alice wants to encode a message $m$ with the key $k$, she performs the following operation between the message and the key, and sends the encoded message $e$ to Bob
\begin{equation}
 {\rm Enc}_k(m)=m \oplus k=e.
\end{equation}

{\em Decoding}:
only Bob, who has also the key, can decode the message (invert the operation) because the key Alice has used is random. Thus, he only needs to use the key again to the encoded message $e$ in order to retrieve the original message $m$
\begin{equation}
 {\rm Dec}_k(e)={\rm Dec}_k [{\rm Enc}_k(m)] =e\oplus k=d=m.
\end{equation}
Let us illustrate Vernam cypher with an specific example. Alice wants to communicate to Bob, in a secure way, a message $m$ (in a binary string) of {\em e.g.} 9 bits. They share the key $k$ of the
same size as the message (9 bits). Alice encodes her message by applying a $XOR$ (exclusive $OR$)~\footnote{Also known as $AND$ or $\oplus \hspace{-1.5mm} \mod (2)$.} operation between the message $m$ and the key $k$.

In the next table we summarize the $XOR$ operation

\begin{equation}
 \begin{tabular}{c|cc}
 $\oplus$ & 0 & 1\\
 \hline
 0 & 0 & 1\\
 1 & 1 & 0
 \end{tabular}
\end{equation}

As a result Alice has the encoded message $e$, that will send to Bob in a public way.

\begin{equation*}
 \begin{tabular}{c|l}
 message & $m=010011101\,$\\
 \hline
 key & $k=110100011$\\
 \hline
 \hline
 encoded message & $e=1000111110$\\
 \end{tabular}
\end{equation*}

Then Bob wants to readout the message, and thus, performs the inverse operation (which is again a $XOR$) between the encoded message $e$ and the key $k$.\\

As a result Bob has the decoded message $d$ that coincides with the message Alice wanted to communicate to him.

\begin{equation*}
 \begin{tabular}{c|l}
 encoded message & $e=1000111110\,$\\
 \hline
 key & $k=110100011$\\
 \hline
 \hline
 decoded message & $d=010011101$\\
 \end{tabular}
\end{equation*}

\subsection{Public key distribution: The RSA algorithm}

Thus far, Classical Cryptography has not a solution for the distribution of the private key needed to perform Vernam cypher encryption. Such distribution can, however be done in a public and practical secure way with the RSA algorithm. The RSA cryptography algorithm proposed in 1977 by {\em R}ivest, {\em S}hamir, and {\em A}dleman from the MIT is the most commonly used algorithm of a public key. It is widely used in bank's security, electronic commerce and internet, relying on the fact that to decrypt this algorithm one needs to solve the factorization problem which is an NP problem. Even now, there is no efficient classical algorithm known to solve the factorization problem. This means that even though computational resources increase constantly, one simply needs to exploit the NP character of the algorithm to make the solution harder to find. Let us illustrate how it works with the following example.

\begin{enumerate}[i)]
\item The sender, Alice, chooses two ``big'' different prime numbers, say $p=61$ and $q=53$, she computes its product $n=p\,q=3233$ and also the following quantity $\phi=(p-1)(q-1)=3120$.
\item She chooses a positive integer $l$ smaller and coprime with $\phi$, in the example above $l=17$.
\item As a secret key, Alice gives to Bob the number $k$ such that $k \, l=1 \hspace{-1.5mm} \mod (\phi)$, take for example\ $k=2753$. Alice makes public $l$ and $n$, this is what is called a public key.
\item With the public key ($l$ and $n$) anyone can encrypt a message $m$ and send it to Bob, but only Bob, who is in possession of the secret key $k$, is able to decrypt the message. This method
can thus be used to perform Classical Key Distribution. Any time Alice wants to communicate with Bob, she sends a private key encrypted with $l$ and $n$ and only Bob will be able to retrieve it. Once Bob has the private key, Alice can send messages to Bob via the Vernam cypher using this key.
\item Encryption proceeds as follows. Alice wants to distribute a key encoded in a message $m=123$. She uses the public key and computes the encryption ${\rm Enc}_{l,n}(m)=m^l \hspace{-1.5mm}\mod(n)=e=855$.
\item Bob now wants to decrypt the message $e$ to extract a private key, so he calculates ${\rm Dec}_{k,n}(e)={\rm Dec}_{k,n} [{\rm Enc}_{l,n}(m)] =e^k \hspace{-1.5mm}\mod(n)=d=m=123$. Bob is
the only one in possession of the secret key $k$ and so the only one that can decrypt the message $e$ to find the private key Alice is going to use with the Vernam cypher to communicate securely with him.
\end{enumerate}

The private key Alice and Bob share can be used more than once to distribute secure keys. In such way, to use the Vernam cypher, one no longer needs to share a long private key because we can distribute many of them in a secure way. Security relies in the fact that, from the encrypted message $e$, and the public keys $l$ and $n$ it is very difficult to find the private key $k$ (and so $m$) or the original two prime numbers $q$ and $p$, even in the case we are reusing the key $k$. This is because factorization is a NP problem, whose efficient classical solution is not known yet. Security in the RSA algorithm relies on the fact that independently of the computer used, if it is classical, the problem is NP and thus hard to solve.

\section{The quantum solution to the distribution of the key}

The ``Quantum Computer'' arises here first as a menace for Classical Key Distribution methods and then as the solution for the security in Cryptography. Based on the quantum nature of the microscopic world, such ``computers'', still in a theoretical stage, are known to be able to solve some hard mathematical problems rapidly. In 1994 Peter Shor proposed a quantum protocol to solve the factorization problem in an efficient way, known as the Shor's algorithm. If such a computer can be realized, current cryptographic protocols will not be anymore secure. Can Quantum Mechanics then offer a solution for a secure Cryptography method? 
The answer is yes. Since the end of the 80s protocols relying on Quantum Mechanics exists permitting to perform Quantum Cryptography in an unconditional secure way.

At present, Quantum Cryptography is the most important real implementation of Quantum Information. It offers an absolute secure distribution of random keys which combined with the Vernam cypher guarantees completely secure Cryptography. Thus, the Quantum Cryptography problem is in fact the problem of distributing, between two distant parties, a secure random key relying on the laws of Quantum Mechanics, {\em i.e.} the Quantum Key Distribution (QKD) problem. If the key is securely distributed, the algorithms used to encode and decode any message can be made public without compromising security. The key consists typically in a random sequence of bits which both, Alice and Bob, share as a string of classically correlated data. The superiority of Quantum Cryptography comes from the fact that the laws of Quantum Mechanics permit to the legitimate users (Alice and Bob) to infer if an eavesdropper has monitored the distribution of the key and has gained information about it. If this is the case, Alice and Bob will both agree in withdrawing the key and will start the distribution of a new one. In contrast, Classical Key Distribution, no matter how difficult the distribution from a technological point of view is, can always be intercepted by an eavesdropper without Alice and Bob realizing it.

In Quantum Cryptography two seemingly independent main schemes exist for QKD. The first one, denoted as ``Prepare and Measure'' scheme, originally proposed by C.H. Bennett and G. Brassard in 1988 and known as BB84 \cite{Bennett1984IEEE}, does not use entangled states shared between Alice and Bob. The key is established by sending non-orthogonal quantum states between the parties and communicating classically the result of some measurements. Security is guaranteed by the quantum nature of the measurements which avoids each party measuring simultaneously non-commuting observables. The second scheme (''Entanglement based``), uses as a resource shared entanglement, like the one originally proposed by A. Ekert in 1991 known as Ekert91 \cite{Ekert1991PRL}. Here entanglement is explicitly distributed and the security is guaranteed by the nature of quantum correlations and proved by Bell inequalities. However, the two schemes have been shown to be completely equivalent \cite{Bennett1992PRL}, and specifically entanglement stands as a precondition for any secure key distribution \cite{Curty2004PRL}. Let us briefly detail these two schemes.

\subsection{``Prepare and Measure'' scheme}

Like in any cryptographic protocol, there is always an eavesdropper. The protocol does not avoid an eavesdropper from intercepting the key, but it permits Alice and Bob know if the key has been intercepted, and so that it can be discarded. We sketch here the steps of the BB84 protocol.

\begin{enumerate}[i)]
\item Alice prepares a secret sequence of random bits and encodes them in the state of a 2-level system {\em e.g.} a spin-1/2 system by choosing randomly between two bases (Z and X). Alice encodes  $0/1$ in $\ket \pm$ ($\ket \pm_x$) according to basis Z (basis X). Then, she sends Bob the states she has prepared. For example:

\begin{equation*}
 \begin{tabular}{c|ccccccccc}
 Alice random bits & 0 & 1 & 1 & 0 & 0 & 1 & 1 & 0 & 0\\
 \hline
 Alice random bases & Z & X & X & X & Z & X & Z & X & X\\
 \hline
 Alice prepared states & $\ket{+}$ & $\ket{-}_x$ & $\ket{-}_x$ & $\ket{+}_x$
 & $\ket{+}$ & $\ket{-}_x$ & $\ket{-}$ & $\ket{+}_x$ & $\ket{+}_x$\\
 \end{tabular}
\end{equation*}

\item Bob receives the states without knowing on which basis have been prepared and measures in another random choice of bases. The outcome of the measurements is going to be retained as the bits received while the state collapses in a certain eigenstate.

\begin{equation*}
 \begin{tabular}{c|ccccccccc}
 Alice prepared states & $\ket{+}$ & $\ket{-}_x$ & $\ket{-}_x$ &
 $\ket{+}_x$ & $\ket{+}$ & $\ket{-}_x$ & $\ket{-}$ & $\ket{+}_x$
 & $\ket{+}_x$\\
 \hline
 Bob random bases & X & Z & X & X & X & X & X & X & X\\
 \hline
 Bob received states & $\ket{+}_x$ & $\ket{+}$ & $\ket{-}_x$ & $\ket{+}_x$ &
 $\ket{+}_x$ & $\ket{-}_x$ & $\ket{-}_x$ & $\ket{+}_x$ &
 $\ket{+}_x$\\
 \hline
 Bob received bits & 0 & 0 & 1 & 0 & 0 & 1 & 1 & 0 & 0\\
 \end{tabular}
\end{equation*}

\item Bob communicates to Alice his choice of bases in a public way.
\item Alice identifies the set of bits for which they have both performed the measurement in the same basis, {\em i.e.} outcomes 3, 4, 6, 8 and 9 in the above example. Alice and Bob discard the set of data in which they did not agree (the rest).
\item Bob sends part of his data (received bits) to Alice by a public channel. Alice checks the correlation between the data and establishes an error rate. The error rate can come from an eavesdropper or noise effects.
\item If the error rate is too high they can assume that an eavesdropper has act and they can restart the protocol from the beginning. One can show that, in the case of free noise effect, and under individual attacks (the eavesdropper intercepts, measures, and resents the states) the error rate is bounded to be (for a sufficiently high amount of data) 25\%. If Alice deduces that there is not an eavesdropper present, she communicates it to Bob. Alice and Bob use the set of remaining data as a private key. Later on, they can improve security by performing information reconciliation and privacy amplification on the private key before using it to encrypt messages with the Vernam cypher.
\end{enumerate}
 
Note that the security of the protocol relies in the quantum nature of the measurements. The two bases, Z and X, are associated with eigenbasis of non-commuting observables, $\hat \sigma_z$ and $\hat \sigma_x$. The eavesdropper cannot measure simultaneously both observables on the same state. Additionally the nocloning theorem prevents her from being able to distinguish with certainty between non-orthogonal quantum state performing cloning, and avoids hence the possibility of resending them to Bob without leaving trace of her intrusion. 

\subsection{``Entanglement based'' scheme}

A second type of protocols demand as a fundamental resource, shared entanglement between Alice and Bob, like Ekert91. In the same way as in BB84, these protocols permit a secure distribution of a secret key. This can be done as far as the protocol ensures if there has been an interception of the key. We sketch below the steps of this well-known protocol. 

\begin{enumerate}[i)]
\item The first step consists on distributing (along the $z$-direction) singlet states of a spin-1/2 system between Alice and Bob. Thus Alice and Bob share many copies of a Bell state $\ket {\Psi^-}=\frac{1}{\sqrt2}(\ket 0 \ket 1-\ket 1 \ket 0)$.
\item Alice and Bob are going to measure in the $x-y$ plane in one of the three directions given by unit vectors $\vec A_i = (\cos{\phi_i^A}, \sin{\phi_i^A})$ and $\vec B_j = (\cos{\phi_j^B}, \sin{\phi_j^B})$ respectively, where the azimuthal angles are fixed to $\phi_i^A=(0,\pi/4,\pi/2)_i$ and to $\phi_j^B=(\pi/4,\pi/2,3\pi/4)_j$. Each time they will choose the basis randomly and independently for each pair of incoming particles.
\item After the measurement has taken place, Alice and Bob can announce in public the directions they have chosen for each measurements and divide them into two separated groups. A first group for which they coincide and a second group for which they do not. The second group of outcomes is made public and it is used to establish the presence or absence of an eavesdropper.
\item One defines correlation coefficients to test security
\begin{equation}
 \mathcal E(\vec A_i,\vec B_j) = \mathcal P_{++}(\vec A_i,\vec
 B_j) + \mathcal P_{--}(\vec A_i,\vec B_j) - \mathcal P_{+-}(\vec
 A_i,\vec B_j) - \mathcal P_{-+}(\vec A_i,\vec B_j)
\end{equation}
which is the correlation coefficient of the measurements performed by Alice along $\vec A_i$ and by Bob along $\vec B_j$. Here $\mathcal P_{\pm \pm}(\vec A_i,\vec B_j)$ denotes the probability
the results $\pm1$ has been obtained along $\vec A_i$ and $\pm1$ along $\vec B_j$. Straightforward calculations give rise to

$\mathcal P_{++}(\vec A_i,\vec B_j) = \frac{1}{2} \sin^2(\phi_i^A-\phi_j^B)$,

$\mathcal P_{--}(\vec A_i,\vec B_j) = \frac{1}{2} \sin^2(\phi_i^A-\phi_j^B)$,

$\mathcal P_{+-}(\vec A_i,\vec B_j) = \frac{1}{2} \cos^2(\phi_i^A-\phi_j^B)$,

$\mathcal P_{-+}(\vec A_i,\vec B_j) = \frac{1}{2} \cos^2(\phi_i^A-\phi_j^B)$.\\
Thus, according to the Quantum rules $\mathcal E(\vec A_i,\vec B_j) = - \vec A_i \vec B_j = - \cos\left[2(\phi_i^A-\phi_j^B)\right]$. As expected, if they choose the same orientation, Quantum Mechanics predicts a total anticorrelation in the outcomes $\mathcal E(\vec A_i,\vec B_j)=-1$ as it should be for the rotational invariant singlet state $\ket {\Psi^-}$.

Finally, let us define here a quantity composed of those correlation coefficients for which Alice and Bob have measured in different directions,
\begin{equation}
 \mathcal S = |\mathcal E(\vec A_1,\vec B_1) + \mathcal E(\vec
 A_3,\vec B_3) - \mathcal E(\vec A_1,\vec B_3) + \mathcal E(\vec
 A_3,\vec B_1)|.
\end{equation}
Again, Quantum Mechanics requires, $\mathcal S = 2\sqrt{2} > 2$.

\item The CHSH (Clauser, Horne, Shimony, and Holt) inequality, is a generalization of Bell inequalities and asserts that $\mathcal S \leq 2$ for any theory compatible with local realism. But Quantum Mechanics, and in particular with Bell states CHSH inequalities are violated. If this is the case, {\em i.e.} the value of $\mathcal S$ that they find is exactly $ 2\sqrt{2}$, they know that their states have not been disturbed and so the first group of outcomes, that are random, are totally anticorrelated and can be converted into a secret string of bits (provided Bob flips all the bits). This ends the distribution of the private key. One can now, as in the BB84 protocol, perform information reconciliation and privacy amplification on the private key before using it to encrypt messages with Vernam cypher to achieve an absolute secure communication.
\end{enumerate}

Note that security relies in the quantum correlations {\em i.e.} entanglement and it is guaranteed by the violation of Bell inequalities. In contrast to BB84, Alice do not need in advance a string of random bits because the randomness comes from the measurement process.

\section{Quantum Key Distribution with Continuous Variable Gaussian states}

Quantum Cryptography can be implemented using systems of Continuous Variable {\em i.e.} using quantum states on infinite dimensional Hilbert spaces. Among all Continuous Variable states, Gaussian states and Gaussian operations, have been the preferred to experimentally implement Quantum Cryptography using ``Prepare and Measure'' schemes {\em e.g.} with either squeezed or coherent states \cite{Gottesman2001PRA,Grosshans2002PRL,Silberhorn2002PRL}. Those schemes do not demand entanglement between the parties. Here we address the problem of the Quantum Key Distribution with entangled Continuous Variable using an ``Entanglement based'' protocol.

Notice that if Alice and Bob share a collection of distillable entangled states, they can always obtain a smaller number of maximally entangled states from which they can establish a secure key \cite{Deutsch1996PRL} using {\em e.g.} Ekert91 protocol. The number of singlets (maximally entangled states) that can be extracted from a quantum state using only Local Operations and Classical Communication (LOCC) is referred to as the Entanglement of Distillation $E_D$. In order to establish a key, another important concept is the number of secret bits $K_D$, that can be extracted from a quantum state using LOCC. As a secret bit can always be extracted from maximally entangled states, $K_D \geq E_D$.

There are quantum states which cannot be distilled in spite of being entangled, {\em i.e.}, they have $E_D=0$. These systems are usually referred to as bound entangled states since its entanglement is bound to the state. Nevertheless, for some of those states it has been shown that $K_D \neq 0$ and thus, they can be used to establish a secret key \cite{Horodecki2005PRL}.

A particular case of states that cannot be distilled by ``normal'' procedures are Continuous Variable Gaussian states, {\em e.g.}, coherent, squeezed and thermal states of light. By ``normal'' procedures we mean operations that preserve the Gaussian character of the state (Gaussian operations), they
correspond {\em e.g.} to beam splitters, phase shifts, mirrors, squeezers, etc. Thus, in the Gaussian scenario all entangled Gaussian states posses bound entanglement.
Navascu{\'e}s {\em et al} \cite{Navascues2005PRL} have shown that it is also possible using only Gaussian operations to extract a secret key {\em \`a la} Ekert91 from entangled Gaussian states, in spite the fact that these states are not distillable. In other words, it has been proven that in the Gaussian scenario all entangled Gaussian states fulfill $GK_D>0$ (where the letter $G$ stands for Gaussian) while $GE_D=0$.

The above result implies that in principle any entangled Gaussian state can be used for implementing Quantum Cryptography. However, any real implementation should address the optimization of the resources needed {\em i.e.} the efficiency of the protocol. The proposed protocol presented in \cite{Navascues2005PRL} suffers precisely from an efficiency problem because the success probability of the protocol is vanishingly small. Here we will study the consequences of relaxing the protocol to a more realistic scenario preserving the security.
 
\subsection{Distributing bits from Gaussian states by digitalizing output measurements}

Extracting bits from discrete variables systems can be easily implemented by measuring for example spin observables and associating the up/down orientation, in a fixed axis, as the bit 0/1. In this sense we are digitalizing the output measurements on states. If one uses many copies of a bipartite entangled state it is possible to distribute among two separated parties a pair of bit strings. Such strings of bits possess in general a degree of correlations due to the entanglement present in the state. In the CV scenario and in particular with Gaussian states, the outcome measurements (of canonical variables) fill the continuum, and one needs to digitalize the results in some way. Before proceeding further we detail how to extract correlated string of bits from entangled Gaussian states.

We consider a bipartite CV system of two bosonic modes, $A$ and $B$ (see Fig.~\ref{crypto}). Quadratures on each mode can be efficiently measured by standard homodyne detection. The probability that measuring the position quadrature $\hat x_A$ in mode $A$ results in an outcome $x_A$ with uncertainty $\sigma$ is given by 
\begin{equation}
 \mathcal{P}_A(x_A) = \tr[\hat \rho_A \hat \sigma(x_A) ],
\end{equation}
where $\hat \sigma(x_A)$ is a single-mode Gaussian (squeezed) state with first moments $\{ x_A,0 \}$ and covariance matrix ${\rm diag}\{\sigma^2,1/\sigma^2\}$. In a similar way we define $\mathcal{P}_B(x_B)$ for mode $B$. The probability distribution associated to a joint measurement of the quadratures $\hat x_A$ and $\hat x_B$, is given by $\mathcal{P}_{AB}(x_A, x_B) = \tr[\hat \rho_{AB} (\hat \sigma(x_{A}) \otimes \hat \sigma(x_{B}))]$.

We digitalize the obtained outputs by assigning the bits $+(-)$ or $0 (1)$ to the positive(negative) values of the measured quadratures. This digitalization transforms each joint quadrature measurement into a pair of classical bits. Let us adopt a compact notation by denoting $\mathcal{P}^{\pm}_A \equiv \mathcal{P}_A(\pm|x_A|)$, and $\mathcal{P}^{\pm \mp}_{AB} \equiv \mathcal{P}_{AB}(\pm|x_A|,\mp|x_B|)$. The probability that at a given string index the bits of the corresponding two modes coincide is given by $\mathcal{P}^=_{AB} \equiv (\mathcal{P} ^{++}_{AB}+\mathcal{P}^{--}_{AB})/\sum_{\{\alpha=\pm,\beta=\pm\}}\mathcal{P}^{\alpha\beta}_{AB}$. Correspondingly, the probability that they differ is $\mathcal{P}^{\neq}_{AB} \equiv (\mathcal{P}^{+-}_{AB}+\mathcal{P}^{-+}_{AB})/\sum_{\{\alpha=\pm,\beta=\pm\}}\mathcal{P}^{\alpha\beta}_{AB}$. Trivially, $\mathcal{P}^=_{AB} + \mathcal{P}^{\neq}_{AB} = 1$. If $\mathcal{P}^=_{AB} > \mathcal{P}^{\neq}_{AB}$ the measurement outcomes display correlations, otherwise they display anticorrelations. Notice that, if the two modes are completely uncorrelated, $\mathcal{P}^=_{AB} = \mathcal{P}^{\neq}_{AB} = 1/2$.

\begin{figure}[h]
 \centering
 \psfrag{A}{$x_A$}\psfrag{B}{$p_A$}\psfrag{C}{$x_B$}\psfrag{D}{$p_B$}\psfrag{E}{Alice}\psfrag{F}{Bob}\psfrag{G}{$+$}\psfrag{H}{$-$}
 \includegraphics[width=10cm]{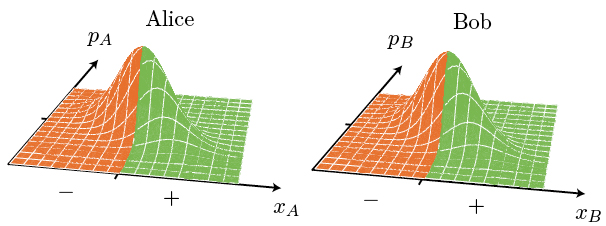}
 \caption{\small{Alice and Bob $1\times 1$ mode Gaussian state. The volume below the green zone denotes the probability for positive outcomes measurements of the $x$ quadrature and the orange zone for the negatives outcomes. We have considered the global state with zero displacement vector.}}\label{crypto}
\end{figure}

\subsection{Efficient Quantum Key Distribution using entanglement}

It has been noticed that the degree of bit correlations are strongly related with the entanglement present in the states used to distribute the pair of strings of classical bits. We ask ourselves which is the possibility that Alice and Bob distribute total correlated strings of bits (a key) in a secure and efficient way by means of bipartite entangled Gaussian states.

The most general scenario we consider is that the state that Alice and Bob share is mixed, letting thus the possibility that an eavesdropper (Eve) has access to some degrees of freedom entangled with Alice and Bob before their distribution. In this general scenario, as the state of Alice and Bob is mixed it always admit a purification (see lemma~\ref{puri}). Thus, any mixed Gaussian state of 2 modes can be expressed as the reduction of a pure Gaussian state of 4 modes in such a way that the bipartite mixed Gaussian state (Alice and Bob) can be obtained after tracing out the 2 other modes (Eve). For what follows it is also important the fact that any NPPT Gaussian state of $N \times M$ can be mapped by Gaussian Local Operations and Classical Communication (GLOCC) to an NPPT symmetric state of $1 \times 1$ modes {\em i.e.} preserving the amount of entanglement. Finally, for $1 \times N$ Gaussian states NPPT and entanglement are equivalent concepts. By virtue of the above properties of Gaussian states, it is sufficient to consider the case in which Alice and Bob share many copies of a quantum system of $1 \times 1$ modes symmetric mixed NPPT Gaussian state $\hat \rho_{AB}$.

While the states that Alice and Bob share correspond to the reduction of a pure 4-mode state, Eve has access to an entangled reduction of two modes. We consider two types of attacks (i) {\em individual} (or incoherent) attacks, where Eve performs individual measurements, possibly non-Gaussian, over her set of states and (ii) {\em finite coherent} (or collective) attacks, where Eve waits until the distribution has been performed and, decides which collective measurement gives her more information on the final key.

We define the error probability as the probability that at a given string index the bits of the corresponding two modes differ {\em i.e.} $\epsilon_{AB}=\mathcal{P}^{\neq}_{AB}$. Having fixed a string of $M$ classical correlated bits, Alice and Bob can apply Classical Advantage Distillation \cite{Maurer1993IEEE} to establish a secret key. To this aim, Alice generates a random bit $b$ and encodes her string of $M$ classical bits ($\vec b_A$) into a vector $\vec b$ of length $M$ such that $b_{Ai}+b_i=b \hspace{-1.5mm} \mod (2)$. The vector $\vec b$ is made public. Bob checks that for his bits all results $b_{Bi}+b_i=b' \hspace{-1.5mm} \mod (2)$ are consistent, and in this case accepts the bit $b$ as the first bit of the secret key. The new error probability is given by \cite{Acin2003PRL}

\begin{equation}
 \epsilon_{AB,M} = \frac{(\epsilon_{AB})^M}{(1 - \epsilon_{AB})^M+(\epsilon_{AB})^M} < \left( \frac{\epsilon_{AB}}{1-\epsilon_{AB}} \right)^M,
\end{equation}
which tends to zero for sufficiently large $M$.

Security with respect to individual attacks from the eavesdropper Eve, can be established if \cite{Acin2003PRL}

\begin{equation}
 \left(\frac{\epsilon_{AB}}{1 - \epsilon_{AB}}\right)^M < |\braket{e_{++}}{e_{--}}|^M,
\end{equation}
where $\ket{e_{\pm \pm}}$ denotes the state of Eve once Alice and Bob have projected their states onto $\ket{\pm |x_{0A}|, \pm |x_{0B}|}$.

Notice that it is favorable for Alice and Bob to have a high degree of success ($\epsilon_{AB}$ small) while Eve can gain information if the overlap between her states after Alice and Bob have measured coincident results ($|\braket{e_{++}}{e_{--}}|$) is sufficiently small. The above inequalities come from the fact that in the case of individual attacks the error on Eve's estimation of the final bit $b$ is bound from below by a term proportional to $|\braket{e_{++}} {e_{--}}|^M$ \cite{Acin2003PRL}. Therefore, Alice and Bob can establish a key if

\begin{equation}\label{Security}
 \frac{\epsilon_{AB}}{1 - \epsilon_{AB}} < |\braket{e_{++}}{e_{--}}|.
\end{equation}

In \cite{Navascues2005PRL} it was shown that any $1 \times 1$ NPPT Gaussian state fulfills the above inequality and thus any NPPT Gaussian state can be used to establish a secure key in front of individual eavesdropper attacks. If we assume that Eve performs more powerful attacks, namely finite coherent attacks, then security is only guaranteed if the much more restrictive condition

\begin{equation}\label{Security2}
 \frac{\epsilon_{AB}}{1 - \epsilon_{AB}} < |\braket{e_{++}}{e_{--}}|^2,
\end{equation}
is fulfilled. This new inequality is violated by some NPPT states. Notice that this implies that the analyzed protocol is not good for these states in this more general scenario. Nevertheless, using the recent techniques of \cite{Christandl2004arXiv}, one can find states for which the presented protocol allows to extract common bits also secure against this attack.

The analyzed protocol \cite{Navascues2005PRL}, results in an inefficient success since it relies on an exact matching between Alice and Bob outputs as a requirement for security. Notice that since security relies on the fact that Alice and Bob have better correlations than the information the eavesdropper can learn about their state, perfect correlation is not a requirement to establish a secure key. We look for a constructive method to improve efficiency without compromising security. By denoting Alice's outputs by $x_{0A}$, we calculate which are the outputs $x_{0B}$ Bob can accept so that the correlation established between Alice and Bob can still be used to extract a secret bit. Specifically, we relax the differences between their output quadrature measurements but imposing correlations in the sign of their quadratures {\em i.e.} we keep correlations in the sign of their output quadratures irrespectively of their numerical value. We thus demand that Alice and Bob associates the bits (+/-) or 0/1 to the positive(negative) value of their respective outcome measurement on each run. Alice will announce publicly the modulus of her outcome measurement each time. We study the efficiency and the security of the protocol if Bob accepts only outcomes measurements that lie in the nearest of Alice's outcomes such that security is guaranteed. Furthermore we analyze the performance of the protocol in terms of the entanglement sharing.

We use the standard form of a bipartite $1\times 1$ mode mixed Gaussian state, (see lemma~\ref{sform}) for the states shared between Alice and Bob
\begin{equation}
\gamma_{AB} = \begin{pmatrix}
 \lambda_a & 0 & c_x & 0\\
 0 & \lambda_a & 0 & -c_p\\
 c_x & 0 & \lambda_b & 0\\
 0 & -c_p & 0 & \lambda_b\\
 \end{pmatrix},
\end{equation}
where, without loss of generality, we flip the sign of $c_p$ and adopt the convention $c_x \geq |c_p| \geq 0$ (we fix also the displacement vector to 0). For simplicity we can deal with mixed symmetric (as said above being totally general) and so $\lambda_a = \lambda_b = \lambda \geq 1$. The positivity condition, see lemma~\ref{sform}, reads $(\lambda - c_x)(\lambda + c_p) \geq 1$, while the entanglement NPPT condition, see lemma~\ref{GNPPTH}, is given by $(\lambda - c_x)(\lambda - c_p) < 1$. We impose that the global state including Eve is pure (she has access to all degrees of freedom outside Alice an Bob) while the mixed symmetric
state, shared by Alice and Bob is just its reduction (see lemma~\ref{puri}), thus

\begin{equation}
 \gamma_{ABE} =
 \begin{pmatrix}
 \gamma_{AB} & C\\
 C^T & \theta_{AB} \gamma_{AB} \theta_{AB}^T
 \end{pmatrix},
\end{equation}

\begin{equation}
 C = \mathcal{J}_{AB} \sqrt{-(\mathcal{J}_{AB} \gamma_{AB})^2 - \id_4} \, \theta_{AB}
 = \begin{pmatrix}
 0 & -X & 0 & -Y\\
 -X & 0 & -Y & 0\\
 0 & Y & 0 & -X\\
 -Y & 0 & -X & 0\\
 \end{pmatrix},
\end{equation}

\begin{equation}
 \theta_{AB} = \theta_A \oplus \theta_B, \quad \quad \mathcal{J}_{AB} = \mathcal{J}_A \oplus \mathcal{J}_B,
\end{equation}
where
\begin{equation}
X=\frac{\sqrt{a+b} + \sqrt{a-b}}{2},\nonumber
\end{equation}

\begin{equation}
Y=\frac{\sqrt{a+b} - \sqrt{a-b}}{2},
\nonumber
\end{equation}
and $a = \lambda^2 -c_x c_p - 1$, $b = \lambda(c_x - c_p)$.

Performing a measurement with uncertainty $\sigma$, the probability that Alice finds $\pm |x_{0A}|$ while Bob finds $\pm |x_{0B}|$, is given by the overlap between the state of Alice and Bob, $\hat\rho_{AB}$, and a pure product state $\hat \rho_{A,i} \otimes \hat \rho_{B,j}$ (with $i,j=0,1$) of Gaussians centered at $\pm |x_{0A}|$ and $\pm |x_{0B}|$ respectively with $\sigma$ width (notice $\hat\rho_{A,0} \equiv \ketbra{+|x_{0A}|}{+|x_{0A}|}$) which gives

\begin{equation}\label{Prob00}
 \begin{split}
 \mathcal{P}^{++}_{AB,\sigma} &= \mathcal{P}^{--}_{AB,\sigma} = \tr[\hat \rho_{AB} (\hat \rho_{A,0} \otimes \hat \rho_{B,0})] =\\
 &= (2\pi)^4 \int d^4 \zeta_{AB} \, \mathcal{W}_{\rho_{AB}} (\zeta_{AB}) \mathcal{W}_{\rho_{A,0} \otimes \rho_{B,0}} (\zeta_{AB}) =\\
 &= K(\sigma) \exp \left( \frac{2|x_{0A}| |x_{0B}| c_x - (\lambda + \sigma^2)(x_{0A}^2 + x_{0B}^2)}{(\lambda + \sigma^2)^2 - c_x^2} \right),
 \end{split}
\end{equation}
for the probability that their symbols do coincide and,

\begin{equation}\label{Prob01}
 \mathcal{P}^{-+}_{AB,\sigma} =\mathcal{P}^{+-}_{AB,\sigma} = K(\sigma) \exp \left( \frac{-2|x_{0A}| |x_{0B}| c_x - (\lambda + \sigma^2)(x_{0A}^2 + x_{0B}^2)}{(\lambda + \sigma^2)^2 - c_x^2} \right),
\end{equation}
for the probability that they do not coincide, where
\begin{equation}
 K(\sigma) = \frac{4\sigma^2}{\sqrt{(\lambda + \sigma^2)^2 - c_x^2}\sqrt{(\lambda \sigma^2 + 1)^2 - c_p^2 \sigma^4}}.
\end{equation}
Their error probability for $\sigma \rightarrow 0$ reads

\begin{equation}\label{Errab}
 \epsilon_{AB} = \lim_{\sigma \to 0} \mathcal{P}^{\neq}_{AB,\sigma} = \frac{1}{1 + \exp \left(\frac{4c_x |x_{0A}||x_{0B}|}{\lambda^2 - c_x^2} \right)}.
\end{equation}
Let us calculate the bipartite state of Eve $\ket{e_{\pm \pm}}$ after they have mesured (Alice has projected onto $\ket{\pm |x_{0A}|}$ and Bob onto $\ket{\pm |x_{0B}|}$)

\begin{equation}
 \gamma_{++} = \gamma_{--} =
 \begin{pmatrix}
 \gamma_{x} & 0\\
 0 & \gamma_{x}^{-1}
 \end{pmatrix}, \quad \quad \gamma_{x} =
 \begin{pmatrix}
 \lambda & c_x\\
 c_x & \lambda
 \end{pmatrix},
\end{equation}

\begin{equation}
 d_{\pm\pm} = \mp
 \begin{pmatrix}
 0\\
 0\\
 A \delta x_0 - B \Delta x_0\\
 A \delta x_0 + B \Delta x_0
 \end{pmatrix},
\end{equation}
where $A = \frac{\sqrt{a+b}}{\lambda + c_x}$, $B =\frac{\sqrt{a-b}}{\lambda - c_x}$, $\Delta x_0 = |x_{0B}| - |x_{0A}|$ and $\delta x_0 = |x_{0B}| + |x_{0A}|$. The overlap between the two states of Eve, which gives a direct quantification of the distinguishability, is given by

\begin{eqnarray}\label{Eveov}
 |\braket{e_{++}}{e_{--}}|^2 = \frac{1}{\sqrt{\det (\frac{\gamma_{++} + \gamma_{--}}{2})}} e^{-(d_{--}-d_{++})^T (\frac{1}{\gamma_{++} + \gamma_{--}}) (d_{--}-d_{++})} =\nonumber \\
 = \exp \Bigg(\frac{-4}{\lambda^2 - c_x^2} \Bigg[ \left( \frac{x_{0A}^2 + x_{0B}^2}{2} \right) (\lambda^2 - c_x^2-1)\lambda + |x_{0A}| |x_{0B}| \left(c_x - c_p(\lambda^2-c_x^2) \right) \Bigg] \Bigg).
\end{eqnarray}
Substituting Eqs.~\eqref{Errab} and \eqref{Eveov} into \eqref{Security} one can check, after some algebra, that the last inequality reduces to

\begin{equation}\label{SecurityIneq}
 \left(\frac{x_{0A}^2 + x_{0B}^2}{2}\right)(\lambda^2 - c_x^2 - 1)\lambda + |x_{0A}| |x_{0B}| \left( -c_x - c_p (\lambda^2 - c_x^2) \right) < 0.
\end{equation}

Notice that condition \eqref{SecurityIneq} imposes both, restrictions on the parameters defining the state ($\lambda, c_x, c_p$), and on the outcomes of the measurements ($x_{0A}, x_{0B}$). The constraints on the state parameters are equivalent to demand that the state is NPPT and satisfies

\begin{equation}\label{Constrain}
 (\lambda - c_x)(\lambda + c_x) \geq 1.
\end{equation}
Nevertheless, as $c_x \geq c_p$, any positive state fulfills this condition. Hence for any NPPT symmetric state, there exists, for a given $x_{0A}$, a range of values of $x_{0B}$ such that secret
bits can be extracted (Eq.~\eqref{Security} is fulfilled) efficiently. This range is given by

\begin{equation}
 \Delta x_0 = |x_{0B}| - |x_{0A}| \in {\mathfrak D}_\alpha =
 \left[ \frac{2}{-\sqrt\alpha-1} , \frac{2}{\sqrt\alpha-1}
 \right] |x_{0A}|,
\end{equation}
where

\begin{equation}
 \alpha = \left( \frac{c_x - \lambda}{c_x + \lambda} \right) \left[ \frac{1 - (\lambda + c_x)(\lambda + c_p)}{1- (\lambda - c_x)(\lambda - c_p)} \right].
\end{equation}
After Alice communicates $|x_{0A}|$ to Bob (the signs obtained are kept in secret), he will accept only measurement outputs within the above interval $\Delta x_0$. The interval is well defined if $\alpha \geq 1$, which equals to fulfill Eq.~\eqref{Constrain}. Notice also that the interval is not symmetric around $|x_{0A}|$ because the probabilities calculated in Eqs.~\eqref{Prob00} and \eqref{Prob01} do depend on this value in a non-symmetric way. The length $D_\alpha$ of the interval of valid measurements outputs for Bob is given by

\begin{equation}
 D_\alpha = \frac{4 \sqrt\alpha}{\alpha-1} |x_{0A}|.
\end{equation}
It can be observed that maximal $D_\alpha\rightarrow\infty$ ($\alpha=1$) corresponds to the case when Alice and Bob share a pure state (Eve is disentangled from the system) and thus condition \eqref{Security} is always fulfilled. On the other hand, any mixed NPPT symmetric state ($\alpha > 1$) admits a finite $D_\alpha$. This ensures a finite efficiency on establishing a secure secret key in front of individual attacks.

If we assume that Eve performs more powerful attacks, namely finite coherent attacks, then security is only guaranteed if

\begin{equation}
 \frac{\epsilon_{AB}}{1 - \epsilon_{AB}} < |\braket{e_{++}} {e_{--}}|^2.
\end{equation}
This condition is more restrictive than \eqref{Security}. With a similar calculation as before we obtain that now security is not guaranteed for all mixed entangled symmetric NPPT states, but only for those that also satisfy

\begin{equation}\label{Constrain2}
 \lambda - (\lambda+c_x)(\lambda-c_x)(\lambda-c_p) > 0.
\end{equation}
For such states, and given a measurement result $x_{0A}$ of Alice, Bob will only accept outputs within the range

\begin{equation}
 \Delta x_0 = |x_{0B}| - |x_{0A}| \in {\mathfrak D}_\beta = \left[ \frac{2}{-\sqrt\beta-1} , \frac{2}{\sqrt\beta-1} \right] |x_{0A}|,
\end{equation}
where

\begin{equation}
 \beta = \frac{2\lambda(\lambda^2-c_x^2-1)}{\lambda-(\lambda+c_x)(\lambda-c_x)(\lambda-c_p)} \geq 1.
\end{equation}
As before, $\beta \geq 1$ is fulfilled by conditions \eqref{Constrain} and \eqref{Constrain2}.

Let us now focus on the efficiency issue. We define the efficiency $E(\gamma_{AB})$ of the protocol for a given state $\gamma_{AB}$, as the average probability (over the range of secure outcomes, ${\mathfrak D}$) of obtaining a classically correlated bit. Explicitly,

\begin{equation}\label{protocolefficiency}
 E(\gamma_{AB}) = \int_{\Delta x_0 \in {\mathfrak D}} dx_{0A}dx_{0B}(1-\epsilon_{AB}) \tr (\hat\rho_{AB} \ketbra{x_{0A}, x_{0B}}{x_{0A},x_{0B}}).
\end{equation}
The marginal distribution in phase-space is easily computed by integrating the corresponding Wigner function in momentum space

\begin{equation}
 \begin{split}
 \tr (\hat\rho_{AB} \ketbra{x_{0A},x_{0B}}{x_{0A},x_{0B}})
 &= \int \int dp_A dp_B \mathcal{W}_{\rho_{AB}} ({\zeta}_{AB})=\\
 &= \frac{\exp \left( \frac{2 c_x x_{0A} x_{0B} - \lambda (x_{0A}^2+x_{0B}^2)}{\lambda^2-c_x^2} \right) } {\pi \sqrt{\lambda^2-c_x^2}},
 \end{split}
\end{equation}
but the final expression of Eq.~\eqref{protocolefficiency} has to be calculated numerically. Note that if Alice and Bob share as a resource $M$ identical states (NPPT state for individual attacks, and NPPT fulfilling condition \eqref{Constrain2} for finite coherent attacks), the number of classically correlated bits that can be extracted from them is $\sim M \times E(\gamma_{AB})$. The efficiency Eq.~\eqref{protocolefficiency} increases with increasing $D$. In particular, for the protocol given in \cite{Navascues2005PRL}, $D=0$, and therefore $E(\gamma_{AB})=0$ for any state.

We investigate now the dependence of $E(\gamma_{AB})$ with the entanglement of the NPPT mixed symmetric state used for the protocol as well as with the purity of the state. In order to fix one parameter of the states we fix the energy of the states at a fixed value ($\lambda$ cte). As a measure of the entanglement between Alice and Bob, see \eqref{lognega}, we compute the logarithmic negativity
\begin{equation}
 {\rm LN} (\gamma_{AB}) = -\log_2[\min(\tilde\mu_-,1)]  -\log_2[\min(\tilde\mu_+,1)] = \log_2 \left( \frac{1} {\sqrt{(\lambda-c_x)(\lambda-c_p)}} \right) > 0.
\end{equation}
because $\tilde\mu_+ >1$ and $1>\tilde\mu_- = \sqrt{(\lambda-c_x)(\lambda-c_p)}$.

\begin{figure}[h]
 \centering
 \psfrag{A}{${\rm LN}$}\psfrag{B}[][][1][-90]{$E$}
 \includegraphics[width=10cm]{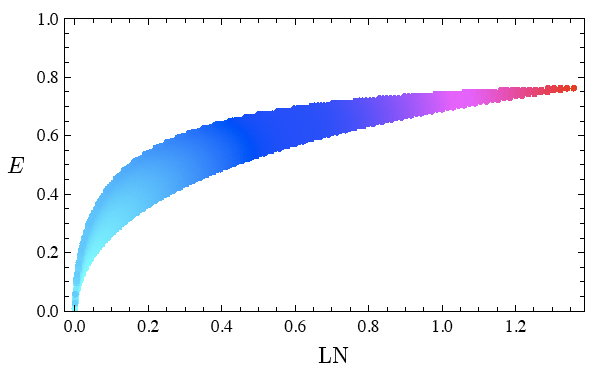}
 \caption{\small{Protocol efficiency (quantified by $E$) versus the entanglement measured by logarithmic negativity ${\rm LN}$. The shading from cyan to red corresponds to purity from zero to one.}}\label{effLN}
\end{figure}
In Fig.~\ref{effLN}, we display the efficiency of the protocol (assuming individual attacks) versus entanglement shared between Alice and Bob for different states $\gamma_{AB}$. There is not a one-to-one correspondence between efficiency and entanglement, since states with the same entanglement can have different purity, which can lead to different efficiency. This is so because there are two favorable scenarios to fulfill Eq.~\eqref{Security}. The first one is to demand large correlations so that the relative error $\epsilon_{AB}$ of Alice and Bob is small. The second scenario happens when Alice and Bob share a state with high purity, {\em i.e.}, Eve is very disentangled. In this case,
independently of the error $\epsilon_{AB}$, Eq.~\eqref{Security} can be fulfilled more easily. Despite the fact that efficiency generally increases with increasing entanglement and increasing purity, this enhancement, as depicted in the figure, is a complex function of the parameters involved. Nevertheless, one can see that there exist an entanglement threshold (around ${\rm LN} \simeq 0.2$) below which the protocol efficiency diminishes drastically no matter how mixed are the states shared between Alice and Bob.

It is also illustrative to examine the dependence of $\alpha$ (which determines the interval length $D_\alpha$) on the entanglement of the states shared by Alice and Bob.

\begin{figure}[h]
 \centering
 \psfrag{A}{$\alpha$}\psfrag{B}[][][1][-90]{${\rm LN}$}
 \includegraphics[width=10cm]{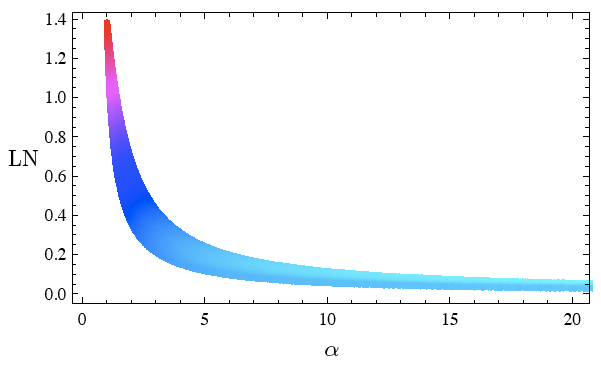}
 \caption{\small{Entanglement of the states shared between Alice and Bob measured in terms of the logarithmic negativity ${\rm LN}$ as a function of the parameter $\alpha$ under individual attacks. The shading from cyan to red corresponds to purity from zero to one.}}\label{LNalpha}
\end{figure}
In Fig.~\ref{LNalpha} we plot the logarithmic negativity of a given state versus the parameter $\alpha$. States with the same entanglement but different purity are associated to quite different values of $\alpha$, specially for states with low entanglement (high purity). Nevertheless states with high entanglement permit a large interval length (small $\alpha$) and, thus, high efficiency.

In both, Fig.~\ref{effLN} and Fig.~\ref{LNalpha}, we have observed that states with different entanglement give the same efficiency. However it is important to point out that to extract the key's bits, Classical Advantage Distillation \cite{Maurer1993IEEE} stills needs to be performed. The efficiency of Maurer's protocol, strongly increases with decreasing $\epsilon_{AB}$, and, therefore, the states with higher entanglement will provide a higher key rate.

\section{Conclusions}

Efficiency is a key issue for any experimental implementation of Quantum Cryptography since available resources are not unlimited. In this chapter, we have shown that the sharing of entangled Gaussian variables and the use of only Gaussian operations permits efficient Quantum Key Distribution against individual and finite coherent attacks.

We have used the fact that all mixed NPPT symmetric states can be used to extract secret bits to design an algorithm, that efficiently succeeds for a secure extraction of a key. Whereas under individual attacks all mixed NPPT symmetric states admit a finite efficiency, for finite coherent attacks an additional condition constrains the parameters of the states. We have introduced a figure of merit (the efficiency $E$) to quantify the number of classical correlated bits that can be used to distill a key from a sample of $M$ entangled states. We have observed that this quantity grows with the entanglement shared between Alice and Bob. This relation it is not one-to-one due to the fact that states with less entanglement but with more purity (eavesdropper more disentangled) can be equally efficient. Nevertheless we have point out that, these states would be inefficient, when performing the Classical Advantage Distillation of the key.


\chapter{Byzantine agreement problem with Continuous Variable}\label{byzantine}

\section{Introduction}

One of the aims of Quantum Information is to provide new protocols and algorithms (set of rules for solving a problem in a finite number of steps) which exploit quantum resources to find a solution to problems which either lack a solution using classical resources or the solution is extremely hard to implement. In this chapter we analyze a multipartite protocol, the Byzantine agreement problem, by means of multipartite Gaussian entanglement.

The term ``Byzantine Agreement'' was originally coined by Lamport and Fischer \cite{Lamport1982SRI} in the context of computer science to analyze the problem of fault tolerance when a faulty processor is sending inconsistent information to other processors. In a cryptographic context, it refers to distributed protocols in which some of the participants might have malicious intentions and could try to sabotage the distributed protocol inducing the honest parties to take contradictory actions between them. This problem is often reformulated in terms of a Byzantine army where there is a general commander who sends the order of attacking or retreating to each one of his lieutenants. Those can also communicate pairwise to reach a common decision concerning attacking or retreating, knowing that there might be traitors among them including the general commander. A traitor could create fake messages to achieve that different parts of the army attack while other retreat, which would put the army at a great disadvantage. The question hence is whether there exists a protocol among all the officials involved that, after its termination, satisfies the following conditions: The commanding general sends an order to his $N-1$ lieutenants such that: (i) All loyal lieutenants obey the same order, (ii) If the commanding general is loyal, then every loyal lieutenant obeys the order he sends. It is assumed that the parties cannot share any previous setup.

Lamport {\em et al} \cite{Lamport1982ACM} proved that if the participants only shared pairwise secure classical channels, then Byzantine Agreement or broadcast is only possible iff $t<n/3$ where $n$ is the number of players and $t$ the number of traitors among them. In \cite{Fitzi2001PRL} Fitzi and co-workers introduced a weaker nevertheless important version of Byzantine Agreement known as {\em detectable broadcast}. Detectable broadcast is said to be achieved if the protocol satisfies the following conditions: (i) If no player is corrupted, then the protocol achieves broadcast and (ii) If one or more players are corrupted, then either the protocol achieves Byzantine Agreement or all honest players abort the protocol. Thus, in a detectable broadcast protocol, cheaters can force the protocol to abort, {\em i.e.} no action is taken. In such cases all honest players agree on aborting the protocol so that contradictory actions between the honest players are avoided. In the same paper \cite{Fitzi2001PRL}, Fitzi, Gisin and Maurer devised a solution to the detectable broadcast problem using multipartite entanglement as a quantum resource. Later on Fitzi {\em et al} \cite{Fitzi2002Proc} and Iblisdir and Gisin \cite{Iblisdir2004PRA} showed that a Quantum Key Distribution (QKD) protocol, which guarantees a private sequences of classical data shared between pairs of parties, suffices to solve detectable broadcast. This situation is reminiscent of quantum cryptography, in which two seemingly independent main schemes exist for QKD, the prepare-and-measure BB84 scheme \cite{Bennett1984IEEE} which does not use entangled states shared between Alice and Bob, and the Ekert91 scheme \cite{Ekert1991PRL} where indeed entanglement is explicitly distributed and the security is guaranteed by the violations of Bell inequalities. However, the two schemes have been shown to be completely equivalent \cite{Bennett1992PRL}, and specifically entanglement stands as a precondition for any secure key distribution \cite{Curty2004PRL}.

Detectable broadcast might be regarded in a similar view as Quantum Key Distribution and, arguably, the same reasoning applies to the different protocols advanced for its solution, making explicit or implicit use of multipartite entanglement. In this paper, we adopt an approach {\em \`a la} Ekert91, guided by the physical motivation of studying the performance and the usefulness of multipartite entangled states as operational resources to achieve detectable broadcast. While protocols for this task exist for
qutrits \cite{Fitzi2001PRL} and qubits \cite{Cabello2003PRA,Gaertner2008PRL}, in this paper we investigate the possibility of solving detectable broadcast with Continuous Variable (CV) systems, namely with Gaussian states and performing Gaussian operations only. The motivations for this approach are manifold. On the practical side, the recent progresses in CV QKD \cite{Grosshans2007ICP} has shown that the use of efficient homodyne detectors, compared to photon counters employed in BB84 schemes, enables the distribution of secret keys at faster rates over increasingly long distances \cite{Grosshans2003N,Lodewyck2007PRA}. On the theoretical side, multipartite entanglement is a central concept whose understanding and characterization (especially in high-dimensional and CV systems), despite recent efforts, is far from being complete. It is important, therefore, to approach this task {\em operationally}, {\em i.e.} by connecting entanglement to the success or to the performance of diverse quantum information and communication tasks \cite{Plenio2007QIC}, while exploiting the differences between discrete-variable and continuous-variable  scenarios. For Gaussian states of light fields, which are presently the theoretical and experimental pillars of Quantum Information with CVs \cite{Cerf2007ICP}, an important result in this respect is due to van Loock and Braunstein \cite{vanLoock2000PRL}. They introduced a scheme to produce fully symmetric (permutation-invariant) $n$-mode Gaussian states exhibiting genuine multipartite entanglement. Furthermore, they devised a communication protocol, the ``quantum teleportation network'', for the distribution of quantum states exploiting such resources, which has been experimentally demonstrated for $n=3$ \cite{Yonezawa2004N}. The optimal fidelity characterizing the performance of such a protocol yields an operational quantification of genuine multipartite entanglement in symmetric Gaussian states. This quantification is equivalent to the information-theoretic ``residual contangle'' measure, a Gaussian entanglement monotone, emerging from the monogamy of quantum correlations \cite{Adesso2007PRL,Adesso2006NJP}. Other applications of multipartite Gaussian entanglement have been advanced and demonstrated, and the reader may find more details in \cite{Braunstein2005RMP,Cerf2007ICP}, as well as in the two more recent complementary reviews \cite{Adesso2007JPA} (theoretical) and \cite{Lian2007NJP} (experimental).

We have proposed a novel protocol to solve detectable broadcast with symmetric multimode entangled Gaussian states and homodyne detection. This provides an alternative interpretation of multipartite
Gaussian entanglement as a resource enabling this kind of secure communication. We concentrate on the case of three parties, and remarkably find that not all three-mode symmetric entangled Gaussian states are useful to achieve a solution: to solve detectable broadcast there is a minimum threshold in the multipartite entanglement. This is at variance with the two-party QKD counterpart: in there, all two-mode entangled Gaussian states are useful to obtain a secure key using Gaussian operations \cite{Navascues2005PRL}. We eventually discuss how our protocol can be implemented in realistic conditions, namely considering detectors with finite efficiency and yielding not perfectly matched measurement outcomes, and Gaussian resources which are not ideally pure, but possibly (as it is in reality) affected by a certain amount of thermal noise. We show that under these premises the protocol is still efficiently applicable to provide a robust solution to detectable broadcast over a broad range of the involved parameters (noise, entanglement, measurement outcomes and uncertainties), paving the way towards a possible experimental demonstration in a quantum optical setting.

\section{Detectable broadcast protocol}\label{protocol}

The protocols to solve detectable broadcast, in the discrete scenario, using entanglement as a resource are based on three differentiated steps:
\begin{enumerate}[i)]
 \item Distribution of the quantum states.
 \item Test of the distributed states.
 \item Protocol by itself.
\end{enumerate}
Step iii) is fundamentally classical, since it uses the outputs of the different measurements of the quantum states to simulate a particular random generator (``primitive''). Let us consider the simplest case, in which only three parties are involve and at most one traitor is allowed, for which no classical solution exists. The parties are traditionally denoted by $S$ (the sender, {\em i.e.} commander general) and the receivers $R_0$ and $R_1$ ({\em i.e.} lieutenants), and at most, only one is a traitor. In this case, the primitive generates for every invocation a random permutation of the elements $\{0,1,2\}$ with {\em uniform distribution}, {\em i.e.} $(t_S, t_{R_0}, t_{R_1}) \in \{(0,1,2), (0,2,1), (1,0,2), (1,2,0), (2,0,1), (2, 1, 0)\}$. In this primitive, no single player $n$ can learn more about the permutation than the value $t_n$ ($n=\{S, R_0, R_1\}$) which she/he obtain {\em i.e.}, each player ignores how the other two values are assigned to the other two players. Furthermore, nobody else (besides the parties) have access to the sequences.

Entanglement is used in the protocol to distribute classical private random variables with a specific correlation between the players, in such a way that any malicious manipulation of the data can be
detected by all honest parties allowing them to abort the protocol. In the discrete variable case such a primitive can be implemented with qutrits using {\em e.g.} Aharonov states $\ket{\mathcal{A}}=\frac{1}{\sqrt{6}} (\ket{0,1,2} + \ket{1,2,0} + \ket{2,0,1}-\ket{0,2,1} - \ket{1,0,2} - \ket{2,1,0})$. This choice allows the distribution and test part to be secure \cite{Fitzi2001PRL}. Whenever the three qutrits are all measured in the same basis, all the three results are different. Hence -after discarding all the states used for the testing of the distributed states (step ii))- the players are left with a sequence of outputs that reproduces the desired primitive. We schematically represent this primitive by the table below
\begin{equation*}
 \begin{tabular}{|c|c|c|c|c|c|c|c|c|c|c|}
 \hline $j$ & 1 & 2 & 3 & 4 & 5 & 6 & 7 & 8 & 9 & $\ldots$\\
 \hline
 \hline $S$ & 2 & 0 & 0 & 1 & 2 & 1 & 0 & 2 & 1 & $\ldots$\\
 \hline $R_0$ & 1 & 1 & 2 & 0 & 1 & 0 & 1 & 0 & 2 & $\ldots$\\
 \hline $R_1$ & 0 & 2 & 1 & 2 & 0 & 2 & 2 & 1 & 0 & $\ldots$\\
 \hline
 \end{tabular}
\end{equation*}
After accomplishing the distribution and test part of the protocol (steps i) and ii)) detailed in section \ref{disandtes}, the sender $S$ will broadcast a bit $b \in \{0,1\}$ ($0\stackrel{\wedge}{=}$``attack'', $1\stackrel{\wedge}{=}$``retreat'') to the two receivers using the mentioned distributed and tested primitive and classical secure channels.

Following \cite{Fitzi2001PRL} the broadcast (step iii)) proceeds as follows:

\noindent iii-1) We denote by $b_i$ the bits received by $R_i$, $i=0,1$ (notice that if the sender is malicious, the broadcasts bits $b_i$ could be different). Each receiver $R_i$ demands to $S$ to send him the indices $j$ for which $S$ got the result $b_i$ (on the primitive). Each player $R_i$ receives a set of indices $J_i$.

\noindent iii-2) Each $R_i$ test consistency of his own data, {\em i.e.} checks weather his output on the set of indices he receives ($J_i$) are all of them different from $b_i$. If so the data is consistent and he settles his flag to $c_i=b_i$, otherwise his flag is settled to $c_i=\perp$.

\noindent iii-3) $R_0$ and $R_1$ send their flags to each other. If both flags agree, the protocol terminates with all honest participants agreeing on $b$.

\noindent iii-4) If $c_i=\perp$, then player $R_i$ knows that $S$ is dishonest, the other player is honest and accepts his flag.

\noindent iii-5) If both $R_0$ and $R_1$ claim to have consistent data but $c_0 \neq c_1$, player $R_1$ demands from $R_0$ to send him all the indices $k \in J_0$ for which $R_0$ has the results $1-c_0$. $R_1$ checks now that (i) all indices $k$ from $R_0$ are not in $J_1$ and (ii) the output $R_1$ obtains from indices $k$ correspond to the value 2. If this is the case, $R_1$ concludes that $R_0$ is honest and changes his flag to $c_0$. If not, $R_1$ knows that $R_0$ is dishonest and he keeps his flag to $c_1$. Detectable broadcast is in this way achieved.

\section{Continuous variable primitive}\label{CVpri}

We first review briefly the basic tools needed to describe multipartite Gaussian states and measurements. Building on the ideas presented in the previous section, we construct a protocol adapted to the Continuous Variable case.

To achieve the primitive presented in the CV set up we consider quantum systems of 3 canonical degrees of freedom, {\em i.e.} 3 modes (one for each player), associated to a Hilbert space $\mathcal{H}=\mathcal{L}^2(\mathbb{R}^6)$.
Following the discussion of the protocol (step iii)) for the implementation of the primitive in the discrete case, we choose to use as a resource a pure, fully inseparable tripartite Gaussian state, completely symmetric under the interchange of the modes \cite{vanLoock2000PRL}.

In contrast to the bipartite case where one kind of entanglement exist, the tripartite case is much richer  \cite{Acin2001PRL} and offers several classes of entanglement. In \cite{Giedke2001PRA} a classification for Gaussian states, in terms of the non-positivity of the partial transpose across several bipartite partitions (NPPT criterium), was studied. Namely for tripartite Gaussian states described by a covariance matrix $\gamma$, 5 classes of states can be distinguished.

\begin{enumerate}[i)]
\item Class 1 (Fully inseparable states or genuine entangled) iff
\begin{eqnarray*}
 \theta_A \gamma \theta_A^T + \im \mathcal{J} \ngeq 0,\\
 \theta_B \gamma \theta_B^T + \im \mathcal{J} \ngeq 0,\\
 \theta_C \gamma \theta_C^T + \im \mathcal{J} \ngeq 0.
\end{eqnarray*}

\item Class 2 (One-mode biseparable) iff
\begin{eqnarray*}
 \theta_i \gamma \theta_i^T + \im \mathcal{J} \geq 0,\\
 \theta_j \gamma \theta_j^T + \im \mathcal{J} \ngeq 0,\\
 \theta_k \gamma \theta_k^T + \im \mathcal{J} \ngeq 0,
\end{eqnarray*}
any permutation of modes ($i,j,k$) must be considered.

\item Class 3 (Two-mode biseparable) iff
\begin{eqnarray*}
 \theta_i \gamma \theta_i^T + \im \mathcal{J} \geq 0,\\
 \theta_j \gamma \theta_j^T + \im \mathcal{J} \geq 0,\\
 \theta_k \gamma \theta_k^T + \im \mathcal{J} \ngeq 0,
\end{eqnarray*}
any permutation of modes ($i,j,k$) must be considered.

\item Class 4 (Three-mode biseparable or bound entanglement)

or Class 5 (Fully separable) iff
\begin{eqnarray*}
 \theta_A \gamma \theta_A^T + \im \mathcal{J} \geq 0,\\
 \theta_B \gamma \theta_B^T + \im \mathcal{J} \geq 0,\\
 \theta_C \gamma \theta_C^T + \im \mathcal{J} \geq 0,
\end{eqnarray*}
\end{enumerate}
with $\mathcal{J}=\oplus_{i=1}^3 \mathcal{J}$.
The condition for full inseparability (truly multipartite entanglement) across the $1\times1\times1$ mode partition can be rewritten and reads as $\gamma+ \im\mathcal{J}_A \ngeq 0$, $\gamma+ \im\mathcal{J}_B \ngeq 0$, $\gamma+ \im\mathcal{J}_C \ngeq 0$ with $\mathcal{J}_A=\mathcal{J}^T\oplus \mathcal{J} \oplus \mathcal{J}$, analogously for $B$ and $C$.

A  tripartite Gaussian state with such properties (pure, fully inseparable, completely symmetric under the interchange of the modes) with covariance matrix $\gamma(a)$ accepts the following parametrization \cite{Giedke2001PRA}

\begin{equation}\label{egamma}
 \gamma(a) =\left(
 \begin{array}{cccccc}
 a&0&c&0&c&0 \\
 0&b&0&-c&0&-c \\
 c&0&a&0&c&0 \\
 0&-c&0&b&0&-c \\
 c&0&c&0&a&0 \\
 0&-c&0&-c&0&b
 \end{array}\right),
\end{equation}
with $a \geq 1$ and

\begin{eqnarray}
 b=\frac{1}{4}(5a-\sqrt{9a^2-8}),\\
 c=\frac{1}{4}(a-\sqrt{9a^2-8}).
\end{eqnarray}
It follows that $\gamma(a)$ is fully inseparable as soon as $a>1$. Quantitatively, the genuine tripartite entanglement of the states of \eqref{egamma}, as measured
by the residual contangle \cite{Adesso2006NJP}, is a monotonically increasing function of $a$ and diverges for $a \rightarrow \infty$.

So far $\gamma(a)$ appears as the ``equivalent'' CV version of the discrete Aharonov state $\ket {\mathcal{A}}$, {\em i.e.} pure, fully inseparable and completely symmetric under exchange of the 3 players. One is tempted to infer, therefore, that the discussed primitive of the discrete case can be straightforwardly generalized to the continuous one. A standard way of transforming the correlations of the shared entangled quantum states $\hat \rho_a$ into a sequence of classically correlated data between the 3 players is to perform a homodyne measurement of the quadratures of each mode. Denoting by $\hat x_S$, $\hat x_{R_0}$, $\hat x_{R_1}$ the position (or momentum) operator of each mode, and by $x_{S}$, $x_{R_0}$, $x_{R_1}$ the output of the respective measurements, the players after quadrature measurement on their modes end up with classical correlated data according to the entanglement sharing. In order to proceed with the classical part of the protocol (step iii)) one need to digitalize the outcome measurements into trit correlated data. A standard way to proceed is the following, the 3 players communicate classically with each other and agree, for instance, only on those outcome values for which either $\left| x_S \right|=\left| x_{R_0} \right|=\left| x_{R_1} \right|=\{0,x_0\}$ with $x_0>0$. In this way each player can associate the logical trit ($t=0,1,2$) to a positive, negative or null result respectively, willing to map the quadrature correlations into trit correlations. To verify the success of such procedure we define the probability distribution that measuring the quadratures $\hat x_S$, $\hat x_{R_0}$, $\hat x_{R_1}$ an outcome $t_S=j$, $t_{R_0}=k$, $t_{R_1}=l$ is produced with uncertainty $\sigma$, being $(j,k,l)\in \{0,1,2\}$. Such probability is given by

\begin{eqnarray}\label{overlap}
 \mathcal{P}(j,k,l)&=&\tr\hat\rho_\sigma^{\{j,k,l\}}\hat\rho_a=\\
 &=&(2\pi)^n\int \mathcal{W}_\sigma^{\{j,k,l\}}(\xi)\mathcal{W}_a(\xi)d^{2n}\xi\nonumber.
\end{eqnarray}
Here $\hat\rho_\sigma^{\{j,k,l\}} = \hat\rho_S^{t_S=j} \otimes \hat\rho_{R_0}^{t_{R_0}=k} \otimes \hat\rho_{R_1}^{t_{R_1}=l}$ describes the separable state of the 3 modes obtained after each party has measured its corresponding quadrature and obtained an output $j,k,l$. Thus, the state  $\hat\rho_\sigma$ of the system after the measure can be described by a fully separable covariance matrix
\begin{equation}
\gamma_\sigma=\begin{pmatrix}
 \sigma^2 & 0 \\
 0 & 1/\sigma^2
\end{pmatrix}
\oplus
\begin{pmatrix}
 \sigma^2 & 0 \\
 0 & 1/\sigma^2
\end{pmatrix}
\oplus
\begin{pmatrix}
 \sigma^2 & 0 \\
 0 & 1/\sigma^2\nonumber
\end{pmatrix},
\end{equation}
(where each party has a pure one mode Gaussian state) and by a displacement vector $d_\sigma^T=(f_{j,k,l}^{(1)}|x_0|,0)\oplus(f_{j,k,l}^{(2)}|x_0|,0)\oplus(f_{j,k,l}^{(3)}|x_0|,0)$ being $\vec f_{j,k,l}=(\pm1/0,\pm1/0,\pm1/0)$ where $\{+1/-1/0\}$ corresponds to the trit values $\{0/1/2\}$. On the other hand $\mathcal{W}_a$ is the Wigner function of the initial tripartite entangled state $\hat\rho_a$ described by the covariance matrix $\gamma(a)$ and displacement vector $d$ (to be fixed).

Since Gaussian states are symmetric with respect to their displacement vector, and quadrature output measurements fill the continuum, it is easy to see that a mapping into classical trits such that all possible outputs $\{0,1,2\}$ occur with the same and non-vanishing probability is not possible. Therefore, the primitive discussed in the discrete case has to be modified when adapted to the CV scenario. The way we find to overcome this asymmetry of the output's probabilities relies on
joining the correlations from pairs of quantum states. First we map the quantum correlations involved in each single quantum state just to classical bits by a ``sign binning'' (as in CV QKD \cite{Navascues2005PRL}). That is, first we keep only the results for which the 3 players obtain a coincident output $|x_S|=|x_{R_0}|=|x_{R_1}|= x_0>0$, and associate the logical bit $b_n=0 (1)$, where $n=\{S,R_0,R_1\}$, to positive (negative) value of the coincident quadrature $+x_0 (-x_0)$. In every measurement the sender makes public his outcome result $x_0$, whatever the output is (in each run the sender will obtain a different outcome result), but not its sign. In this step, all states are used in the protocol (we will relax the ``idealization'' of coincident outputs in section~\ref{effreaimp}).
Second we construct an appropriate primitive consisting of a random permutation of the elements $(i,j,k) \in \{(0,1,1),(1,0,1),(1,1,0)\}$. Compulsory for the implementation of the primitive is that every element of the primitive appears with equal probability and any other combination of the outputs not regarded by the primitive is exceedingly small compared to the allowed permutations. In other words, denoting by $p \equiv \mathcal{P}(0,1,1)=\mathcal{P}(1,0,1)=\mathcal{P}(1,1,0)$, and, denoting by $\delta_i \equiv \mathcal{P}(else)$, we require that the corresponding conditional probabilities,
\begin{equation}
 \tilde{\mathcal{P}}(b_S,b_{R_0},b_{R_1})=\frac{\mathcal{P}(b_S,b_{R_0},b_{R_1})}{\sum_{i,j,k=\{0,1\}}
 \mathcal{P}(i, j, k)}\,,\nonumber
\end{equation} fulfill $\tilde p \rightarrow\frac{1}{3}$ and $\tilde\delta_i\rightarrow0$.

If the above conditions are met, it is possible then by invoking two consecutive times the previous generator ({\em i.e.} using a pair of quantum states of the class \eqref{egamma}), to map pairs of bit values
to trit values (0,1,2) plus an additional undesired element ``$u$''. For instance, the players, after keeping only those bits obtained by coincident quadrature outputs $\pm x_0$, use two consecutive bits $m$ and $m+1$ for the following association

\begin{eqnarray*}
 &&(1,0)\rightarrow {\mathbf 0}\\
 &&(0,1)\rightarrow {\mathbf 1}\\
 &&(1,1)\rightarrow {\mathbf 2}\\
 &&(0,0)\rightarrow {\boldsymbol u}.\\
\end{eqnarray*}
Thus, by concatenating two invocations and using the above association, one generates the permutations corresponding to the primitive $\{(0,1,2),(0,2,1),(1,0,2),$ $(1,2,0),(2,0,1),(2,1,0)\}$ plus a permutation of the undesired element ``$u$'' {\em i.e.} $(u,2,2)$, $(2,u,2)$ and $(2,2,u)$ which will be discarded during the protocol. With all these tools at hand, optimal results for the desired probabilities are achieved for a displacement vector of the form
$d^T=-\frac{x_0}{3}(1,0,1,0,1,0)$, and yield

\begin{eqnarray}
 \delta_1&=&\mathcal{P}(1,1,1)=C(a,\sigma)\exp\left(-\frac{4}{3}\frac{x_0^2}{K_1}\right),\nonumber\\
 \delta_2&=&\mathcal{P}(0,0,0)=C(a,\sigma)\exp\left(-\frac{16}{3}\frac{x_0^2}{K_1}\right),\\
 \delta_3&=&\mathcal{P}(0,0,1)=\mathcal{P}(0,1,0)=\mathcal{P}(1,0,0)=\nonumber\\
 &=&C(a,\sigma) \exp \left[ \frac{-4 x_0^2\left(\sigma^2+\frac{1}{4}\left[5a-\sqrt{9a^2-8}\right]\right)}{K_1 K_2}\right],\nonumber
\end{eqnarray}
and

\begin{equation}
 \begin{split}
 p\ &=\ \mathcal{P}(0,1,1)=\mathcal{P}(1,0,1)=\mathcal{P}(1,1,0)=\\
 &=\ C(a,\sigma)\exp\left(-\frac{8}{3}\frac{x_0^2}{K_2}\right),
\end{split}
\end{equation}
with coefficients $K_1=\sigma^2+\frac{1}{2}\left[3a-\sqrt{9a^2-8}\right]$, $K_2=\sigma^2+\frac{1}{4}\left[3a+\sqrt{9a^2-8} \right]$, and pre-factor

\begin{equation*}
\begin{split}
C&(a,\sigma)=\left[ \det{\left( \frac{\gamma_M+\gamma(a)}{2}\right) }\right]^{-\frac{1}{2}}=\\
=&\frac{8}{\left(a-c+\sigma^2\right) \left(b+c+\frac{1}{\sigma^2}\right) \sqrt{\left(a+2c+\sigma^2 \right) \left(b-2c+\frac{1}{\sigma^2}\right)}}.
\end{split}
\end{equation*}
The probability distribution depends on the parameters $a$, $x_0$ and $\sigma$. In the case $a\gg1$, it can be seen that the conditional probabilities satisfy the following

\begin{equation}
 \tilde{p}=\frac{1}{3}-\frac{4}{9}k+O(k^2), \quad  \tilde{\delta}_i\rightarrow 0,
\end{equation}
with $k=\exp{\left[-\frac{4}{3}(\frac{x_0}{\sigma})^2\right]}$. In the limit of a perfect, zero-uncertainty homodyne detection ($\sigma \rightarrow 0$), $\tilde p$ converges exactly to $1/3$. In general, there exists a large region in the parameters space for which $\tilde p\rightarrow \frac{1}{3}$ and $\tilde \delta_i\rightarrow 0$, as depicted in Fig.~\ref{pideal}.

\begin{figure}[h]
 \centering
 \psfrag{A}{$a$}\psfrag{B}{$x_0$}\psfrag{C}{$\tilde p$}
 \includegraphics[width=8cm]{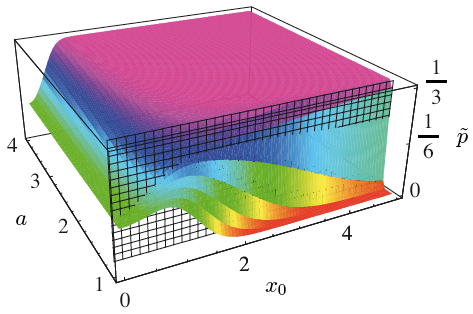}
 \caption{\small{Plot of the conditional probability $\tilde p$ as a function of the measurement outcome $x_0$ and the shared entanglement $a$, for pure symmetric tripartite Gaussian resource states. Detectable broadcast is ideally solvable in the huge, unbounded region of $x_0 \gg 0$, $a \gg 1$, where $\tilde p \rightarrow 1/3$. The entanglement threshold $a=a_{\rm{thresh}} = 5\sqrt2/6$ is depicted as well (wireframe surface). All the quantities plotted are adimensional.}}\label{pideal}
\end{figure}

However, there is a {\em lower} bound on the entanglement content of the symmetric Gaussian states of \eqref{egamma} in order to fulfill the above conditions. Only for $a \geq a_{\rm{thresh}} = \frac{5\sqrt{2}}{6} \approx 1.18$, one has that $\tilde p > \tilde \delta_i$, which is a necessary condition to implement the primitive. This indicates that not all pure 3-mode symmetric fully entangled Gaussian states can be successfully employed to solve detectable broadcast via our protocol. This
entanglement threshold is an {\em a priori} bound which does not depend on the specific form of the employed resource states. For any parametrization of the covariance matrix of $\hat\rho_a$, which is obtainable from $\gamma(a)$ by local unitary squeezing transformations (hence at fixed amount of tripartite entanglement), the same condition discriminating useful resource states is analytically recovered. We remark that this bound becomes tight in the limit $x_0 \rightarrow \infty$, meaning that $a \geq a_{\rm{thresh}}$ becomes necessary {\em and} sufficient for the successful implementation of the primitive.

\section{Distribution and test}\label{disandtes}

We move now to the distribution and test part of the protocol which represents the first step in the execution of the appropriate primitive and has only two possible outputs: global success or global failure. In the case of failure, a player assumes that something went wrong during the execution of the protocol and aborts any further action. In the case of global success, each of the parties ends up with a set $\{K\}$ of classical data, and the protocol proceeds classically, according to the steps explained in section \ref{protocol}.

From now on we assume that the players share pairwise secure classical channels and secure (noiseless) quantum channels. The secure distribution and test part uses correlations to validate the
fairness of the other parties. Therefore, quantum states are sent through noiseless quantum channels and measures are performed massively. In doing so, it is possible to detect manipulation of the data on a statistical basis and abort the protocol if necessary. In this section we mostly focus on the issue of security in the distribution and security test of the data. This should permit the detection of any malicious manipulation of the data. The explicit effects of sabotage actions will be reported in section \ref{errest}. We postpone the important issue of how well the protocol will succeed in the case that the outputs of the measurements are not perfectly correlated considering the efficiency issue while asserting security, as well as the realistic practical implementation considering noise in the preparation of the states to section \ref{effreaimp}.

The distribution and test part proceeds as follows:

\noindent i-1) Without loosing generality, let us assume that $R_1$ prepares a large number, $M$, of tripartite systems in state $\hat\rho_a$ ({\em i.e.} with covariance matrix $\gamma(a)$ and displacement $d$) and sends one subsystem to $S$ and another to $R_0$.

\noindent i-2) $R_1$ wants to check if the distribution of states is faithfully achieved. To this aim she/he chooses randomly a set of indices for player $S$, $\{K_S\}$, and a disjoint set of indices for player $R_0$, $\{K_{R_0}\}$, and sends these two sets over secure classical channels to the corresponding players. Player $S$ sends his/her $K_S$ subsystems to player $R_0$. For each $m \in \{K_S\}$, $R_0$ measures the two subsystems in his/her possession and $R_1$ measures his subsystem. After communication of their results over secure classical channels, they agree on those indices
$\{\tilde K_S\} \subseteq \{K_S\}$ for which $|x_{R_0}|=|x_{R_1}|=x_0$. $R_1$ and $R_0$ check now whether the correlations predicted by the primitive occur: $\tilde p=\tilde{\mathcal{P}}(011)=\tilde{\mathcal{P}}(101)=\tilde{\mathcal{P}}(110)=\frac{1}{3}$, $\tilde \delta_i=0$. If the test was successful, {\em i.e.} if the measurement results were consistent with the assumption that the states have been distributed correctly, the players $i\in \{R_0, R_1\}$ set the flag $f_i=1$, otherwise $f_i=0$. In an analogous way, the test is performed for $S$.

\noindent i-3) Players $S$, $R_0$ and $R_1$ send their flags to each other. Every player who receives a flag ``$0$'', sets his flag also to ``$0$''. Every player with flag ``$0$'' aborts the protocol. Otherwise the execution of the protocol proceeds. This step terminates the distribution and test of the quantum systems.

In the second phase of the protocol a selection of the distributed systems is chosen to establish the bit sequences which will be used to implement the quantum primitive. In this phase, again honest
parties may abort the protocol if malicious manipulations occur.

\noindent ii-1) The players $S$, $R_0$ and $R_1$ agree upon a set of systems which have not been discarded during the distribution and test part.

\noindent ii-2) Player $S$ chooses (randomly) two disjoint sets of subsystems labeled by indices $L_S^i \subset \tilde M$. He/she sends the set $L_S^i$ to player $i$ and demands player $i$ to send via a noiseless quantum channel his/her subsystems $m \in L_S^i$ to him/her. In each case, the (random) choice $L_S^i$ is secret to party $j$, {\em i.e.} player $R_1$ has no information whatsoever about the set $L_S^{R_0}$. An analogous procedure is adopted by $R_0$ and $R_1$.

\noindent ii-3) After measuring their whole sequence of subsystems, player $i\in\{S,R_0,R_1\}$ announces publicly, the set of indices $\{\hat M^m_i\} \in \hat M_i$ for which the output of the quadrature measurement was $|x_0|$. The order in which the players announce their measurement results can be specified initially and based {\em e.g.} on a rotation principle. (Notice that the announcement of $\left|x_i\right|=x_0$ during the actual measurement phase would make possible an effective traitor
strategy, since he could manipulate combinations on a systematic basis).

\noindent ii-4) Without loss of generality, let us explicitly describe this step of protocol for player $S$. From the following sets $L_S^{R_0} \cap \hat M^m_{R_1}=:U^{R_1}_S$ and $L_S^{R_1} \cap \hat M^m_{R_0}=:U^{R_0}_S $ let $\tilde U^{R_i}_S \subseteq U^{R_i}_{S}$ for $i\in\{0,1\}$ be the index set for which the player $S$ measured $\pm x_0$ twice. Analogously to the first phase of the protocol, player $S$ can test if the outputs of his measures agree with the correlations of a proper primitive ($\tilde{\mathcal{P}}(00)=\tilde{\mathcal{P}}(10)=\tilde{\mathcal{P}}(01)=1/3$ and $\tilde{\mathcal{P}}(11)=0$) If the test is successful, the player sets his flag $f_S=1$, otherwise $f_S=0$. The same procedure is analogously performed by $R_0$ and $R_1$. From this step on, the players deal exclusively with the outputs of their measures, {\em i.e.} classical data and secure classical channels.

\noindent ii-5) $S$ checks correlation on his outputs in the following set $\hat M:=\hat M^m_S \cap \hat M^m_{R_0} \cap \hat M^m_{R_1} \subseteq \hat M^m_S$. If the test is successful, $S$ sets his flag $f_S=1$, otherwise $f_S=0$. $R_0$ and $R_1$ do the equivalent step.

\noindent ii-6): From a randomly chosen set $V^S \subset \hat M$ player $S$ demands from $R_0$ and $R_1$ their measurement results. $S$ tests this control sample for the assumed primitive. If
the test is successful, $S$ sets his flag $f_S=1$, otherwise $f_S=0$. $R_0$ ($R_1$) perform this step with a set $V^{R_0}\subset\hat M \backslash V^S$ ($V^{R_1} \subset \hat M \backslash (V^S \cup V^{R_0})$) respectively.

\noindent ii-7) Every player with a flag $0$ aborts the execution of the protocol. Otherwise the players agree upon a set $W:=\hat M \backslash(V^S \cup V^{R_0} \cup V^{R_1})$ as the result of the invocation of the primitive, which consists in an even number of elements. This step concludes the distribution part of the protocol.

\section{Primitive: Errors and manipulations}\label{errest}

We analyze here the effects of malicious manipulations of the data by dishonest parties and its detection by the honest ones. We examine here two possible sources of error, which can occur during the implementation of the primitive with Gaussian states. While the first source is inherent to the system, the second is caused by an active adversary intervention of a participating party. The inherent error is caused by the occurrence of non-consistent combinations on the invocation of the primitive, that is, the occurrence of outputs $(1,1,1)$, $(0,0,0)$, $(0,0,1)$, $(0,1,0)$ and $(1,0,0)$ has to be considered. This error will propagate along the protocol, so that the probability of finding a combination which is not appropriate
is bounded from above from $\eta=1-(3 \tilde p)^2$.

The second source of errors we want to discuss here corresponds to the local actions that one of the players could do in order to manipulate the measurement results of other players. For instance, let us assume that the player $R_0$ has malicious intentions and wants to shift the local component of the displacement vector of the distributed state using local transformations. Please note that a shift of the quadrature output $x_0$, provides the same error in the probability distribution of the outputs as a shift in the corresponding displacement vector. In other words, both types of manipulations produce the same change in the probabilities as calculated in Eq.~\eqref{overlap}. Parameterizing the shift in the displacement vector by the parameter $\lambda$, $d^T \mapsto {(d')^T} = -\frac{x_0}{3}(1,0,1,0,\lambda,0)$, it is interesting to see how that affects the conditional probabilities $\tilde p$ and $\tilde \delta_i$. If probabilities were changed, player $R_0$ could try to determine (via subsequent communication with the other players (step ii-3))), with certain probability, the occurrence of the outputs of the other players, thereby gaining additional information. Notice also that $S$ and $R_1$ cannot realize the local manipulation of $R_0$ without classical communication between them. This can be trivially seen by realizing that the partial trace $\tr_{R_0} (\hat \rho_a(\gamma, d'))=\int \mathcal{W}'_{\xi}dx_{R_0}dp_{R_0}= \mathcal{W}'_{\xi}(\gamma_{S,R_1},d_{S,R_1})=\tr_{R_0}(\hat \rho_a(\gamma,d))$ with

\begin{equation}
 \gamma_{S,R_1}=\left( \begin{array}{cccc}
 a & 0 & c & 0 \\
 0 & b & 0 & -c \\
 c & 0 & a & 0 \\
 0 & -c & 0 & b
 \end{array} \right)\quad {\rm and} \quad d_{S,R_1}=\left( \begin{array}{c}
 x_1 \\
 0 \\
 x_1 \\
 0
 \end{array} \right).
\end{equation}
Thus the most plausible strategy for a traitor could consist on the following: (i) discrediting honest players by manipulating the displacement vector in such a way that non-consistent combinations
appear, (ii) hide successful measurements to the honest players which result in combinations that might be disadvantageous. It is tedious but straightforward to show that by making use of the test
steps ii-4)-ii-6) honest parties can detect the effects of such manipulations.

\section{Efficient realistic implementation}\label{effreaimp}

In this section we focus on the efficiency of the proposed protocol and we extend our results to a realistic practical scenario, relaxing the conditions for the obtention of correlated outputs between the players and assuming noise in the preparation of the input states.

The protocol, as discussed in the previous sections, constitutes a nice proof-of-principle of the fact that detectable broadcast is solvable in the CV scenario using multipartite Gaussian entanglement. However, it suffers from its reliance on two main idealizations which render the practical implementation of the primitive unrealistic, or, better said, endowed with zero efficiency. Specifically, we have requested that (i) a {\em pure} tripartite symmetric Gaussian state is distributed as the entangled resource; and (ii) when the three parties measure (via homodyne detection) the position of their respective modes, their measurement is taken to be ideal, that is not affected by any uncertainty, and moreover all parties have to obtain, up to a sign, {\em the same} outcome $x_0$. In reality, assumption (i) is unjustified as inevitable imperfections and losses result instead in the production of mixed thermalized states; on the other hand, the probability associated to measurements under assumption (ii), and hence the probability of achieving broadcast, is vanishingly small \cite{Rodo2007OSYD}. It is interesting, in view of potential practical implementations of our scheme, to study here how its success is affected, and possibly guaranteed, by relaxing the above two assumptions.

To deal with (i), let us recall that the tripartite entangled Gaussian states of \eqref{egamma} can be produced in principle by letting three independently squeezed beams (one in momentum, and two
in position) interfere at a double beam-splitter, or ``tritter'' \cite{Braunstein1998N}, as proposed by van Loock and Braunstein \cite{vanLoock2000PRL}. In practice, the parametric non-linear process employed to squeeze the vacuum is affected by losses which result in the actual generation of squeezed {\em thermal} states in each single mode. Before the tritter, one then has three independent Gaussian modes with covariance matrices $\gamma_{1}^{\rm{in}}(s,n)={\rm diag}\{n s,\,n/s\}$, $\gamma_{2}^{\rm{in}}(s,n)=\gamma_{3}^{\rm{in}}(s,n)={\rm diag}\{n/s,\,n s\}$, respectively, where $s=\exp(2r)$ (with $r$ the squeezing degree in each single mode) and $n\geq 1$ is the noise parameter affecting each mode. The noise $n$ is related to the initial marginal purity $\mathcal{P}_k^{\rm{in}}$ of each single mode by $n=1/\mathcal{P}_k^{\rm{in}}$, and corresponds to a mean number of thermal photons given by $\bar n^{\rm{th}}=(n-1)/2$. For $n=1$, each mode is in the ideally pure squeezed vacuum state. After the tritter operation, described by the symplectic matrix \cite{vanLoock2000PRL,Braunstein1998N,Adesso2007NJP}
\begin{equation}
 S_{ttt}=\left(
 \begin{array}{llllll}
 \frac{1}{\sqrt{3}} & 0 & \sqrt{\frac{2}{3}} & 0 & 0 & 0 \\
 0 & \frac{1}{\sqrt{3}} & 0 & \sqrt{\frac{2}{3}} & 0 & 0 \\
 \frac{1}{\sqrt{3}} & 0 & -\frac{1}{\sqrt{6}} & 0 & \frac{1}{\sqrt{2}} & 0 \\
 0 & \frac{1}{\sqrt{3}} & 0 & -\frac{1}{\sqrt{6}} & 0 & \frac{1}{\sqrt{2}} \\
 \frac{1}{\sqrt{3}} & 0 & -\frac{1}{\sqrt{6}} & 0 & -\frac{1}{\sqrt{2}} & 0 \\
 0 & \frac{1}{\sqrt{3}} & 0 & -\frac{1}{\sqrt{6}} & 0 & -\frac{1}{\sqrt{2}}
 \end{array}
 \right)\,,\nonumber
\end{equation}
the output covariance matrix of the three modes is given precisely by

\begin{equation}\label{eoutmix}
 \gamma^{\rm{out}}=S_{ttt} \cdot [\gamma^{\rm{in}}_{1}(s,n)\oplus\gamma^{\rm{in}}_{2}(s,n)\oplus\gamma^{\rm{in}}_{3}(s,n)] \cdot S_{ttt}^T = n \gamma (a),
\end{equation}
where $\gamma(a)$ is defined in \eqref{egamma} and we have made the identification $a=(s^2+2)/(3s)$. Eq. \eqref{eoutmix} describes generally mixed, fully symmetric three-mode Gaussian states, with global purity given by $\mathcal{P}=n^{-3}$, thus reducing to the pure instance of \eqref{egamma} for $n=\mathcal{P}=1$. Recall that any additional losses due {\em e.g.} to an imperfect tritter and/or to the distribution and transmission of the three beams can be embedded into the initial single-mode noise factor $n$, so that \eqref{eoutmix} provides a realistic description of the states produced in experiments \cite{Yonezawa2004N,Aoki2003PRL}. We therefore consider this more general family of Gaussian states as resources to implement the CV version of the primitive. It may be interesting to recall that also for the general family of states of \eqref{eoutmix} the genuine tripartite entanglement is exactly computable \cite{Adesso2007NJP} in terms of the residual contangle \cite{Adesso2006NJP}, and as expected, it increases with $a$ and decreases with $n$.\\

Concerning the idealization (ii) discussed above, for an efficient implementation we should first of all consider the realistic case of non-ideal homodyne detections, which means that the outcomes are affected by uncertainties quantified by the parameter $\sigma$, see \eqref{overlap}. Furthermore, we should let the parties measure within a finite range, which means specifically that in each measurement run the expected values for the measurement outcomes of the receivers can be shifted of some quantity $\Delta$ with respect to the corresponding expectation value of the sender's homodyne detection. So, what we should ask for is that the parties (one sender $S$ and two receivers $R_0$ and $R_1$) agree on those outcomes of their measurement results for which $|\hat x_S| =x_0 >0$ and $|\hat x_{R_0}|, |\hat x_{R_1}| = (x_0 + \Delta) > 0$, once the sender has announced his/her output results $x_0$ to the other two parties.

Here, the actual results of each measurement are assumed to distribute according to a Gaussian function centered at $x_0$ (for the sender) and $x_0+\Delta$ (for the receivers), respectively, with a variance $\sigma$. Now the associated events occur with a finite probability. We wish to investigate for which range of the positive shift $\Delta \in [\Delta_{\min},\,\Delta_{\max}]$, being $\Delta_{\max}-\Delta_{\min}$ the range, the conditional probability $\tilde p$ still approaches $1/3$. The wider such range, the higher the probability of success, or efficiency, of the protocol; of course the regions in the space of parameters such that the primitive can be efficiently implemented will depend on the specific boundaries $\Delta_{\min}$ and $\Delta_{\max}$ and not only on their difference. Notice than in order to ensure that the sign of the quadratures does not change when allowing a finite range of valid output values (and therefore, correlations between classical bits are properly extracted (see section \ref{CVpri})), we always demand the shift $\Delta$ to be positive.

This condition can be further relaxed allowing for a non-symmetric range of permitted values around the broadcasted value $x_0$ but constraining only the absolute value of the shift as $|\Delta_{\min}| < x_0$, ensuring that the sign of the correlations remains unchanged. This extra condition, which could be straightforwardly included in our expressions if one aims at maximizing the efficiency of our protocol, does not modify the analysis of the realistic implementation of the protocol we perform in this section. For the sake of clarity, we remark once again the role of the parameter $\Delta$: allowing a non-zero shift $\Delta$ in the computation of the probability $\tilde p$, means considering the realistic case in which the outcomes of the measurements performed by each receiver can be not coincident with each other, and not coincident even with the corresponding broadcasted value of $x_0$. In what if follows we show that the protocol can be run successfully provided that the deviations $|x_{R_0}|-|x_S|$ and $|x_{R_1}|-|x_S|$ (in each run) both lie in between a $\Delta_{\min}$ and a $\Delta_{\max}$, which will be determined in the following.

To this aim, we can perform an analysis similar to the one of section \ref{CVpri}, but considering in full generality a non-unit noise factor $n$, a non-zero uncertainty $\sigma$, and a non-zero shift $\Delta$. The calculations of the probabilities follow straightforwardly along the lines of the special (unrealistic) case previously discussed, obtaining

\begin{eqnarray}\label{eptilde}
 \tilde{p}&=&\Bigg\{3+ 3 \e^{-\frac{4 \left(\Delta+x_0\right) \left[\Delta \left(4 \sigma^2+7 a-3 R\right)+\left(4 \sigma^2+3 a+R\right) x_0\right]}{9n \left(4 \sigma^4+9 a \sigma^2-R \sigma ^2+4\right)}} +\nonumber\\
 & & +\  \e^{\frac{4 x_0 \left[4 (a-R) \Delta +\left(4 \sigma^2+9 a-5 R\right) x_0\right]}{9n \left(4 \sigma^4+9 a \sigma^2-R \sigma^2+4\right)}} +\  \e^{-\frac{32 \left(\Delta +x_0\right) \left[\Delta \left(\sigma^2+a\right)+\left(\sigma^2+R\right) x_0\right]}{9n \left(4 \sigma^4+9 a \sigma^2-R \sigma^2+4\right)}}\,\Bigg\}^{-1},
\end{eqnarray}
with $R=\sqrt{9a^2-8}$. We find that in the parameter space of $a$ (regulating the entanglement), $n$ (regulating the mixedness), $x_0$ (regulating the measurement outcome), $\sigma$ (regulating the measurement uncertainty) and $\Delta$ (regulating the measurement shift), there exists a surface which acts as the boundary for the regime in which our primitive can be faithfully implemented, yielding a feasible, robust solution to the broadcast problem. This surface is obtained by requiring that $\tilde p = 1/3 - \epsilon$, with the deficit $\epsilon$ chosen arbitrarily small. The result is plotted in Fig.~\ref{parameters} for $\epsilon=10^{-7}$ in the three-dimensional space of parameters $x_0/\sqrt{n}$, $\Delta/\sqrt{n}$, and $a$, for different values of $\sigma$. 

\begin{figure}[h]
 \centering
 \psfrag{A}{$x_0/\sqrt{n}$}\psfrag{B}{$\Delta/\sqrt{n}$}\psfrag{C}{$a$}
a) \includegraphics[width=6cm]{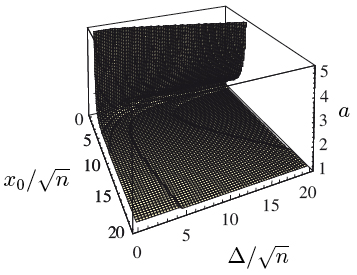}
 \psfrag{A}{$x_0/\sqrt{n}$}\psfrag{B}{$\Delta/\sqrt{n}$}\psfrag{C}{$a$} \hfill
b) \includegraphics[width=6cm]{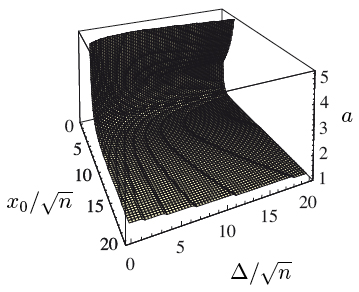} 
 \psfrag{A}{$x_0/\sqrt{n}$}\psfrag{B}{$\Delta/\sqrt{n}$}\psfrag{C}{$a$}\\
c) \includegraphics[width=6cm]{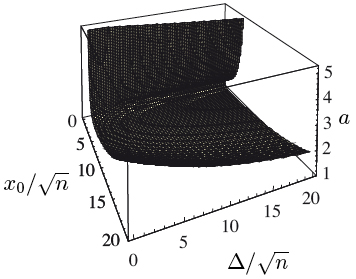}
 \caption{\small{Plots, in the space of $x_0/\sqrt{n}$ (quadrature measurement normalized to noise), $\Delta/\sqrt{n}$ (measurement shift normalized to noise), and $a$ (shared entanglement), of
 the boundary to the useful region for which $\tilde p = 1/3 - \epsilon$ ($\epsilon=10^{-7}$), at different values of the homodyne detection uncertainty (a) $\sigma=0$ (ideal error-free measurement), (b) $\sigma=1$ (fixed error, independent of the outcome), and (c) $\sigma=x_0/10$ (proportional error, corresponding to a 90\% efficiency in the detectors). Detectable broadcast can be solved efficiently, by means of our protocol, for all values of the parameters which lie above the depicted surface. See text for an extended discussion. All the quantities plotted are adimensional.}}\label{parameters}
\end{figure}

We consider the primitive as efficiently implementable in the whole region such that $1/3 - \epsilon \leq\tilde p \leq 1/3$, which spans the volume above the surface of Fig.~\ref{parameters}. Notice that for $\sigma=0$, $n=1$, and $\Delta_{\max} \rightarrow \Delta_{\min} \rightarrow 0$, such an useful volume shrinks to a two-dimensional slice (with $a \geq 5\sqrt2/6$) which represents the parameter range in which the ``ideal'' implementation described in section \ref{CVpri} is successful.

The figure offers several reading keys. Let us investigate independently how the possibility of solving detectable broadcast via our protocol depends on the individual parameters. For simplicity, we will consider $n$ fixed and eventually discuss its role. We will also for the moment keep the idealization of error-free homodyne measurements ($\sigma=0$), corresponding to Fig.~\ref{parameters}(a) which makes the subsequent discussion more tractable. However we will relax such an assumption in the end to show that a realistic description of the homodyne measurement does not significantly affect the performance and the applicability of the protocol. This validates the claims of efficiency that we make in the following.

\subsection{Entanglement role}

Let us henceforth start with the dependence of the solution on $a$. Somehow surprisingly, there exist lower {\em and upper} bounds on the tripartite entanglement such that only for $a_{\min} \leq a \leq
a_{\max}$ the protocol achieves a solution. The bounds naturally depend on $x_0$ and $\Delta$. Specifically, we observe that $a_{\min}$ diverges for $\Delta=0$ and $x_0 \rightarrow 0$, meaning that no feasible solution can be achieved in the low-$\Delta$, low-$x_0$ regime; the reason being that near $x_0=0$ there is not possibility of associating the classical bits "0" and "1" to positive/negative values of the quadratures. The lower bound then goes down reducing to the already devised threshold of $a_{\rm{thresh}} = 5\sqrt2/6$ for $\Delta$ close to zero and $x_0 \gg 0$, and eventually converges to $a=1$ ({\em i.e.} all entangled states are useful) for any finite $x_0 \gg 0$ and $\Delta \rightarrow \infty$. On the other hand, the upper bound on $a$, which surprisingly rules out states with too much entanglement, is obviously diverging at $\Delta=0$ but becomes finite and relevant in the regime of small $x_0$ and large $\Delta$, eventually reducing to $3/2$ for any finite $x_0$ and $\Delta \rightarrow \infty$.

Summarizing, the two extremal regimes corresponding to $\Delta=0$ on one hand and $\Delta \rightarrow \infty$ on the other hand, both allow a solution of detectable broadcast via our protocol: the main difference is that in the former case one needs states with an entanglement above $a_{\min}=5\sqrt2/6$, while in the latter case one needs states whose entanglement is below $a_{\max}=3/2$. The regime of finite shift $\Delta$ interpolates between these two limits. This means, in terms of useful range, and hence of efficient implementations, that if one is able to produce entangled states precisely with $5\sqrt2/6 < a < 3/2$, the protocol is implementable with high efficiency for {\em any} shift $\Delta \in [0,\infty)$ in the measurement outcomes. {\em i.e.} with {\em an infinite range of variability} allowed for the acceptable data resulting from homodyne detections performed by the receivers, for a given outcome $x_0$ of the sender. This information is important in view of practical implementations, and becomes especially valuable since the engineering of the required entangled resources appears feasible: a squeezing between $4,5$ and $6$ dB is required in each single mode, which is currently achieved in optical experiments \cite{Suzuki2006APL,Takeno2007OE,Vahlbruch2007NJP}.

\subsection{Measurement outcomes role}

For a fixed entangled resource, $a$, Fig.~\ref{parameters}(a) alternatively, shows that there exist minimum thresholds both for $x_0$ and $\Delta$ in order to achieve a solution to detectable broadcast. While the useful range for $x_0$ is always unbounded from above, we find that, interestingly, an upper bound $\Delta_{\max}$ exists for $a>3/2$. Precisely, for a given $a>3/2$, with increasing $x_0$ we observe that $\Delta_{\min}$ decreases and $\Delta_{\max}$ increases, {\em i.e.} the useful range spanned by $\Delta$ widens. Conversely, at a given $x_0$, $\Delta_{\max}$ decreases with increasing entanglement $a$, reducing the parameter space in which a solution can be found. Consistently with the previous discussion, we conclude again that in realistic conditions ($\Delta > 0$) it is better to have a moderate amount of shared entanglement to solve broadcast with optimal chances.

\subsection{Thermal noise role}

Now, let us note that the noise parameter $n$ simply induces a rescaling of both $x_0$ and $\Delta$, in such a way that with increasing $n$ the surface bounding the useful range of parameters shrinks, as it could be guessed (the noise degrades the performance of the protocol). Still, with typical noise factors characterizing experimental implementations, {\em e.g.} $n=2$, the protocol appears very robust and the solution is still achievable, for a given amount of entanglement (say $a \lesssim 3/2$), provided that $x_0$ and $\Delta$ exceed $\sqrt2$ times their respective minimum thresholds obtained for ideally pure resources ($n=1$).

\subsection{Homodyne efficiency role}

Finally, let us address the important issue of the uncertainty affecting homodyne detections. Any realistic measurement is characterized by a non-zero $\sigma$. We have explicitly studied two situations, one in which the absolute error is fixed, corresponding to a constant $\sigma$ (see Fig.~\ref{parameters}(b)), and another in which the relative error is fixed to 10\%, corresponding to a $\sigma$ proportional to the measurement outcome $x_0$ (see Fig.~\ref{parameters}(c)). The result is that, in both cases, for not exceedingly high values of the error factor, the useful region is obviously reduced but, crucially, the possibility of achieving detectable broadcast via our protocol is still guaranteed in a broad range of values of the parameters. Specifically, as somehow expected, the error model in which $\sigma$ is proportional to the measurement outcome results in a more consistent modification (shrinkage) of the useful surface, while almost nothing happens in the case of fixed, small $\sigma$. In particular, upper bounds on $x_0$ arise for a practical realization in presence of a proportional error, or in other words due to a limited efficiency in the detection. A significant portion of the parameter space anyway remains valid for a workable implementation of the primitive. Our scheme is thus robust also with respect to the imperfections in the quadrature measurements. We draw the conclusion that the protocol we designed is truly efficient and realistically implementable in non-ideal conditions.

We explicitly depicts $\tilde{p}$, \eqref{eptilde}, for sensible resource values of $a=3/2$ and $n=2$, as a function of $x_0$ and $\Delta$, and according to different values of $\sigma$ like in Fig.~\ref{parameters}.

\begin{figure}[h]
 \centering
 \psfrag{A}{$\Delta$}\psfrag{B}{$x_0$}\psfrag{C}{$\tilde p$}
a) \includegraphics[width=6cm]{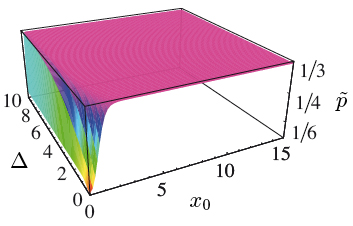}
 \psfrag{A}{$\Delta$}\psfrag{B}{$x_0$}\psfrag{C}{$\tilde p$} \hfill
b) \includegraphics[width=6cm]{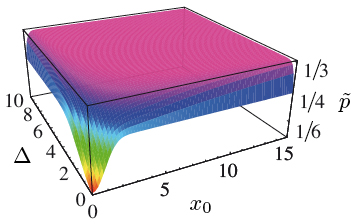} 
 \psfrag{A}{$\Delta$}\psfrag{B}{$x_0$}\psfrag{C}{$\tilde p$}\\
c) \includegraphics[width=6cm]{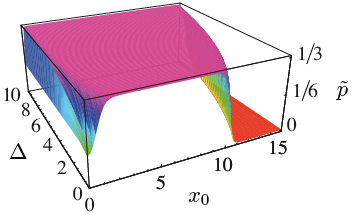}
 \caption{\small{Plots of the conditional probability $\tilde p$ as a function of the measurement outcome $x_0$ for the sender and the shift $\Delta$ of the measurement outcomes for the receivers, for realistically mixed Gaussian resource states with $n=2$ and $a=3/2$. The uncertainty in the homodyne detection is taken to be (a) $\sigma=0$, (b) $\sigma=1$, and (c) $\sigma=x_0/10$. Detectable broadcast is {\em efficiently} solvable in the wide plateau region where $\tilde p \rightarrow 1/3$. All the quantities plotted are adimensional.}}\label{preal}
\end{figure}

We notice a huge (unbounded from above in the ideal case $\sigma=0$) region (see Fig.~\ref{preal}) in which $\tilde p \rightarrow 1/3$, yielding an efficient solution to detectable broadcast via our protocol. We have thus identified an optimal ``work-point'' in terms of the parameters describing the shared Gaussian states, and we have defined the frontiers of application of our scheme in the laboratory practice. In this way, together with the explicit steps of the protocol presented in the previous sections, we obtain a clear-cut recipe for a CV demonstration of detectable broadcast, that we hope it can be experimentally implemented in the near future.

For the sake of completeness, let us mention that from \eqref{eptilde} we analytically find that $\tilde p$ converges {\em exactly} to $1/3$ only in the following limiting cases (for a given finite $n$ and $\sigma=0$): (i) $a \rightarrow \infty$ with $\Delta < 3x_0$; (ii) $x_0 \rightarrow \infty$ with $a\geq 5\sqrt2/6$; (iii) $\Delta \rightarrow \infty$ with $1<a\leq3/2$.

\section{Conclusions}

We have proposed in this chapter a protocol to solve detectable broadcast with entangled Continuous Variable using Gaussian states and Gaussian operations only. Our algorithm relies on genuine multipartite entanglement distributed among the three parties, which specifically have to share two copies of a three-mode fully symmetric Gaussian state. Interestingly, we have found that nevertheless not all entangled symmetric Gaussian states can be used to achieve a solution to detectable broadcast: a minimum threshold exists on the required amount of multipartite entanglement. We have moreover analyzed in detail the security of the protocol.

In its ideal formulation, our protocol requires that the parties share pure resource states, and that the outcomes of homodyne detections are perfectly coincident and not affected by any uncertainty; this however entails that our protocol achieves a solution with vanishing probability. To overcome such a
practical limitation, we have eventually considered a more realistic situation in which the tripartite Gaussian resources are affected by thermal noise, and, more importantly, the homodyne detections are realistically imperfect, and there is a finite range of allowed values for the measurement outcomes obtained by the parties. We have thoroughly investigated the possibility to solve detectable broadcast via our protocol under these relaxed conditions.

As a result, we have demonstrated that there exists a broad region in the space of the relevant parameters (noise, entanglement, range of the measurement shift, measurement uncertainty) in which the protocol admits an efficient solution. This region encompasses amounts of the required resources which appear attainable with the current optical technology (with a legitimate trade-off between squeezing and losses). We can thus conclude that a feasible, robust implementation of our protocol to solve detectable broadcast with entangled Gaussian states may be in reach. This would represent another important demonstration of the usefulness of genuine multipartite Continuous Variable entanglement for communication tasks, coming to join the recent achievement of a quantum teleportation network \cite{Yonezawa2004N}.


\chapter{Classical versus Quantum correlations in Continuous Variable}\label{nongaussian}

\section{Introduction}
In the previous chapters via a digitalization procedure we have been able to distill classical correlations from entangled Gaussian states in such a way that we have been able to implement protocols for either cryptographic or detectable broadcast purpose. In this chapter we want to relate the performance on extracting classical correlations from entangled states to the quantum correlations embedded in the states. While Gaussian states (coherent, squeezed, and thermal states) have been originally the preferred resources for both theoretical and practical implementations, a new frontier emerges with non-Gaussian states (Fock states, macroscopic superposition like Schr\"odinger's cat states, etc.). The latter can be highly non-classical, possess in general ``more entanglement'' than Gaussian states \cite{Wolf2006PRL}, and allow to overcome some limitations of the Gaussian framework like entanglement distillation \cite{Eisert2002PRL,Fiurasek2002PRL,Giedke2002PRA} and universal quantum computation \cite{Menicucci2006PRL}. Therefore, it is of central relevance to provide proper ways to quantify also non-Gaussian entanglement in an experimental approachable way.

At a fundamental level, the difficulty in the investigation of entanglement (quantum correlations) can be traced back to the subtle task of distinguishing it from classical correlations \cite{Groisman2005PRA}. Correlations can be regarded as classical if they can be induced onto the subsystems solely by local operations and classical communication, necessarily resulting in a mixed state. On the other hand, if a pure quantum state displays correlations between the subsystems they are of genuinely quantum nature (entanglement). We adopt here a pragmatic approach: if two systems are {\em in toto} correlated, then this correlation has to be retrieved between the outcomes of some local measurements performed on them. In this chapter, therefore, we investigate quadrature correlations in Continuous Variable (CV) states. We are also motivated by the experimental adequacy: field quadratures can be efficiently measured by homodyne detection, without the need for complete state tomography. Specifically, we study optimal correlations in bit strings obtained by digitalizing the outcomes of joint quadrature measurements on a bipartite two-mode CV system.

\section{Quadrature measurements and bit correlations.}

To study bipartite CV systems we use the well known fact that any NPPT Gaussian state of $N \times M$ modes can be mapped by Gaussian Local Operations and Classical Communication (GLOCC) to an NPPT symmetric state of $1\times1$ modes {\em i.e.} preserving the amount of entanglement. Then, for the most general scenario it is sufficient to consider a bipartite CV system of two bosonic modes, $A$ and $B$. The probability distribution associated to a single/joint measurement of arbitrary rotated quadratures $\hat x_A(\theta)$ and $\hat x_B(\varphi)$ providing outcomes $x_A^\theta$ and $x_A^\varphi$, is given by

\begin{eqnarray}
\mathcal{P}_A(x_A^\theta) &=& \tr[\hat \rho_A \hat U^{A\dag}_\theta \hat\sigma(x_A) \hat U^A_\theta],\\
\mathcal{P}_{AB}(x_A^\theta, x_B^\varphi) &=& \tr[\hat \rho_{AB} (\hat U^A_\theta \otimes \hat U^B_\varphi) (\hat \sigma(x_A) \otimes \hat \sigma(x_B)) (\hat U^{A\dag}_\theta \otimes \hat U^{B\dag}_\varphi)] = \nonumber \\
&=& \tr[(\hat U^{A\dag}_\theta \otimes \hat U^{B\dag}_\varphi) \hat \rho_{AB} (\hat U^A_\theta \otimes \hat U^B_\varphi) (\hat \sigma(x_A) \otimes \hat \sigma(x_B))] = \nonumber \\
&=& \tr[\rho_{AB} (\hat \sigma(x_A^\theta) \otimes \hat \sigma(x_B^\varphi))]= \tr[\hat\rho_{AB}^{\theta,\varphi} (\hat \sigma(x_A) \otimes \hat \sigma(x_B))],
\end{eqnarray}
where $\hat \sigma(x_A)$ is a single-mode Gaussian (squeezed) state with first moments $\{x_A,0 \}$ and covariance matrix ${\rm diag}\{\sigma^2,1/\sigma^2\}$. Here $\hat U^A_\theta$ is a unitary operator describing a rotation of $\theta$ on mode $A$, corresponding to a symplectic transformation given by $S_{\theta,A}=\begin{pmatrix}
 \cos{\theta} & \sin{\theta}\\
 -\sin{\theta} & \cos{\theta}
\end{pmatrix}$. Likewise for $B$. Thus, one can either measure the rotated quadrature ($x_A^\theta$, $x_B^\varphi$) on the state (passive view) or antirotate the state ($\hat\rho_{AB}^{\theta,\varphi}$) and measure the unrotated quadrature (active view).

To digitalize the obtained output quadrature measurements that spread in a continuum we assign the bits $+\ (-)$ to the positive (negative) values of the measured quadrature. This digitalization transforms each joint quadrature measurement into a pair of classical bits. A string of such correlated bits can be used {\em e.g.} to distill a quantum key \cite{Navascues2005PRL,Rodo2007OSYD}.
Let us adopt a compact notation by denoting (at given angles $\theta,\varphi$) $\mathcal{P}^\pm_A \equiv \mathcal{P}_A(\pm|x_A^\theta|)$, and $\mathcal{P}^{\pm \mp}_{AB} \equiv \mathcal{P}_{AB}(\pm|x_A^\theta|,\mp|x_B^\varphi|)$. The conditional probability that the bits of the corresponding two modes coincide is given by $\mathcal{P}^=_{AB} \equiv (\mathcal{P}^{++}_{AB} + \mathcal{P}^{--}_{AB}) / \sum_{\{\alpha=\pm,\beta=\pm\}}\mathcal{P}^{\alpha\beta}_{AB}$. Correspondingly, the probability that they differ is $\mathcal{P}^{\neq}_{AB} \equiv (\mathcal{P}^{+-}_{AB}+\mathcal{P}^{-+}_{AB})/\sum_{\{\alpha=\pm,\beta=\pm\}}\mathcal{P}^{\alpha\beta}_{AB}$. Trivially, $\mathcal{P}^=_{AB} + \mathcal{P}^{\neq}_{AB} = 1$. If $\mathcal{P}^=_{AB} > \mathcal{P}^{\neq}_{AB}$ the measurement outcomes display correlations, otherwise they display anticorrelations. Notice that, if the two modes are completely uncorrelated, $\mathcal{P}^=_{AB} = \mathcal{P}^{\neq}_{AB} = 1/2$. For convenience, we normalize the {\em degree} of bit correlations as

\begin{equation}\label{degree}
 \mathcal{B}(|x_A^\theta|,|x_B^\varphi|) = 2 |\mathcal{P}^=_{AB} - 1/2| = |\mathcal{P}^=_{AB} - \mathcal{P}^{\neq}_{AB}|,
\end{equation}
so that for a completely uncorrelated state $\mathcal{B}(|x_A^\theta|,|x_B^\varphi|)=0$. The interpretation of \eqref{degree} in terms of correlations is meaningful if a {\em fairness} condition is satisfied: locally, on each single mode, the marginal probabilities associated to the outcomes ``$+$'' or ``$-$'' must be the same: $\mathcal{P}^+_A = \mathcal{P}^-_A,\quad \mathcal{P}^+_B = \mathcal{P}^-_B$ otherwise, \eqref{degree} can systematically display false correlations due to individual unbiasedness.

For a two-mode CV system, whose state is described by a Wigner function $\mathcal{W}$, we define the {\em ``bit quadrature correlations''} $Q$ as the maximal average probability of obtaining a pair of classically correlated bit (in the limit of zero uncertainty) optimized over all possible choices of local quadratures

\begin{equation}\label{bitcorr}
 Q(\hat\rho)=\sup_{\theta,\varphi} {\int \int d x_A^\theta d x_B^\varphi \bar{\mathcal{W}}(x_A^\theta, x_B^\varphi) [\lim_{\sigma \rightarrow 0}\mathcal{B}^\sigma(|x_A^\theta|,|x_B^\varphi|)]},
\end{equation}
where $\bar{\mathcal{W}}(x_A^{\theta}, x_B^\varphi) = \int \int d p_A^\theta d p_B^\varphi \mathcal{W}(x_A^\theta, p_A^\theta, x_B^\varphi, p_B^\varphi)$ is the marginal Wigner distribution of the (rotated) positions, and $\{x_A^\theta, p_A^\theta, x_B^\varphi, p_B^\varphi\} = [S_\theta \oplus S_\varphi] \{x_A, p_A, x_B, p_B\}$. We demonstrate numerically (for all clases of Gaussian and non-Gaussian states discussed) that we can write Eq.~\eqref{bitcorr} as

\begin{equation}\label{bitcorr2}
Q(\hat\rho)=\sup_{\theta,\varphi}|E^{\theta,\varphi}_{A,B}(\hat\rho)| = \sup_{\theta,\varphi}|\int\int d x_A^\theta d x_B^\varphi \bar{\mathcal{W}}(x_A^{\theta}, x_B^{\varphi}) {\rm sgn}(x_A^{\theta} x_B^{\varphi})|,
\end{equation}
where $E^{\theta,\varphi}_{A,B}(\hat\rho)=\int\int d x_A^\theta d x_B^\varphi \bar{\mathcal{W}}(x_A^{\theta}, x_B^{\varphi}) {\rm sgn}(x_A^{\theta} x_B^{\varphi})$ is the ``sign-binned'' quadrature correlation function, which has been employed, {\em e.g.}, in proposed tests of Bell inequalities violation for CV systems.

A homodyne Bell test requires measuring two different rotated quadratures per mode, to achieve violation of the bound \cite{Munro1999PRA,Nha2004PRL,GarciaPatron2004PRL,Garcia-Patron2005PRA} $|E^{\theta,\varphi}_{A,B} + E^{\theta',\varphi}_{A,B} + E^{\theta,\varphi'}_{A,B} - E^{\theta',\varphi'}_{A,B}| \leq 2$. Here, we propose to measure a single quadrature per mode, which displays one-shot optimal correlations, unveiling a powerful quantitative connection with Gaussian and non-Gaussian entanglement measured through negativity. While this form is more suitable for an analytic evaluation on specific examples, the definition \eqref{bitcorr} is useful to prove the general properties of $Q$~\footnote{Fair states have necessarily zero first moments, which can be assumed without loss of generality. In general, one could take in \eqref{bitcorr} the difference between $\mathcal{B}$ computed on $\hat \rho$ and $\mathcal{B}$ computed on $\hat\rho_A \otimes \hat\rho_B$. The latter is zero on fair states.} that we analyze here.

\begin{lemma}
{\em Normalization:}\label{lemnorm}
$0\leq Q(\hat\rho) \leq 1$.

{\em Proof.} It follows from the definition of $Q(\hat\rho)$, as both $\mathcal{B}$ and the marginal Wigner distribution range between $0$ and $1$. \hfill $\Box$
\end{lemma}

\begin{lemma}
{\em Zero on product states:}\label{lemprod}
$Q(\hat\rho_A \otimes \hat\rho_B)=0$.

{\em Proof.} For a product state the probabilities factorize {\em i.e.} $\mathcal{P}_{AB}^{\alpha \beta} = \mathcal{P}_A^\alpha \mathcal{P}_B^\beta$ and so $\mathcal{P}_{AB}^= = \mathcal{P}_{AB}^{\neq}$, where we have used the fairness condition. Namely $\mathcal{B} = 0$, hence the integral in \eqref{bitcorr} trivially vanishes. \hfill $\Box$
\end{lemma}

\begin{lemma}
{\em Local symplectic invariance:}\label{lemloc}
Let $\hat U_{A,B}$ be a unitary operator amounting to a single-mode symplectic operation $S_{A,B}$ on the local phase-space of mode $A,B$. Then $Q[(\hat U_A \otimes \hat U_B) \hat\rho (\hat U_A^\dag \otimes \hat U_B^\dag)] = Q(\hat\rho)$.

{\em Proof.} Any single-mode symplectic operation $S$ can be decomposed in terms of local rotations and local squeezings (Euler decomposition). By definition \eqref{bitcorr} is invariant under local rotations, so we need to show that local squeezing transformations, described by symplectic matrices of the form $S_r={\rm diag}\{1/r,r\}$, also leave $Q(\hat\rho)$ invariant, {\em i.e.} $Q[(\hat U^A_s \otimes \hat U^B_{t}) \rho (\hat U^{A\dag}_s \otimes \hat U^{B\dag}_{t})]=Q(\hat\rho)$. Adopting the passive view, the action of local squeezings on the covariance matrix of each Gaussian state $\hat \sigma(x_{A,B})$ is irrelevant, as we take eventually the limit $\sigma \rightarrow 0$. The first moments are transformed as $d_{A,B} \mapsto S^{-1}_{s,t} d_{A,B}$, so that $\mathcal{B}_{AB}^{\sigma\rightarrow 0}(|x_{A}|,|x_{B}|) \mapsto \mathcal{B}_{AB}^{\sigma\rightarrow 0}(|s x_{A}|,|t x_{B}|)$. On the other hand, the Wigner distribution is transformed as $\mathcal{W}(\xi) \mapsto \mathcal{W}[(S^{-1}_s \oplus S^{-1}_t) \xi)]$. Summing up, local squeezings transform $\xi=\{x_A,p_A,x_B,p_B\}$ into $\xi_{s,t}=\{s x_A, p_A/s, t x_B, p_B/t\}$. As \eqref{bitcorr} involves integration over the four phase-space variables $d^4\xi$, we change the variables noting that $d^4 \xi = d^4 \xi_{s,t}$, to conclude the proof. \hfill $\Box$
\end{lemma}

It follows from lemmas~\ref{lemnorm} and \ref{lemprod} that if $Q > 0$, then the state necessarily possesses correlations between the two modes. lemma~\ref{lemloc}, moreover, suggests that $Q$ embodies not only a qualitative criterion, but might be interpreted as {\em bona fide} operational quantifier of CV correlations. We will now show that this is the case for various important classes of states.

First, we apply our procedure to Gaussian states (GS), finding that bit quadrature correlations provide a clear-cut quantification of the total correlations between the two modes. They are monotonic with the entanglement on pure states, and can be arbitrarily large on mixed states, the latter possibly containing arbitrarily strong additional classical correlations. We then address non-Gaussian states (NGS), for which the exact detection of entanglement generally involves measurements of high-order moments \cite{Shchukin2005PRL}.
The underlying idea is that for NGS obtained by de-gaussifying an initial pure GS and/or by mixing it with a totally uncorrelated state, our measure based entirely on second moment is still expected to be a (quantitative) witness of the quantum part of correlations only, {\em i.e.} entanglement. We show that this is indeed the case for relevant NGS including photonic Bell states, photon-subtracted states, and mixtures of Gaussian states. Notably, the complete entanglement picture in a recently demonstrated coherent single-photon-subtracted state \cite{Ourjoumtsev2007PRL} via a de-gaussification procedure is precisely reproduced here in terms of quadrature correlations only. Our results render non-Gaussian entanglement significantly more accessible in a direct, practical fashion.

\section{Gaussian states}

Even though entanglement of GS is already efficiently accessible via their covariance matrix, we use such states as ``test-beds'' for understanding the role of $Q$ in discriminating CV correlations. The covariance matrix $\gamma$ of any two-mode GS $\hat \rho$ can be written in standard form as

\begin{equation*}
{\gamma}=\left(\begin{array}{cc}
A&C\\
C^T&B \end{array} \right), \quad \begin{array}{c}
A=\lambda_a \id_2,\,\,B=\lambda_b \id_2, \\
C={\rm diag}\{c_x,-c_p\}, \end{array}
\end{equation*}
where, without loss of generality, again we flip the sign of $c_p$ and adopt the convention $c_x \geq |c_p| >0$. We fix also the displacement vector to zero, this fact together with the symmetries of GS is sufficient to fulfill the fairness condition. The covariance matrix $\gamma$ describes a physical state if, in terms of the four invariants (2 local purities, global purity and serelian), the following relations holds $\mathcal{P}_A, \mathcal{P}_B \geq 1$, $\mathcal{P} \geq 1$ and $1+\frac{1}{{\mathcal{P}}^2} \geq \Delta$. In standard form these relations transforms, in terms of the parameters as $\lambda_a, \lambda_b \geq 1$, $(\lambda_a \lambda_b-c_x^2)(\lambda_a \lambda_b-c_p^2)\geq1$ and $1+(\lambda_a \lambda_b-c_x^2)(\lambda_a \lambda_b-c_p^2)\geq\lambda_a^2+\lambda_b^2+2c_xc_p$ (see details in lemma~\ref{sform}). The negativity see \eqref{nega}, quantifying entanglement between the two modes, reads $N(\hat \rho)=\frac{1-\tilde\mu_-}{2\tilde\mu_-}$ because $\tilde\mu_+ > 1 >\tilde\mu_-$ for all two mode entangled GS, where $\tilde\mu_{\pm}^2 = [\tilde\Delta\pm(\tilde\Delta^2 - 4/\mathcal{P}^2)^{1/2}]/2$. For two-mode GS, \eqref{bitcorr} evaluates to

\begin{eqnarray}\label{qpure}
 Q(\hat \rho_\gamma) &=& (2/\pi) \arctan \left(c_x/\sqrt{\lambda_a \lambda_b -c_x^2} \right) =\nonumber\\
 &=& (2/\pi) \arctan \left([(\frac{4}{\mathcal{P}_A\mathcal{P}_B[a_- +a_+]})^2-1]^{-\frac{1}{2}}\right),
\end{eqnarray}
where the optimal quadratures are the standard unrotated positions ($\theta=\varphi=0$) and $a_{\pm} =\sqrt{\left[ \Delta-(\mathcal{P}_A\pm\mathcal{P}_B)^2/(\mathcal{P}_A\mathcal{P}_B)^2 \right]^2 - 4/\mathcal{P}^2}$. First, we notice that $Q = 0 \Leftrightarrow \hat\rho$ describes a product state: for GS, $Q > 0$ is then {\em necessary and sufficient} for the presence of correlations. Second, we observe that for pure GS, reducible up to local unitary operations, to the two-mode squeezed states $\hat \rho_r = \ket{\phi_r}\!\bra{\phi_r}$ characterized by $\lambda_a=\lambda_b=\cosh(2r)$ and $c_x=c_p=\sinh(2r)$, Eq.~\eqref{qpure} yields a monotonic function of the negativity. We compute explicitly $Q$ and expressed in terms of $\tilde\mu_-$ for comparison
\begin{eqnarray}
 Q(\hat \rho_r) &=& (2/\pi) \arctan (\sinh{2r}) =\nonumber\\
 &=& (2/\pi) \arctan \left(\frac{1-\tilde\mu_-^2}{2 \tilde\mu_-} \right),
\end{eqnarray}
see also Fig.~\ref{QGS}. $Q$ is thus, as expected, an operational entanglement measure for pure two-mode GS. Third, we find that for mixed states $Q$ {\em majorizes} entanglement. Given a mixed GS $\hat \rho_N$ with negativity $N$, it is straightforward to see that $Q(\hat \rho_N)$ via \eqref{qpure}, is always greater than $Q(\ketbra{\psi_N}{\psi_N})$, with $\ket{\psi_N}$ being a pure two-mode squeezed state with the same negativity $N$.

\begin{figure}[h]
 \centering
 \psfrag{A}{\hspace{-10mm}$2N/(2N+1)$}\psfrag{B}[][][1][-90]{$Q$}
 \includegraphics[width=6cm]{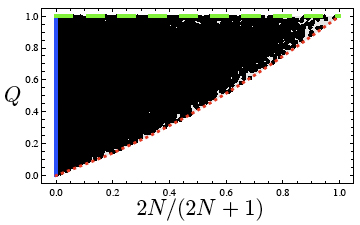}
 \caption{\small{Bit quadrature correlations vs normalized negativity $N$ for 18000 random two-mode Gaussian states. The lowermost dotted (red) curve accommodates pure states. The leftmost solid (blue) vertical line denotes separable states parametrized by $c_p=0$, $c_x=\epsilon (\delta^2-1)/\delta$, with $0 \leq \epsilon \leq 1$ and $\delta \rightarrow \infty$. The uppermost dashed (green) horizontal line denotes perfectly correlated states with an arbitrary degree of entanglement, parametrized by $c_x=(\delta^2-1)/\delta$, $c_p=\epsilon c_x$ (and $\epsilon$, $\delta$ as before). Product states (totally uncorrelated) lie at the origin.}}\label{QGS}
\end{figure}

Hence $Q$ quantifies {\em total} correlations, and the difference $Q(\hat\rho_N)-Q(\ketbra{\psi_N}{\psi_N})$ (where the first term is due to total correlations and the second to quantum ones) can be naturally regarded as an operational measure of {\em classical} correlations~\footnote{The emerging measure of classical correlations is special to Gaussian states, where the correlated degrees of freedom are the field quadratures. Different approaches to the quantification of classical vs quantum correlations were proposed \cite{Groisman2005PRA}.}. We have evaluated $Q$ on random two-mode GS as a function of their entanglement, conveniently scaled to $2N/(1+2N)$, as shown in Fig.~\ref{QGS}. Note that for any entanglement content there exist maximally correlated GS with $Q=1$, and also that separable mixed GS can achieve an arbitrary $Q$ from 0 to 1, their correlations being only classical.

\section{Non-Gaussian states}

Let us now turn our attention towards the even more interesting arena represented by NGS. The description of non-Gaussian states requires an infinite set of statistical moments, so does the entanglement. The best approach to the separability problem of arbitrary bipartite CV states was introduced by Shchukin and Vogel \cite{Shchukin2005PRL}. They provide necessary and sufficient condition for the negativity of the partial transposition {\em i.e.} entanglement through an infinite series of inequalities based on determinants of successively increasing size matrices containing high order moments of the state.
Our aim here is to compute $Q$ on such state. Bit quadrature correlations either in form Eq.~\eqref{bitcorr} or Eq.~\eqref{bitcorr2} can be computed for arbitrary Gaussian and non-Gaussian states at the level of Wigner functions. To this aim one needs to have a way to express arbitrary density operators corresponding to non-Gaussian states into non-Gaussian Wigner functions in a systematical way. Notice that any pure non-Gaussian state can be written as

\begin{equation}
\ket{\psi}=\sum_{n,m=0}^\infty\psi_{n,m}\ket{n,m}=\sum_{n,m=0}^\infty\psi_{n.m}\frac{(\hat a_1^\dag)^n(\hat a_2^\dag)^m}{\sqrt{n!m!}}\ket{0,0}=f_\psi(\hat a_1^\dag,\hat a_2^\dag)\ket{0,0},
\end{equation}
that corresponds to a density operator $\hat\rho_\psi=f_\psi(\hat a_1^\dag,\hat a_2^\dag)\hat\rho_0f^*_\psi(\hat a_1,\hat a_2)$, where $\hat\rho_0=\ketbra{0,0}{0,0}$ is the bipartite vacuum state.
Then it is easy, by using the map in Eq.~\eqref{map}, to find the Wigner function, $\mathcal{W}_\psi$ corresponding to $\hat\rho_\psi$ in terms of the Wigner function of a vacuum state, $\mathcal{W}_0(\zeta)=\frac{1}{\pi^2}\exp(-\zeta^T \cdot \zeta)$.

The action of an operator on a density operator $\hat\rho$ is always mirrored by the action of a corresponding differential operator on the corresponding phase-space probability distribution function. We summarize here that result

\begin{eqnarray}\label{map}
\hat a \hat\rho \longleftrightarrow (\alpha + \frac{1}{2}\frac{\partial}{\partial \alpha^*})\mathcal{W_\rho}(\alpha, \alpha^*)\nonumber \\
\hat\rho \hat a \longleftrightarrow (\alpha - \frac{1}{2}\frac{\partial}{\partial \alpha^*})\mathcal{W_\rho}(\alpha, \alpha^*)\\
\hat a^\dag \hat\rho \longleftrightarrow (\alpha^* - \frac{1}{2} \frac{\partial}{\partial \alpha})\mathcal{W_\rho}(\alpha, \alpha^*)\nonumber \\
\hat\rho \hat a^\dag \longleftrightarrow (\alpha^* + \frac{1}{2} \frac{\partial}{\partial \alpha})\mathcal{W_\rho}(\alpha, \alpha^*),\nonumber
\end{eqnarray}
similar results holds in the position-momentum base,
\begin{eqnarray}
\hat x \hat\rho \longleftrightarrow (x + \frac{\im \hbar}{2}\frac{\partial}{\partial p})\mathcal{W}_\rho(x, p)\nonumber \\
\hat\rho \hat x \longleftrightarrow (x - \frac{\im \hbar}{2}\frac{\partial}{\partial p})\mathcal{W}_\rho(x, p)\\
\hat p \hat\rho \longleftrightarrow (p - \frac{\im \hbar}{2}\frac{\partial}{\partial x})\mathcal{W}_\rho(x, p)\nonumber \\
\hat\rho \hat p \longleftrightarrow (p + \frac{\im \hbar}{2}\frac{\partial}{\partial x})\mathcal{W}_\rho(x, p). \nonumber
\end{eqnarray}

We focus on the most relevant NGS recently discussed in the literature and lastly experimentally realized. Remarkably, we find for all of them a {\em monotonic} functional dependence between the entanglement (negativity) and the quadrature correlations $Q$ measure of \eqref{bitcorr}. This fact indicates that in the preparation of those states, classical correlations are never induced. From a more practical perspective, this observation makes their entanglement amenable to a direct measurement in terms of quadrature correlations.

\subsection{Photonic Bell states}

We consider Bell-like states of the form $\ket{\Phi^{\pm}}=\sqrt{p}\ket{00}\pm \sqrt{1-p}\ket{11}$ and $\ket {\Psi^{\pm}}=\sqrt{p}\ket{01}\pm\sqrt{1-p}\ket{10}$ (with $0\leq p\le 1$), which are non-trivial examples of superpositions of Fock states, entangled with respect to the (discrete) photon number. Notice in particular how $\Psi^{+}$ for $p=1/2$ can be obtained by photon addition to the vacuum: $\ket{\Psi^{+}_{p=1/2}}=\frac{1}{\sqrt2}(\hat a_A^\dag+\hat a_B^\dag)\ket{00}$.
The negativity of these Bell-like states reads $N_B=\sqrt{p(1-p)}$. For the four of them (and at optimal angles $\theta=\varphi=0$), \eqref{bitcorr} reads

\begin{equation}
 Q_B=(4/\pi) N_B,
\end{equation}
showing a perfect agreement between the quadrature CV correlations and the entanglement.

\subsection{Photon subtracted states}

Most attention is being drawn by those CV states obtained from Gaussian states via subtracting photons \cite{Ourjoumtsev2007PRL,Neergaard-Nielsen2006PRL,Wakui2007OE}: they perform better as resources for protocols like teleportation \cite{Kitagawa2006PRA,DellAnno2007PRA} and allow for loophole-free tests of non-locality \cite{Munro1999PRA,Nha2004PRL,GarciaPatron2004PRL,Garcia-Patron2005PRA}. Let us recall their preparation, following \cite{Kitagawa2006PRA}. The beam $A$ ($B$) of a two-mode squeezed state $\ket{\phi_r}$ (created using an squeezed/antisqueezed state and a balanced beam splitter) is let to interfere, via a beam splitter of transmitivity $T$, with a vacuum mode, see Fig.~\ref{Qphsu}. The output is a four-mode Gaussian state of modes $C A' D B'$. Detection of one photon in each of the two beams $C$ and $D$, conditionally projects the state of modes $A' B'$ into a pure NGS, given in the Fock basis by $\ket{\psi_{ps}^{(T,r)}} = \sum^{\infty}_{n=0} c_n \ket{n,n}$, where $c_n (T, \Lambda) = (n+1) (T \Lambda )^n \left(1-T^2 \Lambda^2\right)^{3/2}/\sqrt{1+T^2 \Lambda ^2}$, and $\Lambda = \tanh r$.

\begin{figure}[h]
 \centering
 \psfrag{A}{$A$}\psfrag{B}{$B$}\psfrag{C}{$C$}\psfrag{D}{$D$}\psfrag{E}{$PD$}\psfrag{F}{$\ket{0}$}\psfrag{G}{$\ket{r}$}\psfrag{H}{$\ket{-r}$}\psfrag{I}{$BS(T)$}\psfrag{J}{$BS(50:50)$}\psfrag{K}{$TMS(r)$}\psfrag{L}{$A'$}\psfrag{M}{$B'$}
 \includegraphics[width=8cm]{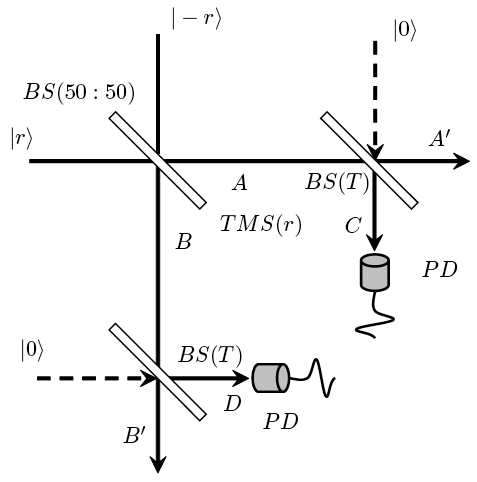}
 \caption{\small{Measurement-induced non-Gaussian operation on the two-mode squeezed vacuum state. First, a squeezed and antisqueezed states are let to interfere in a balanced beam splitter, $BS(50:50)$ to produce a two-mode squeezed state $\ket{\phi_r}$. Second, they are both mixed with the vacuum in a beam splitter with transmitivity $T$ denoted as $BS(T)$. Finally, modes $C$ and $D$ are measured with a photodetector, $PS$.}}\label{Qphsu}
\end{figure}

The limiting case $T\rightarrow 1$, occurring with asymptotically vanishing probability, corresponds to an ideal two-photon subtraction, $\ket{\psi_{ps}^{(1,r)}} \propto (\hat a _A \hat a _B)\ket{\phi_r}$. For any $T$, the negativity is computable as $N(\psi_{ps}^{(T,r)})=2/(1 - T \Lambda) - 1/(1 + T^2 \Lambda^2) - 1$. It increases with $r$ and with $T$ but only in the case $T=1$ it exceeds $N(\phi_r)$ for any $r$, diverging for $r \rightarrow \infty$. For all $T<1$, the entanglement of $\ket{\psi_{ps}^{(T,r)}}$ eventually saturates, and above a squeezing threshold $r_T \gg 1$, the original GS is more entangled than the resulting non-Gaussian one. Following \cite{Munro1999PRA,Nha2004PRL,GarciaPatron2004PRL,Garcia-Patron2005PRA}, the explicit expression of the quadrature bit correlations \eqref{bitcorr} can be analytically obtained

\begin{equation*}
Q(\psi_{ps}^{(T,r)})= \sum_{n>m=0}^{\infty}{\frac{2^{m+n+3} \pi [\mathcal{F}(m,n)-\mathcal{F}(n,m)]^2 c_m c_n}{(m-n)^2 m! n!}},
\end{equation*}
where the $c_k$'s are defined above, $\mathcal{F}(m,n) = \left[\Gamma \left(-\frac{m}{2}\right) \Gamma \left(\frac{1-n}{2}\right)\right]^{-1}$, and $\Gamma$ is the gamma function.
Again optimal quadrature measurements are achieved at $\theta=\varphi=0$. As depicted in Fig.~\ref{QPSS}, the behavior of the entanglement is fully reproduced by $Q(\psi_{ps}^{(T,r)})$ which again is a monotonic function of $N$ {\em i.e.} \eqref{bitcorr} is thus measuring truly quantum correlations of this important class of NGS as well.

\begin{figure}[h]
 \centering
 \psfrag{A}{$r$}\psfrag{B}{\hspace{-5mm}$T$}\psfrag{C}{$Q$}
 (a) \includegraphics[width=6cm]{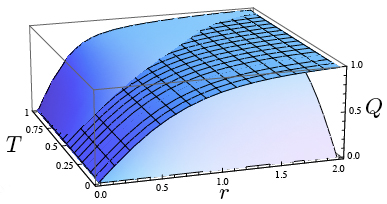} \hfill
 \psfrag{A}{\hspace{-10mm}$2N/(2N+1)$}\psfrag{B}{$Q$}
 (b) \includegraphics[width=6cm]{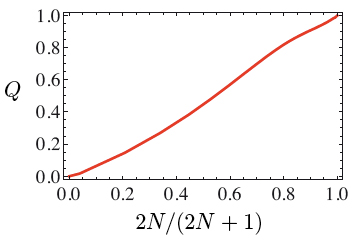}
 \caption{\small{(a) Bit quadrature correlations $Q$ vs squeezing $r$ and beam splitter transmitivity $T$ for photon-subtracted states (shaded surface) and for two-mode squeezed states (wireframe surface). (b) $Q$ vs normalized negativity for photon-subtracted states.}}\label{QPSS}
\end{figure}

\subsection{Experimental de-gaussified states}

Up to now we considered nearly-ideal non-Gaussian situations. We apply now our definition to the coherent photon-subtracted state $\hat \rho_{exp}$ recently studied and demonstrated by Ourjoumtsev {\em et al} \cite{Ourjoumtsev2007PRL}. Following their paper for details on the state preparation, we recall that each of the two beams of the two-mode squeezed state $\ket{\phi_r}$ was let to interfere with the vacuum at a beam splitter with reflectivity $R \ll 1$, and by using an avalanche photodiode a single photon was subtracted from the state in a de-localized fashion. The realistic description of the obtained highly mixed state involves many parameters. We fix all to the values obtained from the theoretical calculations and/or experimental data of \cite{Ourjoumtsev2007PRL}, but for the reflectivity $R$ and the initial squeezing $r$ which are kept free. We then evaluate \eqref{bitcorr} as a function of $r$ for different values of $R$. Unlike the previous cases, optimal correlations in the state $\hat \rho_{exp}$ occur between momentum operators ($\theta=\varphi=\pi/2$).

Also for this realistic mixed case, the correlation measure $Q$ reproduces precisely the behavior of the negativity, as obtained in \cite{Ourjoumtsev2007PRL} after full Wigner tomography of the produced state $\hat \rho_{exp}$. In particular, the negativity (and $Q$) increases with the squeezing $r$, and decreases with $R$. Below a threshold squeezing which ranges around $\sim 3$ dB, the NGS exhibits more entanglement (larger $Q$) than the original two-mode squeezed state. Our results depicted in Fig.~\ref{QdGS} compare extremely well to the experimental results (Fig.~6 of \cite{Ourjoumtsev2007PRL}) where the negativity is plotted there as a function of $r$ for different $R$'s.
Thus, the results give an indication of the intimate relation between $Q$ and the negativity of non-Gaussian states beyond idealizations.

\begin{figure}[h]
 \centering
 \psfrag{A}{$r$[dB]}
 \psfrag{B}[][][1][-90]{$Q$}
 \includegraphics[width=8cm]{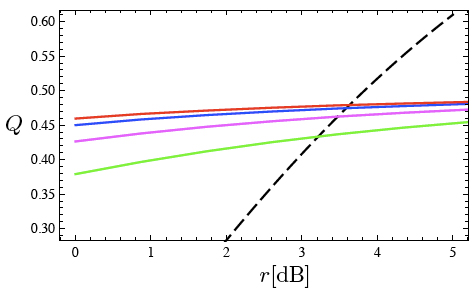}
 \caption{\small{Bit quadrature correlations $Q$ vs squeezing $r$ (in decibels) for the de-gaussified states demonstrated in \cite{Ourjoumtsev2007PRL} with beam splitter reflectivity $R$ equal to (from top to bottom); 3\% (red), 5\% (blue), 10\% (magenta), 20\% (green). The dashed black curve depicts $Q$ for a two-mode squeezed state. See Fig.~6 of \cite{Ourjoumtsev2007PRL}.}}\label{QdGS}
\end{figure}

\subsection{Mixtures of Gaussian states}

Recent papers \cite{Lund2006arXiv,Mista2002PRA} dealt with mixed NGS of the form $\hat \rho_m = p \ket {\phi_r}\!\bra{\phi_r} + (1-p) \ket{00} \bra{00}$, with $0 \le p \le 1$. They have a positive Wigner function yet they are NGS (but for the trivial instances $p=0,1$). Clearly, the de-gaussification here reduces entanglement and correlations in general, as it involves a mixing with the totally uncorrelated vacuum state.
The negativity of such states reads $N_m=N(\hat \rho_m)=p N(\phi_r)= p(\e^{2r}-1)/2$ and is increasing both with $r$ and with $p$. The same dependency holds for the bit correlations, 

\begin{equation}
Q_m=Q(\hat \rho_m)=(2p/\pi) \arctan\left(N_m \left[\frac{1}{2 N_m+p}+\frac{1}{p}\right]\right),
\end{equation}
which again is a monotonic function of $N_m$ for any $p$.

A similar result holds for mixtures of photon-subtracted states with the vacuum. We further studied other NGS including photon-added and squeezed Bell-like states \cite{DellAnno2007PRA}, and their mixtures with the vacuum, for all we found a direct match between entanglement and $Q$.
Interestingly, this is not true for {\em all} CV states. By definition, $Q$ quantifies correlations encoded in the second canonical moments only. We have realized that there exist also states {\em e.g.} the photonic qutrit state $\ket{\psi_h} = \ket{00}/\sqrt{2}+(\ket{02}+\ket{20})/2$] which, though being totally uncorrelated up to the second moments ($Q=0$), are strongly entangled, with correlations embedded only in higher moments. The characterization of such states is an intriguing topic for further study. Finally, we have checked that for all states analyzed, the bit correlations obtained in the conjugate quadrature of the optimal one, {\em i.e.} $Q(\theta+\pi/2, \varphi+\pi/2)$ are either zero or negligible but a possible ``heterodyne'' generalization of our approach, involving measurements of two conjugate quadratures per mode, also deserves further attention.

\section{Conclusions}

In this chapter, by analyzing the maximal number of correlated bits ($Q$) that can be extracted from a CV state via quadrature measurements, we have provided an operational quantification of the entanglement content of several relevant NGS (including the useful photon-subtracted states). Crucially, one can experimentally measure $Q$ by direct homodyne detections (of the quadratures displaying optimal correlations only), in contrast to the much more demanding full tomographical state reconstruction. One can then easily invert the (analytic or numeric) monotonic relation between $Q$ and the negativity to achieve a direct entanglement quantification from the measured data. Our analysis demonstrates the rather surprising feature that entanglement in the considered NGS can thus be detected and experimentally quantified with the same complexity as if dealing with GS.
In this respect, it is even more striking that the measure considered in this paper, based on (and accessible in terms of) second moments and homodyne detections only, provides such an exact quantification of entanglement in a broad class of pure and mixed NGS, whose quantum correlations are encoded nontrivially in higher moments too, and currently represent the preferred resources in CV Quantum Information. We focused on optical realizations of CV systems, but our framework equally applies to collective spin components of atomic ensembles \cite{Julsgaard2001N}, and radial modes of trapped ions \cite{Serafini2009NJP}.
Finally, it is also surprising that for all these family of states we have studied, the optimization of just one quadrature scales monotonically with the negativity of the state. Although this could expected for pure Gaussian states, our study demonstrated that the non-Gaussian states obtained either as de-gaussifications of pure Gaussian states or mixings with uncorrelated states preserve this property. 


\chapter{Measurement induced entanglement in Continuous Variable}\label{mesoscopic}

\section{Introduction}

In the previous chapters we have presented several implementations of Continuous Variable (CV) systems which demand in advance, as a resource, shared entanglement between the parties involved. In particular, for the Byzantine agreement protocol we demanded a pure, fully inseparable tripartite Gaussian state and completely symmetric under the interchange of the modes with a specific degree of entanglement for an efficient realistic implementation of the protocol. In this chapter our aim is to provide a realistic optical scheme for the creation and manipulation of multipartite Gaussian entanglement of arbitrary modes between atomic gas samples. We concentrate on a matter-light interaction between Gaussian polarized light and polarized atomic samples. For convenience we analyze quantum non-demolition (QND) matter-light interfaces.

Matter-light quantum interfaces refer to those interactions that lead to a faithful transfer of correlations between atoms and photons. The interface, if appropriately tailored, generates an entangled state of matter and light which can be further manipulated (for a review see \cite{Hammerer2008arXiv,Sherson2006AAMP} and references therein). To this aim, a strong coupling between atoms and photons is a must. A pioneering method to enhance the coupling is cavity QED, where atoms and photons are made to interact strongly due to the confinement imposed by the boundaries \cite{Haroche2006OUP}. An alternative approach to reach the strong coupling regime in {\em free space} is to use optically thick atomic samples.

Atomic samples with internal degrees of freedom (collective spin) can be made to interact with light via the Faraday effect, which refers to the polarization rotation that is experienced by a linearly polarized light propagating inside a magnetic medium. At the quantum level, the Faraday effect leads to an exchange of fluctuations between matter and light. As demonstrated by Kuzmich and co-workers \cite{Kuzmich1997PRL}, if an atomic sample interacts with a squeezed light whose polarization is measured afterwards, the collective atomic state is projected into a spin squeezed state (SSS). Furthermore, to produce a long lived SSS, Kuzmich and co-workers \cite{Kuzmich2000PRL} proposed a QND measurement, based on off-resonant light propagating through an atomic polarized sample in its ground state.

A step forward within this scheme is measurement induced entanglement between two macroscopic atomic ensembles. As proposed by Duan {\em et al} \cite{Duan2000PRLBis} and demonstrated by Polzik and co-workers \cite{Julsgaard2001N}, the interaction between a single laser pulse, propagating through two spatially separated atomic ensembles combined with a final projective measurement on the light, leads to an EPR state of the two atomic ensembles. Due to the QND character of the measurement, the verification of entanglement is done by a homodyne measurement of a second laser pulse that has passed through the samples. From such measurements, atomic spin variances inequalities can be checked, asserting whether the samples are entangled or not. A complementary scheme for measurement induced entanglement is also introduced in \cite{DiLisi2004PRA,DiLisi2002PRA}.

The quantum Faraday effect can also be used as a powerful spectroscopic method \cite{Sorensen1998PRL}. Tailoring the spatial shape of the light beam, provides furthermore, a detection method with spatial resolution which opens the possibility to detect phases of strongly correlated systems generated with ultra-cold gases in optical lattices \cite{Bruun2009PRL,Eckert2008NP,Roscilde2009NJP}.

Here, we analyze the suitability of the Faraday interface in the multipartite scenario. In contrast to the bipartite case, where only one type of entanglement exists, the multipartite case offers a richer situation \cite{Dur1999PRL,Giedke2001PRA}. Due to this fact, the verification of entanglement using spin variance inequalities \cite{vanLoock2003PRA} becomes an intricate task. We address such problem and provide a scheme for the generation and verification of multipartite entanglement between atomic ensembles. Despite the irreversible character of the entanglement induced by measurement, we find that a second pulse can reverse the action of the first one deleting all the entanglement between the atomic samples. This result has implications in the use of atomic ensembles as quantum memories \cite{Julsgaard2004N}. Finally, we introduce the CV formalism for further analysis. That is, if one prepares both atoms and light as Gaussian states, then due to the linearity of the equations of evolution for atoms and light, the evolution is Gaussian and it is possible to write the states as covariance matrices and the evolution as a symplectic transformation. The explicit use of the covariance matrix enables for an entanglement verification through covariance matrix criteria.

\section{Quantum interface description}

The basic concept, underlying the QND atom-light interface we will use (see appendix~\ref{appendix3}), is the dipole interaction between an off-resonant linearly polarized light with a polarized atomic ensemble, followed by a quantum homodyne measurement of light. In section~\ref{polzik} we review the basic known results that permits, to entangle two separated atomic samples by letting light interact with two atomic samples in a fixed direction. In section~\ref{geo} we detail our new proposal in which light and atoms are free to interact at arbitrary angles.

On one hand, we consider an ensemble of $N_{at}$ non-interacting alkali atoms with individual total angular momentum $\vec F=\vec I+\vec J$ prepared in the ground state manifold $\ket{F,m_F}$. Further we assume that all atoms are polarized along the $x$-direction, which corresponds to preparing them in a certain hyperfine state $\ket{F,m_F}$ ({\em e.g.} in the case of Cesium the hyperfine ground state $6S_{1/2}$ {\em i.e.} $L=0$ and $J=S=1/2$ is split into two hyperfine states with total angular momentum $F=3$ and $F=4$ due to a nuclear spin $I=7/2$). We restrict ourselves to one hyperfine level, $F=4$, which is possible experimentally because the hyperfine splitting is large compared to typical resolutions of the lasers. Furthermore, it is possible to put (almost) all atoms in the outermost state with $m_F=4$, being $x$ the axis of quantization. We describe such sample with its collective angular momentum $\hat{\vec J}=(\hat J_x,\hat J_y,\hat J_z)$ being $\hat J_k = \sum_{n=1}^{N_{at}} \hat F_{k,n}$ the total angular momentum of the ensemble $(k=x,y,z)$. Then, the $\hat J_x$ component of the collective spin can be regarded as a classical number $\hat J_x \approx <\hat J_x> = N_{at} \hbar F$, while the orthogonal spin components encode all the quantum character. Due to the above approximation, the orthogonal collective angular momentum components can be treated as canonical conjugate variables, $\left[\hat J_y/\sqrt{J_x}, \hat J_z/\sqrt{J_x} \right] = \im \hbar$.

On the other hand, the polarization of light propagating along the $z$-direction can be described by the Stokes vector $\hat{\vec s}=(\hat s_x,\hat s_y,\hat s_z)$, whose components correspond to the differences between the number of photons (per time unit) with $x$ and $y$ linear polarizations, $\pm \pi/4$ linear polarizations and the two circular polarizations

\begin{eqnarray}
\hat s_x & = &\frac{\hbar}{2}(\hat n_x - \hat n _y) = \frac{\hbar}{2} (\hat a_x^\dag\hat a_x-\hat a_y^\dag\hat a_y), \nonumber\\
\hat s_y & = &\frac{\hbar}{2}(\hat n_{\nearrow} - \hat n_{\searrow}) = \frac{\hbar}{2}(\hat a_x^\dag\hat a_y+\hat a_y^\dag\hat a_x), \nonumber\\
\hat s_z & = &\frac{\hbar}{2}(\hat n_{\circlearrowleft} - \hat n_{\circlearrowright}) = \frac{\hbar}{2\im}(\hat a_x^\dag\hat a_y-\hat a_y^\dag\hat a_x).
\end{eqnarray}

The above operators have dimension of energy. They are convenient for a microscopic description of interaction between light and atoms, however, we will concentrate on the macroscopic variables, defined as $\hat S_k(z)=\int_0^T \hat s_k(z,t)dt$ $(k=x,y,z)$, where $T$ is the length of the light pulse. Such defined operators correspond now to differences in total number of photons, and obey standard angular momentum commutation rules. For light linearly polarized along the $x$-direction $\hat{S}_x \approx <\hat{S}_x> = N_{\rm{ph}} \hbar/2$. In such case, the orthogonal Stokes components $\hat{S}_y, \hat{S}_z$ are conjugated variables fulfilling canonical commutation relations, $\left[\hat S_y/\sqrt{S_x}, \hat S_z/\sqrt{S_x} \right] = \im \hbar$.

For a light beam propagating through the atomic sample in the $YZ$ plane at a certain angle $\alpha$ with respect to direction $z$ (see Fig.~\ref{angle}), the atom-light interaction can be approximated to the following QND effective interaction Hamiltonian (see appendix~\ref{appendix3} for a detailed derivation)

\begin{equation}\label{halpha}
\hat H^{\rm{eff}}_{\rm{int}}(\alpha)=-\frac{a}{T} \hat{S}_z (\hat{J}_z \cos\alpha+\hat{J}_y \sin\alpha).
\end{equation}
The parameter $a$ is the coupling constant with dimensions of the inverse of an action. The interaction is a linear coupling between the Stokes operator and the collective atomic spin operator, thus, the interaction is a Gaussian interaction between two bosonic modes. Also, the states for atoms and light, by the strong polarization constrain, can be treated as two mode enabling us to tackle the atom-light interaction with a CV formalism.

\begin{figure}[h]
 \centering
 \psfrag{X}{$x$}\psfrag{Y}{$y$}\psfrag{Z}{$z$}\psfrag{A}{$\alpha$}
 \includegraphics[width=5cm]{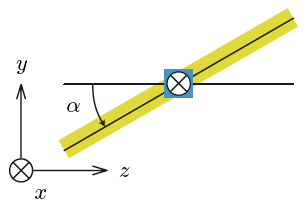}\\
 \caption{\small{A beam passing through an atomic sample at an angle $\alpha$ with respect to $z$.}}\label{angle}
\end{figure}

The equations of motion~\footnote{In the particular configuration of Fig.~\ref{angle}, both the Hamiltonian and the equations of motion can be deduced after an anti $x$-rotation of angle $\alpha$ to the corresponding expressions of appendix~\ref{appendix3}.} for the macroscopic variables for light and atoms are

\begin{eqnarray}
&& \hat J_y^{\rm{out}} = \hat J_y^{\rm{in}}-a\hat S_z^{\rm{in}} J_x \cos{\alpha}\label{Jy},\\
&& \hat J_z^{\rm{out}} = \hat J_z^{\rm{in}}+ a \hat S_z^{\rm{in}} J_x \sin{\alpha}\label{Jz},
\end{eqnarray}
\begin{eqnarray}
&& \hat S^{\rm{out}}_y = \hat S^{\rm{in}}_y - a S_x (\hat J_z^{\rm{in}} \cos\alpha+\hat J_y^{\rm{in}} \sin\alpha)\label{Sy},\\
&& \hat S^{\rm{out}}_z = \hat S^{\rm{in}}_z\label{Sz},
\end{eqnarray}
where the operators $\hat S_k^{\rm{in/out}}$ as the Stokes operators characterizing the pulse entering ($z=0$) and leaving ($z=L$) the atomic sample. Analogously, $\hat J_k^{\rm{in/out}}$ correspond to initial ($t=0$) and final state ($t=T$) of the atomic spin.

From Eq.~\eqref{Sy} it is clear that the polarization of the outgoing light carries information about the collective atomic angular momentum. The quantum character of the interface is reflected at the level of fluctuations, {\em i.e.},

\begin{equation}
\var{\hat S^{\rm{out}}_y} = \var{\hat S^{\rm{in}}_y} + a^2 S_x^2 \var{[\hat J_z^{\rm{in}}\cos{\alpha}+\hat J_y^{\rm{in}}\sin{\alpha}]}.
\end{equation}
At the same time, Eqs.~\eqref{Jy} and \eqref{Jz} show the QND character of the Hamiltonian, {\em i.e.}, the measured combination $\hat J_z^{\rm{in}} \cos{\alpha}+\hat J_y^{\rm{in}}\sin{\alpha}$ is not affected by the interaction since it commutes with the effective Hamiltonian. This fact allows to measure the fluctuations of the atomic spin components with the minimal disruption permitted by Quantum Mechanics.

In the following sections we will generalize the above formalism to the interaction of a light pulse with an arbitrary number, $N_s$, of spatially separated atomic samples. Variables characterizing each sample will be denoted by $\hat J_k^{(i)}$, where $i=1,2,\ldots, N_s$ denotes the samples and $k=x,y,z$.

\section{Bipartite entanglement: Generation and verification}

\subsection{Magnetic field addressing scheme}\label{polzik}

We begin by reviewing the seminal work of Polzik {\em et al} \cite{Julsgaard2001N} leading to the entanglement of two spatially separated atomic samples, as schematically shown in Fig.~\ref{bipPolzik}.

\begin{figure}[h]
 \centering
 \psfrag{X}{$x$}\psfrag{Y}{$y$}\psfrag{Z}{$z$}\psfrag{A}{$S_x$}\psfrag{B}{$J_x$}\psfrag{C}{$J_x$}
 a)\includegraphics[width=5cm]{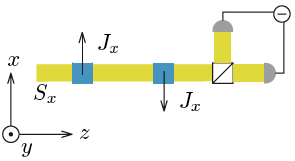} \hfill
 \psfrag{X}{$x$}\psfrag{Y}{$y$}\psfrag{Z}{$z$}\psfrag{A}{$S_x$}\psfrag{B}{$J_x$}\psfrag{C}{$J_x$}\psfrag{D}{$B$}  
 b)\includegraphics[width=5cm]{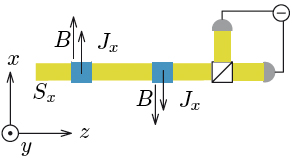}
 \caption{\small{Sketch of the experimental setup applied in \cite{Julsgaard2001N} to generate bipartite entanglement. a) Entangling pulse. b) Verifying pulse followed by homodyne measurement. A local magnetic field is added (parallel to the spin polarization direction) in order to measure two transverse components of the spin with a single light beam.}}\label{bipPolzik}
\end{figure}

In the experimental setup, both light and atomic samples were strongly polarized along the $x$-direction while light propagated along the $z$-direction ($\alpha=0$). It is straightforward to generalize the equations of motion of appendix~\ref{appendix3} for two samples. For example, the collective polarization of atoms along the $z$-direction is still preserved, {\em i.e.}, ${\partial \hat J_z^{(i)}}/{\partial t} = 0$ with $i=1,2$, and Eq.~\eqref{Sy} reads now~\footnote{We will omit the superscripts in/out when it is not necessary.}

\begin{equation}\label{bipout}
 \hat S_y^{\rm{out}}=\hat S_y^{\rm{in}}-a S_x \left(\hat J_z^{(1)}+\hat J_z^{(2)}\right).
\end{equation}
Entanglement between the atomic samples is established {\em as soon as} the $\hat S_y^{\rm{out}}$ component of {\em light is measured}. Moreover, it should be emphasized that entanglement is generated independently of the outcome of the measurement. The real challenge, though, is its experimental verification, since spin entanglement criterium relies on spin variances inequalities of operators of the type $(\hat J_y^{(1)}-\hat J_y^{(2)})$ and $(\hat J_z^{(1)}+\hat J_z^{(2)})$. This is so because the maximally entangled EPR state is a co-eigenstate of such operators. This fact, in turn, imposes an upper bound (see theorem~\ref{GINSEPD}) on the variances of such operators giving rise to a sufficient and necessary condition for separability \cite{Duan2000PRL}

\begin{equation}\label{sepDuan}
\var{\left[|\lambda|\hat J_y^{(1)}+\frac{\hat J_y^{(2)}}{\lambda}\right]}+\var{\left[|\lambda|\hat J_z^{(1)}-\frac{\hat J_z^{(2)}}{\lambda}\right]} \geq \left(\lambda^2+\frac{1}{\lambda^2}\right) \hbar J_x,
\end{equation}
for all $\lambda \in \mathbb{R}$.

The way to experimentally check \cite{Julsgaard2001N} the above equation with $\lambda=-1$ was to add on each sample an external magnetic field, parallel to the spin polarization direction, $x$ (see also \cite{Muschik2006PRA}).
The magnetic field was local, therefore, it did not affect the generation of entanglement. However, it caused a Larmor precession of the collective atomic momenta, which permitted a simultaneous measurement of the appropriately redefined "canonical variables" $\hat J_y^{(1)}+\hat J_y^{(2)}$ and $\hat J_z^{(1)}+\hat J_z^{(2)}$. Notice that this can only be done if the atomic samples are polarized oppositely along the $x$-direction, so that the commutator $[\hat J_z^{(1)}+\hat J_z^{(2)},\hat J_y^{(1)}+\hat J_y^{(2)}]=J_{x}^{(1)}+J_{x}^{(2)}=0$. Therefore, the first light beam was used for creation of EPR-type entanglement, and another one for its verification through Eq.~\eqref{sepDuan}.

Let us detail how the local magnetic field, if measuring the appropriated observables, enables for checking Eq.~\eqref{sepDuan}. As in the appendix~\ref{appendix3}, we derive here the complete equation system for observables of light and both samples. As we will see the linearity of the system enables us to, in a easy way, to jump to the CV formalism via symplectic matrices.

If a constant and homogeneous magnetic field is added in the $x$-direction (parallel to the spin polarization direction) the transverse spin components precess at a Larmor frequency $\Omega$. The magnetic field adds a term~\footnote{Remember that $\vec{B}=\Omega \frac{m}{e}\vec i$ and $\vec{\hat\mu} = -\frac{e}{m}\vec{\hat J}$.} $\hat H = -\vec{\hat\mu} \cdot \vec B= + \Omega J_x$ into the Hamiltonian with same field strength and global sign for each one of the two samples. The action of the magnetic field at a fixed time produces a rotation of the spin vector (active view) that can be mimic as a 
rotation on the system (passive view) via  the substitution $\alpha \to -\Omega t$ on the equations of motion considered in the appendix~\ref{appendix3} {\em i.e.}
\begin{eqnarray}
 && \frac{\partial}{\partial t} \hat j_y^{(1)}(z,t) = -a \hat s_z(t) j_{x}^{(1)} \cos{\Omega t},\label{Jy1}\\
 && \frac{\partial}{\partial t} \hat j_z^{(1)}(z,t) = -a \hat s_z(t) j_{x}^{(1)} \sin{\Omega t},\label{Jz1}\\
 && \frac{\partial}{\partial t} \hat j_y^{(2)}(z,t) = -a \hat s_z(t) j_{x}^{(2)} \cos{\Omega t},\label{Jy2}\\
 && \frac{\partial}{\partial t} \hat j_z^{(2)}(z,t) = -a \hat s_z(t) j_{x}^{(2)} \sin{\Omega t},\label{Jz2}\end{eqnarray}
\begin{eqnarray}
 && \frac{\partial}{\partial z} \hat s_y(z,t) = -\frac{a}{c T} s_x \left(\hat j_z^{(1)}\cos{\Omega t} - \hat j_y^{(1)} \sin{\Omega t}\right),\label{Sy12}\\
 && \hat s_z(z,t) = \hat s_z(t),\label{Sz12}
\end{eqnarray}
Grouping the equations \eqref{Jy1}-\eqref{Jz2} and using the fact that the two samples are opposite polarized, $J_{x}^{(1)}=-J_{x}^{(2)}=J_x > 0$ equations for atoms are
\begin{eqnarray}
 \frac{\partial}{\partial t} \left[\hat j_y^{(1)}(z,t)+\hat j_y^{(2)}(z,t)\right] &=& -a\hat s_z(t)
 \left(j_{x}^{(1)}+j_{x}^{(2)}\right) \cos{\Omega t}=0,\\
 \frac{\partial}{\partial t} \left[\hat j_z^{(1)}(z,t)+\hat j_z^{(2)}(z,t)\right] &=& -a \hat s_z(t)
 \left(j_{x}^{(1)}+j_{x}^{(2)}\right) \sin{\Omega t}=0,\\
 \frac{\partial}{\partial t} \left[\hat j_y^{(1)}(z,t)-\hat j_y^{(2)}(z,t)\right] &=& -a \hat s_z(t)
 \left(j_{x}^{(1)}-j_{x}^{(2)}\right) \cos{\Omega t} =\nonumber\\
 &=& -a \hat s_z(t) 2 j_x \cos{\Omega t},\\
 \frac{\partial}{\partial t} \left[\hat j_z^{(1)}(z,t)-\hat j_z^{(2)}(z,t)\right] &=& -a \hat s_z(t)
\left(j_{x}^{(1)}-j_{x}^{(2)}\right) \sin{\Omega t} =\nonumber\\
 &=& -a \hat s_z(t) 2 j_x \sin{\Omega t}.
\end{eqnarray}
We next integrate in space ($\int_0^L \rho A dz$) and time ($\int_0^T dt$) to introduce the macroscopical variables
\begin{eqnarray}
\hat J_y^{\rm{(1),in}}+\hat J_y^{\rm{(2),in}} &=& \hat J_y^{\rm{(1),out}}+\hat J_y^{\rm{(2),out}},\\
\hat J_z^{\rm{(1),in}}+\hat J_z^{\rm{(2),in}} &=& \hat J_z^{\rm{(1),out}}+\hat J_z^{\rm{(2),out}},\\
\hat J_y^{\rm{(1),out}}-\hat J_y^{\rm{(2),out}} &=& \hat J_y^{\rm{(1),in}}-\hat J_y^{\rm{(2),in}} - 2aJ_x \int_0^T dt \cos{\Omega t} \, \hat s_z(t),\\
\hat J_z^{\rm{(1),out}}-\hat J_z^{\rm{(2),out}} &=& \hat J_y^{\rm{(1),in}}-\hat J_y^{\rm{(2),in}} - 2aJ_x \int_0^T dt \sin{\Omega t} \, \hat s_z(t).
\end{eqnarray}
Eq.~\eqref{Sz12} is also integrated in space ($\int_0^L \rho A dz$) and Eq.~\eqref{Sy12} in space and in time ($\int_0^T dt \cos{\Omega t}$), ($\int_0^T dt \sin{\Omega t}$). Using the fact that the observables $\left[\hat j_y^{(1)}(z)+\hat j_y^{(2)}(z)\right]$ and $\left[\hat j_z^{(1)}(z)+\hat j_z^{(2)}(z)\right]$ do not evolve, equations for light are
\begin{equation}
 \begin{split}
 \int_0^T & dt \cos{\Omega t} \, \hat s^{\rm{out},2}_y(t) = \int_0^T dt \cos{\Omega t} \, \hat s^{\rm{in},1}_y(t) +\\
 &-\frac{a}{T} S_x \int_0^T dt \cos{\Omega t} \left(\left[\hat J_z^{(1)}+\hat J_z^{(2)}\right] \cos{\Omega t} - \left[\hat J_y^{(1)}+\hat J_y^{(2)}\right] \sin{\Omega t}\right),
 \end{split}
 \end{equation}
\begin{equation}
 \begin{split}
 \int_0^T & dt \sin{\Omega t} \, \hat s^{\rm{out},2}_y(t) = \int_0^T dt \sin{\Omega t} \, \hat s^{\rm{in},1}_y(t) +\\
 &-\frac{a}{T} S_x \int_0^T dt \sin{\Omega t} \left(\left[\hat J_z^{(1)}+\hat J_z^{(2)}\right] \cos{\Omega t}  - \left[\hat J_y^{(1)}+\hat J_y^{(2)}\right] \sin{\Omega t}\right),
  \end{split}
 \end{equation}
\begin{equation}
 \hat s^{\rm{out},2}_z(t) = \hat s^{\rm{in},2}_z(t) = \hat s^{\rm{out},1}_z(t) = \hat s^{\rm{in},1}_z(t).
\end{equation}
We introduce here a suitable choice of pairs of macroscopical observables for light and atoms, where the subscript ``$L$'' stands for light while ``$A$'' for atoms
\begin{eqnarray}
\hat Q_{L1}(z)=\sqrt{\frac{2}{S_x \hbar}}\int^T_0 dt \cos{\Omega t} \, \hat s_y(z,t), &\hfill& \hat P_{L1}=\sqrt{\frac{2}{S_x \hbar}}\int^T_0 dt \cos{\Omega t} \, \hat s_z(t),\nonumber \\
 \hat Q_{L2}(z)=\sqrt{\frac{2}{S_x \hbar}}\int^T_0 dt \sin{\Omega t} \, \hat s_y(z,t), &\hfill& \hat P_{L2}=\sqrt{\frac{2}{S_x \hbar}}\int^T_0 dt \sin{\Omega t} \, \hat s_z(t),\nonumber \\
 \hat Q_{A1}(t)=\frac{\hat J_y^{(1)}-\hat J_y^{(2)}}{\sqrt{2J_x\hbar}}, &\hfill& \hat P_{A1}=\frac{\hat J_z^{(1)}+\hat J_z^{(2)}}{\sqrt{2J_x\hbar}},\nonumber \\
 \hat Q_{A2}(t)=-\frac{\hat J_z^{(1)}-\hat J_z^{(2)}}{\sqrt{2J_x\hbar}}, &\hfill& \hat P_{A2}=\frac{\hat J_y^{(1)}+\hat J_y^{(2)}}{\sqrt{2J_x\hbar}},\nonumber
\end{eqnarray}
to finally~\footnote{Taking into account that $\frac{1}{T}\int^T_0 dt\sin^2{\Omega t}=\frac{1}{T}\int^T_0 dt\cos^2{\Omega t}\simeq\frac{1}{2}$ and $\frac{1}{T}\int^T_0 dt\sin{\Omega t}\cos{\Omega t}\simeq0$ if $\Omega T\simeq2\pi$.} obtain the solution of the evolution equations for the set of canonical variables before and after interaction (denoted by in/out, respectively)
\begin{eqnarray}\label{inoutevo}
 && \hat P^{\rm{out}}_{A1} = \hat P^{\rm{in}}_{A1},\label{inoutevop}\\
 && \hat P^{\rm{out}}_{A2} = \hat P^{\rm{in}}_{A2},\\
 && \hat Q^{\rm{out}}_{A1} = \hat Q^{\rm{in}}_{A1} - \kappa \hat P^{\rm{in}}_{L1},\\
 && \hat Q^{\rm{out}}_{A2} = \hat Q^{\rm{in}}_{A2} - \kappa \hat P^{\rm{in}}_{L2},
\end{eqnarray}
\begin{eqnarray}
 && \hat Q^{\rm{out}}_{L1} = \hat Q^{\rm{in}}_{L1} - \kappa \hat P^{\rm{in}}_{A1},\label{QL1}\\
 && \hat Q^{\rm{out}}_{L2} = \hat Q^{\rm{in}}_{L2} - \kappa \hat P^{\rm{in}}_{A2},\label{QL2}\\
 && \hat P^{\rm{out}}_{L1} = \hat P^{\rm{in}}_{L1},\\
 && \hat P^{\rm{out}}_{L2} = \hat P^{\rm{in}}_{L2},\label{inoutevou}
\end{eqnarray}
where $\kappa = a\sqrt{S_x J_x}$ (adimensional) and $\left[\hat Q_{Ai}, \hat P_{Ai} \right] = \im$, $\left[\hat Q_{Li}, \hat P_{Li} \right] = \im$, for $i=1,2$.

Separability condition~\eqref{sepDuan} for $\lambda = -1$
\begin{equation}
\var{\left[\hat J_y^{(1)}-\hat J_y^{(2)}\right]}+\var{\left[\hat J_z^{(1)}+\hat J_z^{(2)}\right]} \geq
2 \hbar J_x
\end{equation}
translates to
\begin{equation}
\var{\hat P_{A1}}+\var{\hat P_{A2}} \geq 1
\end{equation}
taking into account that the sign of the spin $y$-component flips the sign (sample 1 and 2 are symmetric under a $\pi$ rotation around $z$). Thus, checking the inequality can be realized within this scheme by measuring the light quadratures operators $\hat Q^{\rm{out}}_{L1,2}$ as seen in Eqs.~\eqref{QL1} and \eqref{QL2}.

Now, we shall illustrate the power of using a CV phase-space formalism to retrieve the complete covariance matrix of the final states and to detect and quantify the entanglement generation.
The variables describing the system after interaction are expressed as a linear combination of the initial ones. Let us denote by $\mathcal{O}$ the following linear transformation $\mathcal{O}: \{ \hat Q_{Ai}^{\rm{in}},\hat P_{Ai}^{\rm{in}},\hat Q_{Li}^{\rm{in}},\hat P_{Li}^{\rm{in}}\} \mapsto \{ \hat Q_{Ai}^{\rm{out}},\hat P_{Ai}^{\rm{out}},\hat Q_{Li}^{\rm{out}},\hat P_{Li}^{\rm{out}} \}$, which can be straightforwardly obtained from Eqs.~\eqref{inoutevop}-\eqref{inoutevou}. Due to the linear coupling between light and matter under consideration, the evolution equations can be directly translated in phase-space as $\zeta^{\rm{out}}=O \cdot \zeta^{\rm{in}}$ where $\zeta=(Q_{A1},P_{A1},Q_{A2},P_{A2},Q_{L1},P_{L1},$ $Q_{L2},P_{L2})^T$ is a phase-space vector. There is a symplectic transformation $S_{\rm{int}}$, corresponding to the unitary evolution $\mathcal{O}$, relating the state of atoms and light $\gamma$, before (in) and after (out) the interaction fulfilling $\gamma_{\rm{out}}=S_{\rm{int}}^T \gamma_{\rm{in}} S_{\rm{int}}$ and can be reconstructed in the following way
\begin{eqnarray}
\zeta^{\rm{out},T} \cdot \gamma_{\rm{in}}^{-1} \cdot \zeta^{\rm{out}}
 &=&\zeta^{{\rm{in}},T} O^T \cdot \gamma_{\rm{in}}^{-1} \cdot O \zeta^{\rm{in}}=\nonumber\\
 &=&\zeta^{\rm{in},T} \cdot (O^{-1}\gamma_{\rm{in}} (O^T)^{-1})^{-1} \cdot \zeta^{\rm{in}}=\nonumber\\
 &=&\zeta^{\rm{in},T} \cdot \gamma_{\rm{out}}^{-1} \cdot \zeta^{\rm{in}}.
\end{eqnarray}
Therefore the symplectic transformation acting on an initial state with covariance matrix for light and atoms $\gamma_{\rm{in}}$ due to the Hamiltonian \eqref{hamiltonian0} is therefore $S_{\rm{int}}=(O^T)^{-1}$ and can be written in a matricial form. If, as we will consider, one prepares the ensemble of atoms in a Gaussian state and Gaussian light is used, the CV formalism enables us not only to measure fluctuations of specific collective spin variables of atoms but the complete covariance matrix and displacement vector after interaction enabling for verification of entanglement amenable with covariance matrix criteria. We derive explicitly the states and the evolution using this formalism in section~\ref{geo} presenting a new geometric scheme.

\subsection{Geometrical scheme}\label{geo}

Our aim has been to apply the QND atom-light interface to study genuine multipartite entanglement generation with less restrictive conditions, {\em i.e.}, we assume that: (i) individual magnetic field addressing of each atomic ensemble is not allowed and, (ii) the number of atomic ensembles can be made arbitrary. Such experimental setups that can be build, for instance, using optical micro-traps \cite{Lengwenus2007APB,Dumke2002PRL} which allow for isotropic confinement of $10^4$ cold atoms, creating in this way mesoscopic atomic ensembles~\footnote{In these experiments with ultra-cold atoms one can reduce the number of atoms up to $10^4$ still having the same opacity of the medium.}. In these setups, the preparation of each sample in a different initial magnetic state or the addressing of a sample with individual magnetic fields is out of reach. Despite these limitations, an array of micro-traps offers considerable advantages, ranging from its experimental feasibility to the possibility to generate chains and arrays of atomic samples.

\begin{figure}[h]
 \centering
 \psfrag{X}{$x$}\psfrag{Y}{$y$}\psfrag{Z}{$z$}\psfrag{A}{$S_x$}\psfrag{B}{$J_x$}\psfrag{C}{$J_x$}
 a)\includegraphics[width=5cm]{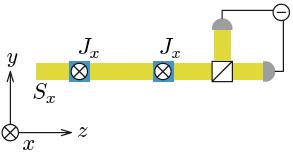} \hfill
 \psfrag{X}{$x$}\psfrag{Y}{$y$}\psfrag{Z}{$z$}\psfrag{A}{$S_x$}\psfrag{B}{$J_x$}\psfrag{C}{$J_x$}
 b)\includegraphics[width=5cm]{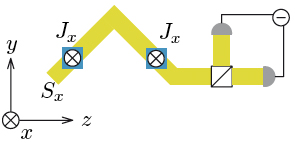}
 \caption{\small{The simplest setup for generation and verification of bipartite entanglement between mesoscopic atomic ensembles. a) First light pulse passing through the samples along direction $z$ entangles the samples. b) Second light pulse passing through the samples at angles $\pi/4$ and $-\pi/4$, respectively, allows for verification of entanglement through a variance inequality (see Eq.~\eqref{sepDuan}).}}\label{bipJulia}
\end{figure}

Assuming (i), the verification of entanglement cannot be done anymore by means of local magnetic fields on each sample, thus we assume the two samples to be parallel polarized. The way to overcome this problem is depicted in Fig.~\ref{bipJulia} {\em i.e.} we mimic the action of local magnetic fields in each sample with light propagating at different angles. As seen in the previous section they are mathematically equivalent local operations. To better understand the dynamics of the interaction in this configuration, we analyze in some detail the setup of Fig.~\ref{bipJulia}. In the first shot, light passes through the two sample. As indicated in Eq.~\eqref{bipout}, the light carries information about $\hat J_z^{(1)}+\hat J_z^{(2)}$ and the measurement of $\hat S_y^{\rm{out}}$ generates entanglement between the atomic samples. Starting from the evolution equations \eqref{Jy},\eqref{Jz} and taking into account the light measurements, one can explicitly derive the variances of the atomic spin samples and interpret them in terms of global or non-local squeezings. In the picture, polarized ensembles of atoms in the ground state will be characterized as a vacuum state while light beams will be described as coherent states. In this view, the final bipartite state of the ensembles is characterized by the following variances

\begin{eqnarray}
 \var{[\hat J_y^{(1)}+\hat J_y^{(2)}]} &=&(1+2 \kappa^2) \hbar J_x, \label{Jyp}\\
 \var{[\hat J_y^{(1)}-\hat J_y^{(2)}]} &=&\hbar J_x \label{Jym},\\
 \var{[\hat J_z^{(1)}+\hat J_z^{(2)}]} &=&\frac{1}{1+2 \kappa^2} \hbar J_x \label{Jzp},\\
 \var{[\hat J_z^{(1)}-\hat J_z^{(2)}]} &=&\hbar J_x \label{Jzm},
\end{eqnarray}
where $\kappa=a \sqrt{S_x J_x}$ is the adimendional coupling interaction. The observables for which the separability criterion (Eq.~\eqref{sepDuan}) is violated correspond to $\hat J_z^{(1)}+\hat J_z^{(2)}$ and $\hat J_y^{(1)}-\hat J_y^{(2)}$ with $\lambda = -1$. Such a measurement induces squeezing on the variances along the $z$-direction below the separability limit, as clearly indicated by Eqs.~\eqref{Jym} and \eqref{Jzp}.

The second part, the verification, it involves a measurement of the sum of the variances corresponding to Eqs.~\eqref{Jym} and \eqref{Jzp}. In order to do this with a single beam we use light propagating at different angles, as schematically depicted in Fig.~\ref{bipJulia}(b). In this case, according to Eq.~\eqref{Sy} we obtain

\begin{equation}
 \hat S_y^{\rm{out}}=\hat S_y^{\rm{in}}-\frac{\kappa}{\sqrt{2}} \sqrt{\frac{S_x}{J_x}}\left[ \left(\hat J_z^{(1)}+\hat J_z^{(2)})+(\hat J_y^{(1)}-\hat J_y^{(2)}\right)\right].
\end{equation}
Since within this scheme
$<\hat J_y^{(i)}\hat J_z^{(j)}>=<\hat J_y^{(i)}><\hat J_z^{(j)}>$, the variance of the output can be written as

\begin{eqnarray}
& &\var{\hat S_y^{\rm{out}}}=\var{\hat S_y^{\rm{in}}}+\nonumber\\
&+&\frac{\kappa^2}{2} \frac{S_x}{J_x}\left(\var{[\hat J_z^{(1)} + \hat J_z^{(2)}]}+\var{[\hat J_y^{(1)}-\hat J_y^{(2)}]}\right).
\end{eqnarray}
For details concerning the experimental measurement of such variances see \cite{Julsgaard2003PhD,Sherson2005NATO}. This shows that entanglement between two identically polarized atomic ensembles can be generated, irrespectively of the value of the coupling constant $\kappa$, and verified using only two beams and no additional magnetic fields, if the second field impinges on the two samples at certain angles.

To increase entanglement between the two samples one should introduce global squeezing in two independent variables. This is schematically depicted in Fig.~\ref{bipJuliaincrease}(a) and \ref{bipJuliaincrease}(b). The first beam introduces squeezing in $\hat J_z^{(1)}+\hat J_z^{(2)}$ variable. Then, a second beam propagating through the first sample at an angle $\alpha=\pi/2$ and through the second one at an angle $\alpha=-\pi/2$ generates squeezing in $\hat J_y^{(1)}-\hat J_y^{(2)}$.

\begin{figure}[h]
 \centering
 \psfrag{X}{$x$}\psfrag{Y}{$y$}\psfrag{Z}{$z$}\psfrag{A}{$S_x$}\psfrag{B}{$J_x$}\psfrag{C}{$J_x$}
 a)\includegraphics[width=5cm]{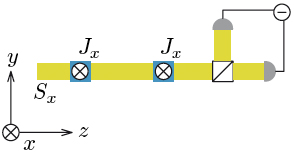} \hfill
 \psfrag{X}{$x$}\psfrag{Y}{$y$}\psfrag{Z}{$z$}\psfrag{A}{$S_x$}\psfrag{B}{$J_x$}\psfrag{C}{$J_x$}
 b)\includegraphics[width=5cm]{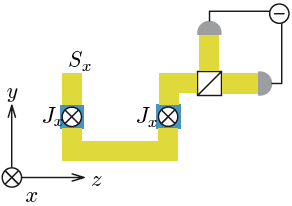}\\
 \psfrag{X}{$x$}\psfrag{Y}{$y$}\psfrag{Z}{$z$}\psfrag{A}{$S_x$}\psfrag{B}{$J_x$}\psfrag{C}{$J_x$}
 c)\includegraphics[width=5cm]{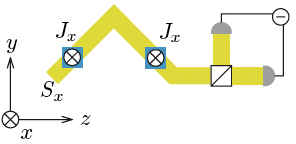}
 \caption{\small{The setup for generation and verification of bipartite entanglement between atomic ensembles in which squeezing is introduced in two variables a) $\hat J_z^{(1)}+\hat J_z^{(2)}$ and b) $\hat J_y^{(1)}-\hat J_y^{(2)}$ increasing entanglement. The third pulse depicted in figure c) allows for verification of entanglement through a variance inequality. It should be emphasized that the first and last step are exactly the same as in Fig.~\ref{bipJulia}.}}\label{bipJuliaincrease}
\end{figure}
Note that these are commuting operators, so the second beam would not change the effect of the first one (squeezing of $\hat J_z^{(1)}+ \hat J_z^{(2)}$). The verification of entanglement (see Fig.~\ref{bipJuliaincrease}(c)) can be done as previously described. Within this scheme one reproduces the results of Julsgaard {\em et al} \cite{Julsgaard2001N} without individual addressing {\em i.e.} EPR entanglement between two atomic samples.

\subsection{Continuous Variable analysis}

We detail here explicitly the setup of Fig.~\ref{bipJulia} at the level of covariance matrix (CM) formalism. The initial state of the two samples and the light is given by the CM for atoms and light $\gamma_{\rm{in}}^{A,L}=\id_4^A\oplus\id_2^L$. The symplectic matrix, $S_{\rm{int}}$, describing the interaction of light passing through the samples at zero angle is

\begin{equation}
S_{\rm{int}}=\left(\begin{array}{cccc|cc}
 1 & 0 & 0 & 0 & 0 & 0\\
 0 & 1 & 0 & 0 & \kappa & 0\\
 0 & 0 & 1 & 0 & 0 & 0\\
 0 & 0 & 0 & 1 & \kappa & 0\\
 \hline
 0 & 0 & 0 & 0 & 1 & 0\\
 \kappa & 0 & \kappa & 0 & 0 & 1\\
 \end{array}\right),
\end{equation}
thus, the state after the interaction takes the form

\begin{equation}
\gamma_{\rm{out}}^{A,L}=S_{\rm{int}}^T \gamma_{\rm{in}}^{A,L} S_{\rm{int}}=\left(\begin{array}{cccc|cc}
 1+\kappa^2 & 0 & \kappa^2 & 0 & 0 & \kappa\\
 0 & 1 & 0 & 0 & \kappa & 0\\
 \kappa^2 & 0 & 1+\kappa^2 & 0 & 0 & \kappa\\
 0 & 0 & 0 & 1 & \kappa & 0\\
 \hline
 0 & \kappa & 0 & \kappa^2 & 1+2\kappa^2 & 0\\
 \kappa & 0 & \kappa & 0 & 0 & 1
 \end{array}\right).
\end{equation}
Both modes, representing the samples, are entangled with light, however their reduced state is separable as one can check applying NPPT criterium in the CM of the lower-right block matrix. Entanglement between atomic samples is not produced until one measures one quadrature on the light. For this aim, one uses the action of an homodyne measurement on the light mode (see section \ref{staandope}). The measurement of light leads to the CM describing the final state of the samples.

\begin{eqnarray}
\gamma_{\rm{out}}^A&=&\left(\begin{array}{cccc}
 1+\kappa^2 & 0 & \kappa^2 & 0\\
 0 & \frac{1+\kappa^2}{1+2 \kappa^2} & 0 & -\frac{\kappa^2}{1+2 \kappa^2}\\
 \kappa^2 & 0 & 1+\kappa^2 & 0\\
 0 & -\frac{\kappa^2}{1+2 \kappa^2} & 0 & \frac{1+\kappa^2}{1+2 \kappa^2}
 \end{array}\right)\to\\
 &\stackrel{S_{\tilde r} \oplus S_{\tilde r}}{\to}&\left(\begin{array}{cccc}
 \frac{1+k^2}{\sqrt{1+2 k^2}} & 0 & \frac{k^2}{\sqrt{1+2 k^2}} & 0 \\
 0 & \frac{1+k^2}{\sqrt{1+2 k^2}} & 0 & -\frac{k^2}{\sqrt{1+2 k^2}} \\
 \frac{k^2}{\sqrt{1+2 k^2}} & 0 & \frac{1+k^2}{\sqrt{1+2 k^2}} & 0 \\
 0 & -\frac{k^2}{\sqrt{1+2 k^2}} & 0 & \frac{1+k^2}{\sqrt{1+2 k^2}}
\end{array}\right).
\end{eqnarray}
being $\tilde r=\frac{1}{4}\ln(1+2\kappa^2)$ a local squeezing. In this way we write the final state of the two atomic samples in standard form.

The resulting bipartite state of the two samples is pure and parametrized by $\kappa$ only, thus it can be written as a two mode squeezed state

\begin{equation}
\gamma_{TMS} = \begin{pmatrix}
 \cosh{2r} & 0 & \sinh{2r} & 0\\
 0 &  \cosh{2r} & 0 & -\sinh{2r}\\
 \sinh{2r} & 0 &  \cosh{2r} & 0\\
 0 & -\sinh{2r} & 0 &  \cosh{2r}
\end{pmatrix},
\end{equation}
with squeezing parameter $r=\frac{1}{2}{\rm arccosh}(\frac{1+\kappa^2}{\sqrt{1+2\kappa^2}})$ (see lemma~\ref{sform}) {\em i.e.} containing pure EPR entanglement with negativity
\begin{equation}
 E_N(\gamma_{TMS})=\cosh^2{r}\log_2\cosh^2{r} - \sinh^2{r}\log_2\sinh^2{r}.
\end{equation} 

\section{Entanglement eraser}

Interesting enough, our geometrical approach also opens the possibility of deleting all the entanglement created by the first light beam, if intensities are appropriately adjusted. The entanglement procedure is intrinsically irreversible because of the projective measurement, so coming ``deterministically'' back to the initial state is not a trivial task. In \cite{Filip2002JOB,Filip2003PRA}, a quantum erasing scheme in Continuous Variable systems was proposed. The measurement of the meter coordinate entangled with the quantum system leads to a back-action on it. The authors shown that it is possible to erase the action of the measurement and restore the original state of the system. Here we are interested in deleting the measurement induced entanglement between two atomic samples, exploiting the squeezing and antisqueezing effects produced by the laser beams.

\begin{figure}[h]
 \centering
 \psfrag{X}{$x$}\psfrag{Y}{$y$}\psfrag{Z}{$z$}\psfrag{A}{$S_x^{(1)}$}\psfrag{B}{$J_x$}\psfrag{C}{$J_x$}
 a)\includegraphics[width=5cm]{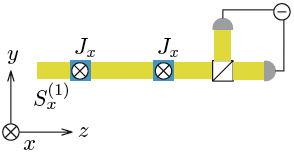} \hfill
 \psfrag{X}{$x$}\psfrag{Y}{$y$}\psfrag{Z}{$z$}\psfrag{A}{$S_x^{(2)}$}\psfrag{B}{$J_x$}\psfrag{C}{$J_x$}
 b)\includegraphics[width=5cm]{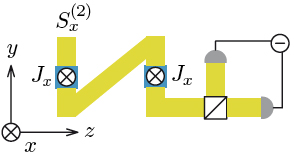}
 \caption{\small{Entanglement eraser scheme realized by two pulses of different intensity, $\kappa_1^2 \propto N_{\rm{ph}}^{(1)}$ and $\kappa_2^2 \propto N_{\rm{ph}}^{(2)}$. See the text for details.}}\label{biperaser}
\end{figure}

Let us assume that the first entangling beam, characterized by a coupling constant $\kappa_1^2 \propto N_{\rm{ph}}^{(1)}$, propagates along the $z$-direction, exactly as it was described before (see Fig.~\ref{biperaser}(a)). The interaction, followed by the measurement of light, creates squeezing in the observable $\hat J_z^{(1)}+\hat J_z^{(2)}$ accompanied by antisqueezing in the conjugate variable $\hat J_y^{(1)}+\hat J_y^{(2)}$ (Eqs.~\eqref{Jyp} and \eqref{Jzp}). Assume a second beam characterized by a coupling constant $\kappa_2^2 \propto N_{\rm{ph}}^{(2)}$ propagates through the samples in an orthogonal direction with respect to the first beam as shown in Fig.~\ref{biperaser}(b). This corresponds to setting $\alpha=\pi/2$ in the Hamiltonian of Eq.~\eqref{halpha}. In this setup the measurement of the variable $\hat S_y^{\rm{out}}$ introduces antisqueezing in the observable $\hat J_z^{(1)}+\hat J_z^{(2)}$ while squeezing in the conjugate variable $\hat J_y^{(1)}+\hat J_y^{(2)}$.

The bipartite state created by propagation and measurement of the first and second beam is characterized by the variances

\begin{eqnarray}
&&\var{[J_y^{(1)}+J_y^{(2)}]}=\frac{2 \kappa_1^2+1}{\left(4 \kappa_1^2+2\right) \kappa _2^2+1}\hbar J_x,\\
&&\var{[J_y^{(1)}-J_y^{(2)}]}=\hbar J_x,\\
&&\var{[J_z^{(1)}+J_z^{(2)}]}=\left(2 \kappa_2^2+\frac{1}{2 \kappa_1^2+1}\right)\hbar J_x,\\
&&\var{[J_z^{(1)}-J_z^{(2)}]}=\hbar J_x.
\end{eqnarray}
A close look at these equations shows that the second beam can lower or even completely destroy entanglement between the samples. This happens when

\begin{equation}\label{kappas}
\kappa_2^2=\frac{\kappa _1^2}{2\kappa_1^2+1}.
\end{equation}
In such case the atomic ensembles are left in a vacuum (uncorrelated) state, however, displaced. Hence, the overall effect of these two beams is simply a displacement of the initial vacuum state. The value of the displacement depends on the coupling constants $\kappa_{1,2}$ and outputs obtained in the measurement of the light polarization component, $	\hat S_y^{\rm{out}}$, of both beams. Therefore, it will vary run to run.

Using negativity, computed by the symplectic eigenvalues of the partial transpose of the covariance matrix \eqref{lognega}, one finds that indeed entanglement diminishes continuously or even disappears depending on the value of $\kappa_2$, as shown in Fig.~\ref{negativity}. Notice that for every fixed value of $\kappa_1$ there always exists a value of $\kappa_2$ for which negativity becomes zero and the state becomes separable even though it was entangled after the first interaction and measurement on the beam.

\begin{figure}[h]
 \centering
 \psfrag{X}{$\kappa_1$}\psfrag{Y}{$\kappa_2$}\psfrag{Z}{$N$}\psfrag{m}{$0$}\psfrag{M}{$\frac{1}{\sqrt{2}}$}
 \includegraphics[width=10cm]{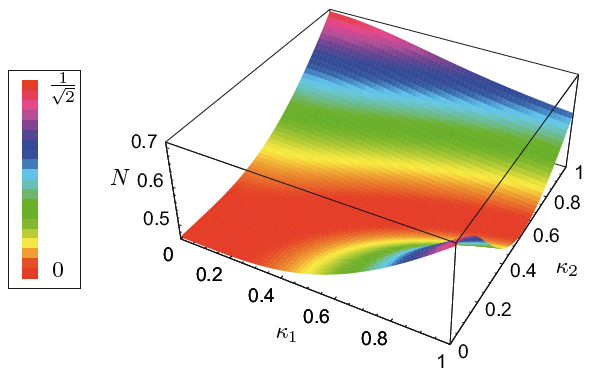}
 \caption{\small{The negativity of a bipartite state of atomic ensembles after passage and measurement of two beams of coupling parameters $\kappa_1$ and $\kappa_2$ (see Fig.~\ref{biperaser}). For specific values of $\kappa_1$ and $\kappa_2$, {\em i.e.} fulfilling \eqref{kappas}, the negativity approaches zero.}}\label{negativity}
\end{figure}

\section{Multipartite entanglement}\label{multi}

In what follows we generalize our study to the multipartite scenario and we present different strategies to achieve multipartite entanglement without individual addressing. The strategies will not depend on the total number of samples but only if this number is odd or even. For the verification part, we shall adopt the criterium for multipartite entanglement, expressed via generalized inequalities for variances of quadratures, derived by van Loock and Furusawa \cite{vanLoock2003PRA}. We rewrite the inequalities for angular momentum variables as follows. If an $N_s$-mode state $\hat \rho$ is separable, then the sum of variances of the following operators

\begin{eqnarray}
\hat{u}&=&h_1 \hat{J}_y^{(1)}+\ldots +h_{N_s} \hat{J}_y^{(N_s)},\nonumber\\
\hat{v}&=&g_1 \hat{J}_z^{(1)}+\ldots +g_{N_s} \hat{J}_z^{(N_s)},
\end{eqnarray}
is bounded from above by a function of the coefficients $h_1,\ldots,h_{N_s},g_1,\ldots,g_{N_s}$. Mathematically the inequality is expressed as

\begin{equation}\label{sepvanLiF}
\var{\hat{u}}+\var{\hat{v}}\geq f(h_1,\ldots,h_{N_s},g_1,\ldots,g_{N_s})\hbar J_x,
\end{equation}
with

\begin{equation}
f(h_1,\ldots,h_{N_s},g_1,\ldots,g_{N_s})=|h_m g_m+\sum_{r\in I} h_r g_ r|+|h_n g_n+\sum_{s\in I'} h_s g_s|.
\end{equation}
In the above formula two modes, $m$ and $n$, are distinguished and the remaining modes are grouped in two disjoint sets $I$ and $I'$. The criterion \eqref{sepvanLiF} holds for all bipartite splittings of a state defined by the sets of indices $\{m\}\cup I$ and $\{n\} \cup I'$, and for every choice of parameters $h_1,\ldots,h_{N_s},g_1,\ldots,g_{N_s}$. For example, in case of three samples we have
$f(h_1,h_2,h_3,g_1,g_2,g_3)=(|h_n g_n|+|h_k g_k+h_m g_m|)$, where $(n,m,k)$ is some permutation of the sequence $(1,2,3)$, and the coefficients $h_1,h_2,h_3,g_1,g_2,g_3$ are arbitrary real numbers.

\subsection{GHZ-like states}

Genuine multipartite entanglement between any number of equally polarized atomic modes can
be obtained with a single beam propagating through all of them followed by a projective measurement of the light. After the measurement, the $N_s$-mode variable $\hat J_z^{(1)}+\ldots+\hat J_z^{(N_s)}$ is squeezed. This is a trivial extension of the bipartite scheme schematically shown in Fig.~\ref{bipJulia}(a).

The phenomenon of destruction of entanglement by squeezing of the conjugate variable, which was discussed in the previous section for two modes, can be also found in the multimode setup. The entanglement prepared with the light beam characterized by the coupling constant $\kappa_1$ can be erased by the second orthogonal beam with appropriately adjusted intensity. The relation between the coupling constants for which entanglement is removed from the system is

\begin{equation}
\kappa_2^2=\frac{\kappa_1^2}{N_s \kappa_1^2 + 1}.
\end{equation}
One can see that with increasing number of samples the value of $\kappa_2$ required to delete entanglement decreases.

To generate a maximally entangled GHZ state with $N_s$-parties, simultaneous squeezing in more independent variables is needed. By independent here we mean commuting linear combinations of atomic spin operators.
The most straightforward way to do it is to generate squeezing in the variable $\hat J_z^{(1)}+\ldots+\hat J_z^{(N_s)}$ and in the pairwise differences of angular momenta: $\hat J_y^{(i)}-\hat J_y^{(j)}$ ($1 \leq i,j\leq N_s$, $i\neq j$) (see \cite{vanLoock2000PRL,Braunstein2005RMP}). An entangled state with such properties can be realized by generalization of the bipartite scheme summarized in Figs.~\ref{bipJuliaincrease}(a) and \ref{bipJuliaincrease}(b). Notice, however, that the last step b) should be repeated for all combinations of $i>j$. The final variances characterizing the state would be

\begin{eqnarray}
\var{[\hat J_z^{(1)}+\ldots+\hat J_z^{(N_s)}]}=\frac{N_s}{2+2 N_s \kappa^2} \hbar J_x,\\
\var{[\hat J_y^{(i)}-\hat J_y^{(j)}]}=\frac{1}{1+N_s \kappa^2} \hbar J_x \quad (i \neq j).
\end{eqnarray}
Thus the samples are in a genuine $N_s$-mode GHZ state. Within this scheme the number of measurements, $\binom{N_s}{2}+1 \sim \frac{N_s^2}{2}$, one has to perform in order to create genuine entanglement, grows quadratically with the number of samples. Also verification implies checking all the inequalities of the type

\begin{equation}
\var{[\hat J_y^{(i)}-\hat J_y^{(j)}]}+\var{[\hat J_z^{(1)}+\ldots+\hat J_z^{(N_s)}]} \geq 2 \hbar J_x \quad (i>j).\nonumber
\end{equation}
While the above procedure works for an arbitrary number of samples, to optimize it we consider separately even and odd $N_s$.

For even number of ensembles $N_s=2M$ the optimal approach generalizes the one proposed for two samples and summarized in Figs.~\ref{bipJuliaincrease}(a) and \ref{bipJuliaincrease}(b). In first step we generate squeezing in $\hat J_z^{(1)}+\ldots+\hat J_z^{(2M)}$. As the second step we squeeze the observable $\hat J_y^{(1)}-\hat J_y^{(2)}+\ldots+(-1)^{2M-1} \hat J_y^{(2M)}$ with the second beam passing through the $i$th sample at an angle $(-1)^{i-1}\pi/2$.
The final state is pure and genuine multipartite entangled. The entanglement can be detected using the criterion \eqref{sepvanLiF} with the two squeezed observables discussed in this paragraph. The measurement of light propagating through the $i$th sample at an angle $(-1)^{i-1}\pi/4$ gives at the level of variances

\begin{eqnarray}
&&\var{\hat S_y^{\rm{out}}}=\var{\hat S_y^{\rm{in}}}+\nonumber\\
&&+\frac{\kappa^2}{2}\frac{S_x}{J_x}\var{[\hat J_y^{(1)}-\hat J_y^{(2)}+\ldots+(-1)^{2M-1} \hat J_z^{(2M)}]}+\nonumber\\
&&+\frac{\kappa^2}{2}\frac{S_x}{J_x}\var{[\hat J_z^{(1)}+\ldots+\hat J_z^{(2M)}]}.
\end{eqnarray}
Therefore, again a single beam can be used for verification of entanglement. The same criterion and the above measurement scheme can be applied not only to detect the entanglement in the above setup but also in those proposed before, {\em i.e.}, (i) the state with squeezing only in $\hat J_z^{(1)}+\ldots+\hat J_z^{(2M)}$ (after interaction and measurement of only the first beam), and (ii) the state with squeezing in $\hat J_z^{(1)}+\ldots+\hat J_z^{(2M)}$ and all combinations $\hat J_y^{(i)}-\hat J_y^{(j)}$ $(i \neq j)$. The reduction in the number of measurements is significant. Moreover, a recently proposed multi-pass technique \cite{Namiki2009arXiv} could lead to a simplification of the geometry.

Optimization of the scheme for odd number of atomic ensembles within this geometric approach is to our knowledge not possible. Even though it is possible to find independent variables involving all the samples, it is not clear what geometry should be applied in order to measure these operators.

A different way to deal with multimode entanglement of odd number of samples is to generalize directly the bipartite scheme of Julsgaard {\em et al}, {\em i.e.}, polarize the samples in such a way that the collective polarization $\sum_i J_x^{(i)}$ is zero. Moreover, each sample should experience a different local magnetic field. In such system it is possible to generate squeezing in appropriately redefined (due to Larmor precession) operators $\sum_i \hat J_y^{(i)}$ and $\sum_i \hat J_z^{(i)}$, using a single light beam. This is possible due to the choice of the initial polarization of the samples making the redefined operators to commute. Analogously to the bipartite case the entanglement test that can be applied involves measurement of variances of the sums of angular momentum components and reads

\begin{equation}
\var{\sum_{i}\hat J_y^{(i)}}+\var{\sum_i \hat J_z^{(i)}} \geq N_s \hbar J_x.
\end{equation}

\subsection{Cluster-like states}

In \cite{Briegel2001PRL} a class of $N$-qubit quantum states generated in an arrays of qubits with an Ising-type interaction were presented, the so-called cluster states. While for pure states of bipartite qubit systems there is a single ``unit'' of entanglement, the one contained in a Bell state, for three or more parties several inequivalent classes exists. Cluster states, generated in optical lattices and similar systems, can be regarded as an entanglement resource since one can generate a family of other multipartite entangled states by simply performing measurement and using classical communication.
Using the scalability properties of cluster states, Hans Briegel {\em et al} presented a scheme for scalable one-way quantum computation \cite{Raussendorf2001PRL,Raussendorf2003PRA}. In there, cluster states are the entire resource for quantum computation while computation consists of a sequence of one-qubit projective measurements on them driving the computation. It is an universal quantum computer since any unitary quantum logic network can be simulated on it efficiently.

We detail here, how the analyzed setup allows for generation of Continuous Variable cluster-like states \cite{vanLoock2007PRA}. We associate the modes of the $N_s$-mode system with the vertices of a graph $G$. The edges between the vertices define the notion of nearest neighborhood. By $N_a$ we denote the set of nearest neighbors of vertex $a$. A cluster is a connected graph. For angular momentum variables, cluster states are defined only asymptotically as those with infinite squeezing in the variables

\begin{equation}\label{clustervariables}
\hat J^{(a)}_z-\sum_{b\in N_a } \hat J^{(b)}_y,
\end{equation}
for all $a \in G$. Cluster-like states are defined when the squeezing is finite.
Given a set of $N_s$ atomic ensembles, it is possible to create a chosen cluster-like state by squeezing the required combinations of variables \eqref{clustervariables}. Since they commute, it is possible to squeeze them sequentially. Hereafter, we will illustrate the procedure by a simple example. The method is general and can be applied to create any cluster-like state.

\begin{figure}[h]
 \centering
 a)\includegraphics[width=5cm]{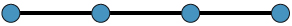}\\
 \psfrag{X}{$x$}\psfrag{Y}{$y$}\psfrag{Z}{$z$}\psfrag{A}{$S_x$}\psfrag{B}{$J_x$}\psfrag{1}{$1$}\psfrag{2}{$2$}\psfrag{3}{$3$}\psfrag{4}{$4$}
 b)\includegraphics[width=5cm]{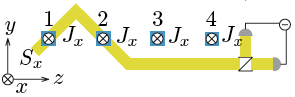} \hfill
 \psfrag{X}{$x$}\psfrag{Y}{$y$}\psfrag{Z}{$z$}\psfrag{A}{$S_x$}\psfrag{B}{$J_x$}\psfrag{1}{$1$}\psfrag{2}{$2$}\psfrag{3}{$3$}\psfrag{4}{$4$}
 c)\includegraphics[width=5cm]{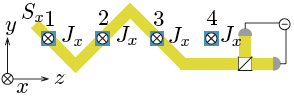}\\
 \psfrag{X}{$x$}\psfrag{Y}{$y$}\psfrag{Z}{$z$}\psfrag{A}{$S_x$}\psfrag{B}{$J_x$}\psfrag{1}{$1$}\psfrag{2}{$2$}\psfrag{3}{$3$}\psfrag{4}{$4$}
 d)\includegraphics[width=5cm]{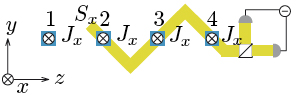} \hfill
 \psfrag{X}{$x$}\psfrag{Y}{$y$}\psfrag{Z}{$z$}\psfrag{A}{$S_x$}\psfrag{B}{$J_x$}\psfrag{1}{$1$}\psfrag{2}{$2$}\psfrag{3}{$3$}\psfrag{4}{$4$}
 e)\includegraphics[width=5cm]{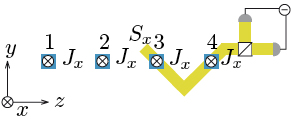}
 \caption{\small{Generation of the cluster state schematically depicted in a). The sequence of beams squeeze the following variables: b) $\hat J^{(1)}_z{}'-\hat J^{(2)}_y{}'$, c) $\hat J^{(2)}_z{}'-\hat J^{(1)}_y{}'-\hat J^{(3)}_y{}'$, d) $\hat J^{(3)}_z{}'-\hat J^{(2)}_y{}'-\hat J^{(4)}_y{}'$, e) $\hat J^{(4)}_z{}'-\hat J^{(3)}_y{}'$.}}\label{cluster}
\end{figure}
In Fig.~\ref{cluster} we show how to create the simplest 4-site (linear) cluster state. Let us introduce the new variables for each sample

\begin{eqnarray}
\hat  J^{(i)}_y{}'&=&\frac{1}{\sqrt{2}}\left( \hat J^{(i)}_y-\hat J^{(i)}_z\right),\nonumber\\
\hat J^{(i)}_z{}'&=&\frac{1}{\sqrt{2}}\left( \hat J^{(i)}_y+\hat J^{(i)}_z\right).
\end{eqnarray}
The squeezing in the combinations of the new variables is produced by passing light as depicted in Figs.~\ref{cluster}(b)-\ref{cluster}(e). For example the squeezing in $\hat J^{(1)}_z{}'-\hat J^{(2)}_y{}'$ is generated when light passes only through samples $1$ and $2$ at angles $\pm \pi/4$ respectively (see Fig.~\ref{cluster}(b)). All the other required combinations are squeezed in a similar way.

In order to verify that the state is entangled it is enough to check the set of variance inequalities given in \cite{Yukawa2008PRA}. This can be done, for example, by repetition of each step as first proposed in \cite{Julsgaard2001N,Julsgaard2003PhD}.

\section{Conclusions}

In this chapter, we have studied multipartite mesoscopic entanglement using a quantum atom-light interface in various physical setups, in particular those in which the ensembles cannot be addressed individually. Exploiting a geometric approach in which light beams propagate through the atomic samples at different angles makes it possible to establish and verify EPR bipartite entanglement and GHZ multipartite entanglement with a minimal number of light passages and measurements, so that the quantum non-demolition character of the interface is preserved. We have also shown how to generate cluster-like states by a similar technique.

Furthermore, we have shown that the multipartite entanglement created by the quantum interface of a single light beam can be appropriately tailored and even completely erased by the action of a second pulse with different intensity. This control widens the possibilities offered by measurement induced entanglement to perform quantum information tasks.

\begin{subappendices}

\section{Appendix: Detailed atom-light interactions}\label{appendix3}

\subsection{Interaction Hamiltonian}

We detail here the derivation of the effective Hamiltonian (see \cite{Hammerer2008arXiv,Sherson2006AAMP,Julsgaard2004N,Kupriyanov2005PRA}) coupling atoms and light in the off-resonant limit, neglecting absorption effects and spontaneous emission which is justified if the detuning from the optical transition is large enough. Dispersion effects can change the polarization state of the light if the sample is birefringent {\em i.e.} the index of refraction is different for orthogonal polarization components. In our scheme, $x$-polarized light propagates in the $z$-direction through the atomic samples which is polarized along the $x$-axis. Thus, $x$ is an optical axis that leaves unchanged the $x$ and $y$ polarizations of light. In what follows, we omit dispersion effects considering that the linear birefringence is zero. We will concentrate more on the continuous description of light and matter since it is convenient for describing the time dynamics of the system.

We consider real cesium atoms with its hyperfine split ground state and excited states, and coupling them off-resonantly to the $6S_{1/2} \to 6P_{3/2}$ dipole transition (see Fig.~\ref{levels}). The interaction Hamiltonian, $\hat H_{\rm{int}} = -\sum_j \hat{\vec d}_j \cdot \hat{\vec E}(\vec r_j)$, in the rotating wave approximation, eliminating adiabatically the excited states turns in

\begin{figure}[h]
 \centering
 \psfrag{A}{$6S_{1/2}$} \psfrag{B}{$6P_{1/2}$} \psfrag{C}{$6P_{3/2}$} \psfrag{D}{$F=2$} \psfrag{E}{$F=3$} \psfrag{F}{$F=4$} \psfrag{G}{$F=5$} \psfrag{H}{$-\Delta$}
 \includegraphics[width=5cm]{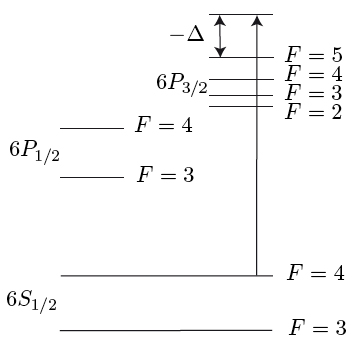}
 \caption{\small{Atomic energy levels for the cesium.}}\label{levels}
\end{figure}

\begin{eqnarray}\label{hamiltonian}
&& \hat H_{\rm{int}}(t) = -a \int_0^L (a_0 \hbar^2 \hat \phi(z,t) + a_1 \hat s_z(z,t) \hat j_z(z,t) +\nonumber \\
&& + a_2 \left[ \hbar \hat \phi (z,t) \hat j_z(z,t) - \frac{1}{\hbar}\hat s_-(z,t) \hat j^2_-(z,t) - \frac{1}{\hbar}\hat s_+(z,t) \hat j^2_-(z,t) \right] ) \rho A dz =\nonumber\\
&& = \int_0^L \hat h_{\rm{int}}(z,t) \rho A dz
\end{eqnarray} coupling the spin degrees of freedom of atoms and the Stokes vector of light.
The parameter $a=\frac{\gamma}{8 A \Delta} \frac{\lambda^2}{2\pi}$ is a coupling constant with $A$ being the cross section, $\lambda$ the wave length of light, $\Delta$ the detuning energy and $\gamma$ the frequency width of the atomic excited states. As one can see from the above expression, the detuning should not be too large for the interaction not to vanish.
The first term proportional to $a_0$ amounts for a Stark shift to all atoms proportional to the photon density $\hat \phi(z,t) = \frac{\hbar}{2}(\hat n_x + \hat n _y) = \frac{\hbar}{2} (\hat a_x^\dag\hat a_x+\hat a_y^\dag\hat a_y)$. The second term proportional to $a_1$, known as the Faraday rotation, rotate around the $z$-axis the spin vector and the Stokes vector. The last terms proportional to $a_2$ are higher orders coupling light and atoms. All these terms conserve the total angular momentum of light and atoms. As usual we have defined ladder operators for light $\hat s_\pm(z,t) = \hat s_x \pm \im \hat s_y = -\frac{\hbar}{2}\hat a^\dag_\pm \hat a_\mp$ and for angular momentum $\hat j_\pm(z,t) = \hat j_x \pm \im \hat j_y$~\footnote{Choosing $z$ as the axis of quantization, we know from any elementary book on Quantum Mechanics that \begin{eqnarray*}
 \hat j_x&=&\frac{1}{2}\sum_{m_F}\sqrt{F(F+1)-m_F(m_F+1)}\left(\ketbra{m_F+1}{m_F}+\ketbra{m_F}{m_F+1}\right),\\
 \hat j_y&=&\frac{1}{2\im}\sum_{m_F}\sqrt{F(F+1)-m_F(m_F+1)}\left(\ketbra{m_F+1}{m_F}-\ketbra{m_F}{m_F+1}\right),\\
 \hat j_z&=&\sum_{m_F} m_F \ketbra{m_F}{m_F}.
\end{eqnarray*}.}.
For the particular case of Cesium on $F=4$ and large detuning, $a_0 \to 4$, $a_1 \to 1$, $a_2 \to 0$. Additionally the Stark shift term can be suppressed by shifting the energy, thus we restrict ourselves to the linear coupling {\em i.e.} we can fix the constants to the corresponding values $a_0=a_2=0$ and $a_1=1$. Performing a time and space integration to write the Hamiltonian in terms of the macroscopical variables $\hat S_k$ and $\hat J_k = \int_0^L \hat j_k(z,t) \rho A dz$, being $\rho$ the spin density, then
\begin{equation}\label{hamiltonian0}
\hat H^{\rm{eff}}_{\rm{int}} = \frac{1}{T} \int_0^T dt \hat H_{\rm{int}}(t) =  -\frac{a}{T} \int_0^T dt \int_0^L \hat s_z(z,t) \hat j_z(z,t)  \rho A dz = -\frac{a}{T}\hat S_z \hat J_z,
\end{equation}
and we recover \eqref{halpha} for $\alpha=0$ {\em i.e.}, when the light propagates along the $z$-axis while the sample is polarized along $x$.

\subsection{Equations of evolution}

The total Hamiltonian $\hat h_{\rm{tot}} = \hat h_A + \hat h_L + \hat h_{\rm{int}}$ contains the Hamiltonian for atoms ($A$), light ($L$) and the interaction (${\rm{int}}$). The effective Hamiltonian governs the atomic dynamics and the evolution equations are derived straight through the Heisenberg equations for matter and Maxwell-Bloch equations (neglecting retardation effects) for light. We derive the equations of evolution for light an atoms in the following. 

Heisenberg evolution for light can be recasted to a Maxwell-Block evolution for the ladder operators as  follows
\begin{eqnarray*}
 \frac{\partial}{\partial t} \hat a(z,t) &=& \frac{1}{\sqrt{2\pi}} \int_{-\infty}^\infty \frac{\partial}{\partial t} \hat a(k,t) e^{\im kz} dk =\\
 &=& \frac{1}{\sqrt{2\pi}} \int_{-\infty}^\infty \frac{1}{\im \hbar} [\hat a(k,t), \hat h_A + \hat h_L + \hat h_{\rm{int}}] e^{\im kz}dk =\\
 &=& \frac{1}{\sqrt{2\pi}} \int_{-\infty}^\infty \frac{1}{\im \hbar} \left( \int_{-\infty}^\infty dk' \hbar c k' [\hat a(k,t), \hat a^\dag(k',t)]\hat a(k',t) \right) e^{\im kz} dk +\\
&+& \frac{1}{\sqrt{2\pi}} \int_{-\infty}^\infty \frac{1}{\im \hbar} [\hat a(k,t), \hat h_{\rm{int}}] e^{\im kz} dk = \frac{1}{\im \hbar}[\hat a(z,t), \hat h_{\rm{int}}] -\\ 
&-& \frac{1}{\sqrt{2\pi}} \int_{-\infty}^\infty \im c k \hat a(k,t)e^{\im kz} dk = \frac{1}{\im \hbar} [\hat a(z,t), \hat h_{\rm{int}}] - c \frac{\partial}{\partial z}\hat a(z,t),
\end{eqnarray*}
finally $(\frac{\partial}{\partial t} + c \frac{\partial}{\partial z})\hat a(z,t)= \frac{1}{\im \hbar} [\hat a(z,t), \hat h_{\rm{int}}]$.
As said, we will neglect retardation effects, {\em i.e.} we do not calculate dynamics on the time scale $L/c$ of propagation across the sample (equivalently we set $c \to \infty$).

In terms of the Stokes components then the evolution for light reads
\begin{equation}\label{MBeq}
 \frac{\partial}{\partial z}\hat s_i(z,t) = \frac{1}{\im \hbar c}[\hat s_i(z,t), \hat h_{\rm{int}}(z,t)],
\end{equation}
while for atoms, Heisenberg evolution equations are
\begin{equation}\label{Heq}
 \frac{\partial}{\partial t}\hat j_i(z,t) = \frac{1}{\im \hbar}[\hat j_i(z,t), \hat h_{\rm{int}}(z,t)].
\end{equation}

Evolution for the ortogonal $y$ and $z$ components using Eqs.~\eqref{MBeq},\eqref{Heq}, the interacting Hamiltonian \eqref{hamiltonian} in the linear regime, {\em i.e.} setting $a_0=a_2=0$ and $a_1=1$ together with the commutation rules~\footnote{For light $\comm{S_i}{S_j} = [\int^T_0 \hat s_i(t) dt, \int^T_0 \hat s_j(t') dt']= \im \hbar\int^T_0\int^T_0 dt dt' \epsilon_{ijk}\hat s_k(t) \delta(t-t') = \im \hbar \epsilon_{ijk}\int^T_0 \hat s_k(t) dt = \im \hbar \epsilon_{ijk}\hat S_k$ and atoms $\comm{j_i(z)}{j_j(z)} = \im \hbar \epsilon_{ijk}\hat j_k(z)$.} gives rise to

\begin{eqnarray}
&& \frac{\partial}{\partial t}\hat j_y(z,t) = -a \hat s_z(z,t) \hat j_x,\\
&& \frac{\partial}{\partial t}\hat j_z(z,t) = 0,
\end{eqnarray}
\begin{eqnarray}
&& \frac{\partial}{\partial z}\hat s_y(z,t) = -\frac{a}{c} \hat s_x \hat j_z(z,t) \delta(t-t'),\\
&& \frac{\partial}{\partial z}\hat s_z(z,t) = 0.
\end{eqnarray}
Strong $x$-polarized atoms and light imposes $\hat j_x(z,t) = j_x$ and $\hat s_x(z,t) = s_x$, thus

\begin{eqnarray}
 && \hat j_x(z,t) = j_x,\\
 && \frac{\partial}{\partial t}\hat j_y(z,t) = -a \hat s_z(t) j_x,\\
 && \hat j_z(z,t) = \hat j_z(z),
\end{eqnarray}
\begin{eqnarray}
 && \hat s_x(z,t) = s_x,\\
 && \frac{\partial}{\partial z}\hat s_y(z,t) = -\frac{a}{c} s_x \hat j_z(z) \delta(t-t'),\\
 && \hat s_z(z,t) = \hat s_z(t).
\end{eqnarray}
Finally, we integrate in space ($\int^L_0 \rho A dz$) and time ($\int^T_0 dt$) to recover the equations of evolution for the macroscopical variables of atoms and light 

\begin{eqnarray}
 && J_x=\int^L_0 \hat j_x(z,t)\rho A dz = j_x,\\
 && \hat J_y(t=T) - \hat J_y(t=0) = -a \hat S_z J_x,\\
 && \hat J_z=\int^L_0 \hat j_z(z)\rho A dz = J_z(t=0) = J_z(t=T),
\end{eqnarray}
\begin{eqnarray}
 && S_x=\int^T_0 \hat s_x(z,t)dt = s_x T,\\
 && \hat S_y(z=L)-\hat S_y(z=0) = -a S_x \hat J_z,\\
 && \hat S_z=\int^T_0 \hat s_z(t) dt= \hat S_z(z=0) = \hat S_z(z=L).
\end{eqnarray}

In the following, it will be useful to think in terms of input/output variables, {\em i.e.} we define the operators $\hat{S}_k^{\rm{in/out}}$ as the Stokes operators characterizing the pulse entering ($z=0$) and leaving ($z=L$) the atomic sample. Analogously, $\hat J_k^{\rm{in/out}}$ correspond to initial ($t=0$) and final state ($t=T$) of the atomic spin. In this way we can write the evolution as a linear equation system

\begin{eqnarray}
&& \hat J_y^{\rm{out}} = \hat J_y^{\rm{in}}-a\hat{S}_z^{\rm{in}} J_x,\\
&& \hat J_z^{\rm{out}} = \hat J_z^{\rm{in}},
\end{eqnarray}
\begin{eqnarray}
&& \hat S^{\rm{out}}_y = \hat S^{\rm{in}}_y - a S_x \hat{J}_z^{\rm{in}},\\
&& \hat S^{\rm{out}}_z = \hat S^{\rm{in}}_z,
\end{eqnarray}
recovering Eqs.~\eqref{Jy}-\eqref{Sz} for $\alpha=0$ {\em i.e.}, when the light propagates along the $z$-axis while the sample is polarized along $x$.

\end{subappendices}


\chapter{Conclusions}\label{conclusions}

Summarizing we have studied several quantum protocols with Continuous Variable (CV) systems giving special importance to the efficiency such protocols can achieve when considering current experimental possibilities. Independently of future technological improvements, noise is intrinsically ascribed to any measurement and manipulation. We have taken into account that, but also we have considered non-ideal Gaussian resources and imperfections on the experimental realization. Regarding Gaussian multipartite entanglement, we have used entanglement induced measurement schemes, for the creation and manipulation of multipartite entanglement while proposing a possible candidate physical system for its realization. Finally, motivated by the performance enhancement offered by non-Gaussian states for communication tasks, we have analyzed and proposed a correlation measure based on quadrature correlations. Our measure, provides an excellent quantification of correlations not only for Gaussian states but also for non-Gaussian states, where determination of entanglement is normally not known. Furthermore, our measure has a low computable cost compared to other methods which require a full tomographic analysis of the state.

Specifically, we have first shown that the sharing of entangled Gaussian variables and the use of only Gaussian operations permits efficient Quantum Key Distribution against individual and finite coherent attacks. We have used the fact that all mixed NPPT symmetric states can be used to extract secret bits to design an algorithm, that efficiently succeeds for a secure extraction of a key. Whereas under individual attacks all mixed NPPT symmetric states admit a finite efficiency, for finite coherent attacks an additional condition constrains the parameters of the states. We have introduced a figure of merit (the efficiency $E$) to quantify the number of classical correlated bits that can be used to distill a key from a sample of $M$ entangled states. We have observed that this quantity grows with the entanglement shared between Alice and Bob. 

Secondly, we have proposed a protocol to solve detectable broadcast with entangled Continuous Variable using Gaussian states and Gaussian operations only. Our protocol relies on genuine multipartite entanglement distributed among the three parties, which specifically have to share two copies of a three-mode fully symmetric Gaussian state. Interestingly, we have found that nevertheless not all entangled symmetric Gaussian states can be used to achieve a solution to detectable broadcast: a minimum threshold exists on the required amount of multipartite entanglement. We have moreover analyzed in detail the security of the protocol. In its ideal formulation, our protocol requires that the parties share pure resource states, and that the outcomes of homodyne detections are perfectly coincident and not affected by any uncertainty; this however entails that our protocol achieves a solution with vanishing probability. To overcome such a practical limitation, we have eventually considered a more realistic situation in which firstly the tripartite Gaussian resources are affected by thermal noise, and, more importantly, the homodyne detections are realistically imperfect, and secondly there is a finite range of allowed values for the measurement outcomes obtained by the parties. We have thoroughly investigated the possibility to solve detectable broadcast via our protocol under these relaxed conditions. As a result, we have demonstrated that there exists a broad region in the space of the relevant parameters (noise, entanglement, range of the measurement shift, measurement uncertainty) in which the protocol admits an efficient solution. This region encompasses amounts of the required resources which appear attainable with the current optical technology (with a legitimate trade-off between squeezing and losses). We can thus conclude that a feasible, robust implementation of our protocol to solve detectable broadcast with entangled Gaussian states may be in reach. 

Motivated by the relation between the entanglement and the distillation of classically correlated bits we analyzed which is the maximal number of correlated bits ($Q$) that can be extracted from a CV state via quadrature measurements. We have provided an operational quantification of the entanglement content of several relevant non-Gaussian states (including the useful photon-subtracted states). Crucially, one can experimentally measure $Q$ by direct homodyne detections (of the quadratures displaying optimal correlations only), in contrast to the much more demanding full tomographical state reconstruction. One can then easily invert the (analytic or numeric) monotonic relation between $Q$ and the negativity to achieve a direct entanglement quantification from the measured data. Our analysis demonstrates the rather surprising feature that entanglement in the considered non-Gaussian states can thus be detected and experimentally quantified with the same complexity as if dealing with Gaussian states. In this respect, it is even more striking that the measure considered in this paper, based on (and accessible in terms of) second moments and homodyne detections only, provides such an exact quantification of entanglement in a broad class of pure and mixed non-Gaussian states, whose quantum correlations are encoded nontrivially in higher moments too, and currently represent the preferred resources in CV Quantum Information. We focused on optical realizations of CV systems, but our framework equally applies to collective spin components of atomic ensembles, and radial modes of trapped ions. Finally, it is also surprising that for all these family of states we have studied, the optimization of just one quadrature scales monotonically with the negativity of the state. Although this could be expected for pure Gaussian states, our study demonstrated that the non-Gaussian states obtained either as de-gaussifications of pure Gaussian states or mixings with uncorrelated states preserve this property. 

Finally, we have studied multipartite mesoscopic entanglement using a quantum atom-light interface in various physical setups, in particular those in which the ensembles cannot be addressed individually. Exploiting a geometric approach in which light beams propagate through the atomic samples at different angles makes it possible to establish and verify EPR bipartite entanglement and GHZ multipartite entanglement with a minimal number of light passages and measurements, so that the quantum non-demolition character of the interface is preserved. We have also shown how to generate cluster-like states by a similar technique. Furthermore, we have shown that the multipartite entanglement created by the quantum interface of a single light beam can be appropriately tailored and even completely erased by the action of a second pulse with different intensity. This control widens the possibilities offered by measurement induced entanglement to perform quantum information tasks.


\bibliography{Tesis}
\bibliographystyle{unsrt}

\addcontentsline{toc}{chapter}{Bibliography}

\end{document}